\documentclass[reqno, 11pt]{amsart}

\usepackage{amsfonts,amssymb,amsbsy,amsmath,amsthm,enumerate}
\usepackage{txfonts}
\usepackage{eucal}
\usepackage{dsfont}
\usepackage{mathrsfs}

\usepackage{bbm}

\usepackage{stackrel}

\usepackage[small,nohug,
]{diagrams}
\diagramstyle[labelstyle=\scriptstyle]
\newarrow{Corresponds}<--->

\usepackage[dvips]{graphicx}
\usepackage{color}

\newtheorem{theorem}{Theorem}[section]
\newtheorem{proposition}[theorem]{Proposition}
\newtheorem{lemma}[theorem]{Lemma}
\newtheorem{corollary}[theorem]{Corollary}
\newtheorem{assumption}[theorem]{Assumption}
\newtheorem{definition}[theorem]{Definition}
\newtheorem{remark}[theorem]{Remark}

\newtheorem{example}[theorem]{Example}

\numberwithin{equation}{section}

\numberwithin{figure}{section}
\numberwithin{table}{section}

\newcommand\beq{\begin{equation}}
\newcommand{\bea}{\begin{eqnarray}}
\newcommand{\eea}{\end{eqnarray}}
\newcommand{\beas}{\begin{eqnarray*}}
\newcommand{\eeas}{\end{eqnarray*}}
\newcommand{\beql}{\begin{equation} \label}
\newcommand{\eeq}{\end{equation}}

\newcommand{\R}{\mathds R}
\newcommand{\N}{\mathds N}
\newcommand{\C}{\mathds C}                           

\newcommand{\Z}{\mathds Z}
\newcommand{\T}{\mathds T}

\newcommand{\s}[1]{\CMcal{#1}}
\newcommand{\f}[1]{\mathcal{#1}}                  
\newcommand{\bb}[1]{\mathscr{#1}}
\newcommand{\rr}[1]{\mathfrak{#1}}
\newcommand{\n}[1]{\mathds {#1}}

\newcommand{\expo}[1]{{\rm e}^{#1}}                 

\newcommand{\ii}{\,{\rm i}\,}
\newcommand{\ncint}{\mathrel{{\ooalign{$\int$\cr\kern+.07em\raise.15ex\hbox{$\pmb{\scriptstyle-}$}\cr}}}}           \newcommand{\ncpartial}{\mathrel{{\ooalign{$\partial$\cr\kern+.29em\raise.79ex\hbox{$\pmb{\scriptstyle-}$}\cr}}}}

\newcommand{\virg}[1]{\lq\lq#1\rq\rq}                
\newcommand{\ie}{{\sl i.\,e.\,}}
\newcommand{\eg}{{\sl e.\,g.\,}}
\newcommand{\cf}{{\sl cf.\,}}

\begin{document}

\title[A cohomological generalization of the Fu-Kane-Mele index]{
The cohomological nature of the Fu-Kane-Mele invariant}


\author[G. De~Nittis]{Giuseppe De Nittis}
\address[De~Nittis]{Facultad de Matem\'aticas \& Instituto de F\'{\i}sica,
Pontificia Universidad Cat\'olica,
Santiago, Chile}
\email{gidenittis@mat.uc.cl}

\author[K. Gomi]{Kiyonori Gomi}
\address[Gomi]{Department of Mathematical Sciences, Shinshu University,  Nagano, Japan}
\email{kgomi@math.shinshu-u.ac.jp}




\begin{abstract}
\vspace{-4mm}
In this paper we generalize the definition of the FKMM-invariant introduced in \cite{denittis-gomi-14-gen}
for the case of \virg{Quaternionic} vector bundles over involutive base spaces endowed with  free involution or with a non-finite fixed-point set. In \cite{denittis-gomi-14-gen} it has already be shown how the  FKMM-invariant provides a cohomological description of the Fu-Kane-Mele index used to classify topological insulators in class AII. It follows that the   
FKMM-invariant described in this paper provides a cohomological generalization of the Fu-Kane-Mele index which is applicable to the classification of protected phases for other type of topological quantum systems (TQS) which are not necesarily related to models for  topological insulators (e.g. the two-dimensional  models of adiabatically perturbed  systems discussed in \cite{gat-robbins-15}).
As a byproduct we provide the complete classification of  \virg{Quaternionic} vector bundles  over a big class of low dimensional involutive spheres and tori. 
\end{abstract}


\maketitle

\vspace{-5mm}
\tableofcontents

\section{Introduction}\label{sect:intro}

In its simplest incarnation, a \emph{Topological Quantum System} (TQS) is a continuous matrix-valued map 
\begin{equation}\label{eq:intro_tqs0}
X\;\ni\;x\;\longmapsto\; H(x)\;\in\; {\rm Mat}(\C^d)
\end{equation}
defined on a topological space $X$ sometimes called \emph{Brillouin zone}. 
 Although the precise definition of TQS requires some more ingredients {and  can be stated in a more general form}  \cite{denittis-gomi-14,denittis-gomi-14-gen,denittis-gomi-15}, 
one can certainly
state that
the most relevant feature of these systems is the nature of the spectrum which is made by continuous (energy) bands.
It is exactly the {family of eigenprojectors emerging from} this peculiar band structure which may encode information that are of topological nature. The study and the classification of the topological properties of TQS have recently risen  to the level of \virg{hot topic} in mathematical physics because its connection with the study of \emph{topological insulators} in condensed matter 
(we refer to the two reviews  \cite{hasan-kane-10} and \cite{ando-fu-15} for a modern overview about topological insulators and an updated bibliography of the most relevant publications on the subject). However, systems like \eqref{eq:intro_tqs0} are  
ubiquitous in
the mathematical physics  and are not necessarily linked with problem coming from condensed matter.

\medskip

For the sake of simplicity we consider here   {a very specific and simple} realization of a TQS:
\begin{equation}\label{eq:intro_tqs1}
H(x)\;:=\;\sum_{j=0}^{2N}F_j(x)\;\Sigma_j\;.
\end{equation}
The $\{\Sigma_0,\ldots,\Sigma_{2N}\}\in{\rm Mat}(\C^{2^N})$  define a (non-degenerate) irreducible representation of the \emph{complex Clifford algebra} ${\rm Cl}_\C(2N+1)$
and the real-valued functions $F_j:X\to \R$, with $j=0,1,\ldots ,2N$, are assumed to be continuous. 
Under the \emph{gap condition}
\begin{equation}\label{eq:intro_tqs2}
{Q}(x)\;:=\;\sum_{j=0}^{2N}F_j(x)^2\;>\;0
\end{equation}
one can associate  the negative (or positive) part of the band spectrum of  $x\mapsto H(x)$ with a complex vector bundle $\bb{E}\to X$ of rank $2^{N-1}$, called \emph{{spectral bundle}}\footnote{{Spectral bundles obtained as a result of a Bloch-Floquet transform, are sometimes called \emph{Bloch bundle} \cite{panati-schulz-baldes-03,panati-07}}}. For the details of this identification we refer to Section \ref{sect:non-trivial_ex} as well as to 
{\cite[Section IV]{denittis-lein-11} or \cite[Section 2]{denittis-gomi-14} (the specific model \eqref{eq:intro_tqs1} is considered in \cite[Section 4]{denittis-lein-13})}. The considerable consequence of the duality between  gapped TQS's  and  spectral  bundles  is that one can classify the possible topological phases of the TQS by means of the  elements of the set ${\rm Vec}_{\C}^{2^{N-1}}(X)$ given by the isomorphism classes of all rank  $2^{N-1}$ vector bundles over $X$. Therefore, in this general setting, the classification problem for the topological phases of a TQS like \eqref{eq:intro_tqs1} can be traced back to a classic problem in topology which has been elegantly solved in \cite{peterson-59}. For instance, in the (physically relevant) case of a \emph{low dimensional} 
base space $X$, the complete classification of the topological phases 
of \eqref{eq:intro_tqs1} is provided by
\begin{equation}\label{eq:intro_tqs3}
{\rm Vec}_{\C}^{2^{N-1}}(X)\;\stackrel{c_1}{\simeq}\;H^2\big(X,\Z\big)\;,\qquad\quad {\rm dim}(X)\leqslant 3
\end{equation}
where on the right-hand side one has the second singular cohomology group of $X$ and the map $c_1$ is the \emph{first Chern class}.

\medskip

The problem of the classification of the topological phases becomes more interesting, and challenging, when the TQS is constrained by the presence of certain symmetries or \emph{pseudo-symmetries}. Among the latter, the \emph{time-reversal symmetry} (TRS) attracted recently a considerable  interest in mathematical physics community. A system like \eqref{eq:intro_tqs1} is said to be time-reversal symmetric if there is an  \emph{involution} $\tau:X\to X$ on the base space and an \emph{anti}-unitary map $\Theta$ such that
\begin{equation}\label{eq:intro_tqs3bis}
\left\{
\begin{aligned}
\Theta\;H(x)\;\Theta^*\;&=\; H\big(\tau(x)\big)\;,&\quad\qquad \forall\ x\in\ X\;\\
\Theta^2\;&=\;\epsilon\;\n{1}_{2^N}& \epsilon=\pm1\;.
\end{aligned}
\right.
\end{equation}

\medskip

The case $\epsilon=+1$ corresponds to an even (or bosonic) TRS. In this event, the spectral 
vector bundle $\bb{E}$ turns out to be equipped with an additional structure named \emph{\virg{Real}} by  {M.~F. Atiyah} in \cite{atiyah-66}. Therefore, in the presence of an even TRS the   classification problem of the topological phases  is reduced to the study of the set  
${\rm Vec}_{\rr{R}}^{2^{N-1}}(X,\tau)$ of isomorphism classes of rank  $2^{N-1}$ vector bundles over $X$ endowed with a \virg{Real} structure. This problem has been analyzed and solved in \cite{denittis-gomi-14}. In particular, in the low dimensional case one has that
\begin{equation}\label{eq:intro_tqs4}
{\rm Vec}_{\rr{R}}^{2^{N-1}}(X,\tau)\;\stackrel{c_1^{\rr{R}}}{\simeq}\;H^2_{\Z_2}\big(X,\Z(1)\big)\;,\qquad\quad {\rm dim}(X)\leqslant 3\;.
\end{equation}
Here, in the right-hand side there is the equivariant Borel cohomology with local coefficient system $\Z(1)$ of the involutive space $(X,\tau)$ (see Section \ref{subsec:borel_cohom} and references therein for more details) and the isomorphism is given by \emph{first \virg{Real} Chern class} $c_1^{\rr{R}}$ introduced for the first time by B. Kahn in \cite{kahn-59} (see also \cite[Section 5.6]{denittis-gomi-14}). Note the close similarity between the equations \eqref{eq:intro_tqs3} and \eqref{eq:intro_tqs4}.

\medskip

The case $\epsilon=-1$ describes an odd (or fermionic) TRS. Also in this situation the spectral 
vector bundle $\bb{E}$ acquires an additional structure called \emph{\virg{Quaternionic}} (or symplectic \cite{dupont-69}) and the topological phases of a TQS with an odd TRS turn out to be labelled  by the set ${\rm Vec}_{\rr{Q}}^{2^{N-1}}(X,\tau)$ of isomorphism classes of rank  $2^{N-1}$ vector bundles over $X$ with \virg{Quaternionic} structure. 

\medskip

The study of   systems with an odd TRS is more interesting, and for several reasons also harder, than the case of an even TRS. Historically, the fame of these fermionic systems begins with the seminal papers \cite{kane-mele-05, fu-kane-mele-95}  by 
L. Fu,  C. L. Kane and E. J. Mele. The central result of these works is the interpretation of a physical phenomenon called \emph{Quantum Spin Hall Effect} as the evidence  of a non-trivial topology for TQS constrained by an odd TRS. Specifically, the papers \cite{kane-mele-05, fu-kane-mele-95} are concerned about the study of systems like \eqref{eq:intro_tqs1} (with $N=2$) where the base space is a torus of dimension 2 or 3 endowed with a \virg{time-reversal} involution. These involutive \emph{TR-tori} are  the quotient spaces $(\R/\Z)^d$ endowed with the quotient involution defined on $\R$ by $x\to-x$ (we  denote these spaces with   ${\T}^{0,d,0}$ according to a general notation which will be clarified in equation \eqref{eq:intro_inv_tori}). The distinctive aspect of the spaces ${\T}^{0,d,0}$ is the existence of a \emph{fixed point set} formed by $2^d$ isolated points. The latter plays a crucial role in the classification scheme proposed in \cite{kane-mele-05, fu-kane-mele-95} where the different topological phases are distinguished by the signs that a particular function $\rr{d}_{\bb{E}}$ defined by  the spectral bundle (essentially the inverse of a normalized pfaffian) takes on the $2^d$ points fixed by the involution.
These numbers are usually known as \emph{Fu-Kane-Mele indices}. 

\medskip

In the last years the problem of the topological classification  of systems with an odd TRS has been discussed with several different approaches.  
As a matter of fact, many (if not almost all) of these approaches focus on the particular cases $\n{T}^{0,2,0}$ and $\n{T}^{0,3,0}$  with the aim of reproducing in different way the $\Z_2$-invariants described by the {Fu-Kane-Mele indices}.
From one hand there are classification schemes based on \emph{K-theory}  and \emph{$KK$-theory} \cite{kitaev-09,freed-moore-13,thiang-16,kellendonk-15,kubota-17,bourne-carey-rennie-16,prodan-schulz-baldes-16}
or \emph{equivariant homotopy} techniques \cite{kennedy-guggenheim-15,kennedy-zirnbauer-16} which are extremely general. In the opposite side there are constructive procedures based on the interpretation of the topological phases as \emph{obstructions} for the construction of continuous time-reversal symmetric frames (see \cite{porta-graf-13} for the case ${\T}^{0,2,0}$ and \cite{fiorenza-monaco-panati-14,monaco-cornean-teufel-16} for the generalization to the case ${\T}^{0,3,0}$)
or as \emph{spectral-flows} \cite{carey-phillips-schulz-baldes-16,denittis-schulz-baldes-13} or as \emph{index pairings} \cite{grossmann-schulz-baldes-15}
{or by means of the \emph{Wess-Zumino amplitudes} \cite{carpentier-delplace-fruchart-gawedzki-15,carpentier-delplace-fruchart-gawedzki-tauber-15,monaco-tauber-17}.}
All these approaches, in our opinion, present some limitations. The $K$-theory 
is unable to distinguish the \virg{spurious phases} (possibly) present outside of the \emph{stable rank} regime. The homotopy calculations are non-algorithmic and usually extremely hard. The use of the $KK$-theory 
up to now is restricted only to  the non-commutative version of the involutive Brillouin tori ${\T}^{0,2,0}$ and 
${\T}^{0,3,0}$. A similar consideration holds for the  recipes for the \virg{handmade} construction of {global continuous} frames which are strongly dependent of the specific form of the involutive spaces ${\T}^{0,2,0}$ and 
${\T}^{0,3,0}$, and thus are  difficult to generalize  to other spaces and higher dimensions. Finally, none of these approaches clearly identifies the invariant which labels the different phases as a \emph{topological (characteristic) class}\footnote{
{The expression \emph{topological (characteristic) class} is used here to denote an element which can be defined for each element in the
 (topological) category of \virg{Quaternionic} vector bundles 
  and which is \emph{natural} with respect to the pullback in this category.}},  as it happens in the cases of systems with broken TRS (\cf eq. \eqref{eq:intro_tqs3}) or with even TRS (\cf eq. \eqref{eq:intro_tqs4}).

\medskip

To overcome these deficiencies we developed  in \cite{denittis-gomi-14-gen} an alternative classification scheme  based on a cohomological description. By adapting an idea of M. Furuta,  Y. Kametani, H. Matsue, and N. Minami \cite{furuta-kametani-matsue-minami-00}, we introduced a cohomological class $\kappa$ called  \emph{FKMM-invariant}, 
that we have shown to be a characteristic class for the category of \virg{Quaternionic} vector bundles. In the low dimensional case this invariant provides an \emph{injection}
\begin{equation}\label{eq:intro_tqs5}
{\rm Vec}_{\rr{Q}}^{2^{N-1}}(X,\tau)\;\stackrel{\kappa}{\longrightarrow}\;H^2_{\Z_2}\big(X|X^\tau,\Z(1)\big)\;,\qquad\quad {\rm dim}(X)\leqslant 3\;.
\end{equation}
The meaning of the cohomology group which appears in right-hand side can be guessed by looking at the exact sequence (\cf Section \ref{subsec:borel_cohom})
$$
\ldots\;\longrightarrow\;\big[X^\tau,\{\pm 1\}\big]\;\stackrel{}{\longrightarrow}\;H^2_{\Z_2}\big(X|X^\tau,\Z(1)\big)\;\stackrel{}{\longrightarrow}\;{\rm Pic}_{\rr{R}}\big(X,\tau\big)\;\stackrel{r}{\longrightarrow}\;{\rm Pic}_{\R}\big(X^\tau\big)\;\longrightarrow\;\ldots\;
$$
where 
$[X^\tau,\{\pm 1\}]$ is the set of homotopy classes of maps from $X^\tau$ into $\{\pm 1\}$,
${\rm Pic}_{\rr{R}}(X,\tau)={\rm Vec}_{\rr{R}}^1(X,\tau)$ is the Picard group of the \virg{Real} line bundles over $(X,\tau)$, ${\rm Pic}_{\R}(X^\tau)\simeq {\rm Pic}_{\rr{R}}(X^\tau,\tau)$ is the Picard group of the real line bundles over $X^\tau$ and $r$ is the 
restriction map. 

\medskip

Despite its similarity with the equations \eqref{eq:intro_tqs3} and \eqref{eq:intro_tqs4}, the equation \eqref{eq:intro_tqs5}  cannot be considered  completely satisfactory, at least as derived in \cite[Theorem 1.1]{denittis-gomi-14-gen}.
The most dramatic weakness
of the   \eqref{eq:intro_tqs5}   in its original form 
is that it has been proved only under the hypothesis of $(X,\tau)$ being an FKMM-space \cite[Definition 1.1]{denittis-gomi-14-gen}. This is a strong restriction since: (i) it excludes involutive spaces with fixed point sets $X^\tau$ of dimension \emph{bigger than zero} and involutive spaces with a \emph{free involution} (\ie $X^\tau=\text{\rm  \O}$); (ii) it introduces by \virg{brute force} the condition ${\rm Pic}_{\rr{R}}\big(X,\tau\big)=0$ which implies the isomorphism \cite[Lemma 3.1]{denittis-gomi-14-gen}
\begin{equation}\label{eq:intro_tqs5bis}
H^2_{\Z_2}\big(X|X^\tau,\Z(1)\big)\;\simeq\;\big[X^\tau,\{\pm 1\}\big] /\big[X,\n{U}(1)\big]_{\Z_2}
\end{equation}
where $[X,\n{U}(1)]_{\Z_2}$ denotes the set of the homotopy classes of $\Z_2$-equivariant maps from $(X,\tau)$ into $\n{U}(1)$ endowed with the involution given by the complex conjugation.
Clearly, the isomorphism \eqref{eq:intro_tqs5bis} fails in general for  involutive spaces $(X,\tau)$ which are not of FKMM-type. On the other hand, \eqref{eq:intro_tqs5bis} is also the principal ingredient  to link the {FKMM-invariant} $\kappa$ with  the (strong) {Fu-Kane-Mele index} \cite[Theorem 4.2]{denittis-gomi-14-gen}  through the well-known formula 
\begin{equation}\label{eq:intro_tqs5bisbis}
\kappa(\bb{E})\;=\;\prod_{x\in X^\tau}\rr{d}_{\bb{E}}(x)\;,\qquad\quad (X,\tau)\ \ \ \text{an FKMM-space}\;.		
\end{equation}
There is also a second, more delicate, reason which makes \eqref{eq:intro_tqs5} weaker
than \eqref{eq:intro_tqs3} or  \eqref{eq:intro_tqs4}. In fact  \eqref{eq:intro_tqs5}
 provides, in general, only an injection while the other two are
 isomorphisms. This topic will be discussed in Section \ref{sect:surject_FKMM}.

\medskip

The main result of this paper  is the extension of the range of validity of \eqref{eq:intro_tqs5}
to  involutive spaces $(X,\tau)$ in full generality. More precisely, we introduce  a \emph{generalized} version of the FKMM-invariant which extends the \virg{old} invariant constructed in  \cite{denittis-gomi-14-gen} to involutive spaces 
which are not necessary of FKMM-type. 
This generalization has a pay-off: The \eqref{eq:intro_tqs5}  becomes meaningful (and valid)  for spaces with fixed point set of any co-dimension. This is, in our opinion, a big step forward in the theory of the classification of topological phases for systems with odd TRS. In fact, motivated by the \eqref{eq:intro_tqs5bisbis}, we can interpret $\kappa$ as the extension of the  Fu-Kane-Mele indices when $X^\tau$   is not simply a collection of isolated points. {Also the case $X^\tau=\text{\rm  \O}$ can be  handled
inside our formalism: Not only is the \emph{generalized}  FKMM-invariant well defined 
for even-rank \virg{Quaternonic} vector bundles over  base spaces with free involution, but we can also define a  FKMM-invariant for odd-rank \virg{Quaternonic} vector bundles
and classify them.}
\medskip

Before discussing some interesting consequences of the  \eqref{eq:intro_tqs5} in its generalized meaning, let us spend few words about the importance of considering spaces different form the TR-tori ${\T}^{0,d,0}$.
If from one hand the spaces ${\T}^{0,d,0}$ emerge naturally as quasi-momentum spaces (a.k.a. Brillouin zone) for $d$-dimensional periodic electronic systems, on the other hand 
TQSs  of type \eqref{eq:intro_tqs1}
are ubiquitous in mathematical physics and  are not necessarily related to model in condensed matter (see e.g. the rich monographs \cite{bohm-mostafazadeh-koizumi-niu-zwanziger-03,chruscinski-jamiolkowski-04}). Just to give few examples let us mention that TQSs  can be used to model
systems subjected to  \emph{cyclic adiabatic processes} in  classical and quantum mechanics \cite{pancharatnam-56,berry-84}, or in the description of the \emph{magnetic monopole} \cite{dirac-31, yang-96}
and  the \emph{Aharonov-Bohm effect} \cite{aharonov-bohm-59} 
or in the molecular dynamics in the context of the  \emph{Born-Oppenheimer approximation} \cite{teufel-03}. 
Usually, the space $X$ plays the role of a \emph{configuration} space for {parameters} which describe an \emph{adiabatic} action of external fields on a system governed by the instantaneous
 Hamiltonian $H(x)$.
The phenomenology of these systems is  enriched by the presence of certain symmetries like a TRS as in \eqref{eq:intro_tqs3bis}. Models of \emph{adiabatic} topological systems of this type have been recently investigated in 
\cite{carpentier-delplace-fruchart-gawedzki-15,carpentier-delplace-fruchart-gawedzki-tauber-15,gat-robbins-15}.
In particular, in the second of these works \cite{gat-robbins-15} the authors 
consider the adiabatically perturbed dynamics of a \emph{classic rigid rotor} and a \emph{classical  particle on a ring}. In the first case the classical phase space turns out to be a two-dimensional sphere endowed with an antipodal involution induced by the
 TRS (we use  $\n{S}^{0,3}$ for this space). In the second case the phase space is a two dimensional torus $\T^2=\n{S}^1\times\n{S}^1$ where the TRS induces a time-reversal involution only on one of its factors (we denote  this space by $\T^{1,1,0}$). The main result of 
\cite{gat-robbins-15} consists in  the classification of \virg{Quaternionic} vector bundles  
over the involutive spaces $\n{S}^{0,3}$ and $\T^{1,1,0}$ and it is based on the analysis of the obstruction for
the \virg{handmade} construction of a {global} frame. As already commented above, this 
technique is hard (and tricky) to extend to higher dimensions and other involutive spaces.
Conversely, the classification provided by the map \eqref{eq:intro_tqs5} turns out to be 
extremely effective and versatile. In fact it is algorithmic! As a matter of fact the formula \eqref{eq:intro_tqs5} allows us to classify \virg{Quaternionic} vector bundles over   a big class of involutive spheres and tori up to dimension three extending, in this way, the results in  \cite{gat-robbins-15}. The cost for such a classification 
amounts to just a little more than the (algorithmic) calculation of the cohomology group in  \eqref{eq:intro_tqs5}. We postpone a synthesis of the results of this classification at the end of this introductory section. As a final remark let us mention that recently the classification technique based on the FKMM-invariant has been successfully used in \cite{thiang-sato-gomi-17} in order {to classify  Weyl semimetals with TRS.}

\medskip

The map \eqref{eq:intro_tqs5} can be used also to obtain information about the possibility of non trivial topological states for  systems of  type \eqref{eq:intro_tqs1} independently of the details of the functions $F_j$. For instance, the study of the case $N=2$ in Section \ref{sect:non-trivial_ex} leads to the following general result:
\begin{theorem}[Necessary condition for non-trivial phases]
Consider a topological quantum system of  type \ref{eq:intro_tqs1} with $N=2$ and endowed with an odd TRS of type \eqref{eq:intro_tqs3bis}. A necessary condition for the existence of non-trivial phases 
is that $F_2$ and \emph{at least} two of the functions $\{F_0,F_1,F_3,F_4\}$ must be  \emph{not} identically zero.
\end{theorem}
\noindent
For sake of precision, one has to refer the statement of the theorem above 
 to the definitions and notations introduced in Section \ref{sect:non-trivial_ex}. Note that this result is just a (partial) summary of the content of Proposition \ref{propo:nec1} and Proposition \ref{propo:nec2}.

\medskip

This paper is organized as follows: In {\bf Section \ref{sect:top_class}} we construct the \emph{generalized} version of the FKMM-invariant and  we prove that this is a characteristic class for the category of \virg{Quaternionic} vector bundles. Moreover, we show that it reduces to the \virg{old version} of the  FKMM-invariant described in \cite{denittis-gomi-14-gen}, and so to the  {Fu-Kane-Mele indices} in the case of spaces with a discrete number of fixed points. {\bf Section \ref{sect:quat_lin_bund}} is devoted to the study of \virg{Quaternionic} line bundles.
In {\bf Section \ref{sect:top_class_low}} we use the \virg{new} FKMM-invariant to classify 
\virg{Quaternionic} vector bundles {of even and odd rank} over low dimensional involutive spaces.  We also discuss the classification over involutive spheres and tori. {\bf Section \ref{sect:non-trivial_ex}} is devoted to the study of TQS of type \eqref{eq:intro_tqs1} with $N=2$. Finally, {\bf Appendices \ref{app:Cohom_comput}, \ref{sect:cohom_sper}} and {\bf \ref{sect:cohom_tori}} contain all the cohomology computations  needed in the paper. These calculations are technical, but also of general validity and may also be useful in other areas of the mathematical research.

\medskip
\noindent
{\bf Acknowledgements.} 
GD's research is supported
 by
the  grant \emph{Iniciaci\'{o}n en Investigaci\'{o}n 2015} - $\text{N}^{\text{o}}$ 11150143 funded  by FONDECYT.	 KG's research is supported by 
the JSPS KAKENHI Grant Number 15K04871.
\medskip

\medskip

\noindent
{\bf Resum\'{e} of the classification over involutive spheres and tori.}
Let us fix some notations:  The \emph{involutive sphere} of type $\n{S}^{p,q}:=(\n{S}^{p+q-1},\theta_{p,q})$ is defined as the 
sphere of dimension $d:=p+q-1$
$$
\n{S}^{d}\;:=\;\left\{(k_0,k_1,\ldots,k_d)\in\R^{d+1}\ |\ k_0^2+k_1^2+\ldots+k_d^2=1\right\}
$$
endowed with the involution
$$
\theta_{p,q}(\underbrace{k_0,k_1,\ldots,k_{p-1}}_{\text{first}\; p\; \text{coord.}},\underbrace{k_{p},k_{p+1},\ldots,k_{p+q-1}}_{\text{last}\; q\; \text{coord.}})\;:=\;(\underbrace{k_0,k_1,\ldots,k_{p-1}}_{\text{first}\; p\; \text{coord.}},\underbrace{-k_{p},-k_{p+1},\ldots,-k_{p+q-1}}_{\text{last}\; q\; \text{coord.}})
$$
{The cases $p=0$ or $q=0$ correspond to the absence of the related  block of coordinates}.
The space $\n{S}^{d+1,0}$ coincides with the sphere
$\n{S}^{d}$ endowed with the \emph{trivial involution} fixing all points. On the opposite side one has the
space $\n{S}^{0,d+1}$ which is the  sphere $\n{S}^{d}$
endowed with the \emph{antipodal} (or \emph{free}) \emph{involution}. The space $\n{S}^{1,d}$ (denoted with the symbol $\tilde{\n{S}}^d$ in \cite{denittis-gomi-14,denittis-gomi-14-gen,denittis-gomi-15})
coincides with the \virg{Brillouin zone} of condensed matter systems invariant under continuous translations and subjected to a  TRS . For this reason one often refers to 
$\theta_{1,d}$ as a \emph{TR-involution}. For more details, the reader can refer to \cite[Section 2]{denittis-gomi-14}. The classification of \virg{Quaternionic} vector bundles over these spaces in low dimension is summarized in Table \ref{tab:01}.1. We notice that the case 
$\n{S}^{0,3}$ studied in \cite{gat-robbins-15} is also included.

With multiple products of involutive one-dimensional spheres one can define several types of \emph{involutive tori}. By using as building blocks the three involutive space 
$\n{S}^{2,0}$, $\n{S}^{1,1}$ and $\n{S}^{0,2}$ one can define the involutive torus of type $(a,b,c)$
by
\begin{equation}\label{eq:intro_inv_tori}
\n{T}^{a,b,c}\;:=\;\underbrace{\n{S}^{2,0}\times\ldots\times \n{S}^{2,0}}_{a-\text{times}}\;\times\;\underbrace{\n{S}^{1,1}\times\ldots\times \n{S}^{1,1}}_{b-\text{times}}\;\times\;\underbrace{\n{S}^{0,2}\times\ldots\times \n{S}^{0,2}}_{c-\text{times}}\;\qquad\quad a,b,c\in\N\cup\{0\}\;.
\end{equation}
Topologically this is a torus of dimension $d=a+b+c$ endowed with the  
 natural product involution. The space $\n{T}^{d,0,0}$ coincides with the torus
$\n{T}^{d}$ endowed with the \emph{trivial involution} fixing all points.  The space $\n{T}^{0,d,0}$ (denoted with the symbol $\tilde{\n{T}}^d$ in \cite{denittis-gomi-14,denittis-gomi-14-gen,denittis-gomi-15})
coincides with the \virg{Brillouin zone} of condensed matter systems invariant under $\Z^d$-translations and subjected to a TRS (for more details, we refer to \cite[Section 2]{denittis-gomi-14}). Notice that $c\neq0$ implies that the involution acts freely.
For a fixed dimension $d$ there are   $\frac{1}{2}(d+1)(d+2)$ possible combinations for $a+b+c=d$ but only $2d+1$
inequivalent involutive tori due to  Proposition \ref{prop:cohom_Tab1-iso} which states the equivalence
\begin{equation}\label{eq:iso_sempl_A}
\n{T}^{a,b,c}\;\simeq\; \n{T}^{a+c-1,b,1}\;\qquad\quad c\geqslant 2.
\end{equation}
The classification in the case of tori with fixed point $(c=0)$ up to dimension three is described in  Table \ref{tab:03}.2. Note that also the case  $\n{T}^{1,1,0}$ discussed in \cite{gat-robbins-15} is included.

 \begin{center}
 \begin{table}[h]\label{tab:01}
 \begin{tabular}{|c||c|c|c|c|c|}
\hline
 \rule[-3mm]{0mm}{9mm}
 $p+q\leqslant 4$ & $q=0$  & $q=1$ & $q=2$&$q=3$&$q=4$\\
\hline
 \hline
 \rule[-3mm]{0mm}{9mm}
 ${\rm Vec}_{\rr{Q}}^{2m+1}(\n{S}^{0,q})$& $\text{\rm  \O}$  & $0$ & $0$ & $2\Z+1$ &  $\text{\rm  \O}$ \\
\hline
 \rule[-3mm]{0mm}{9mm}
 ${\rm Vec}_{\rr{Q}}^{2m}(\n{S}^{0,q})$& $\text{\rm  \O}$ & $0$ & $0$ & $2\Z$ &  0  \\
  \hline
\hline
 \rule[-3mm]{0mm}{9mm}
 ${\rm Vec}_{\rr{Q}}^{2m}(\n{S}^{1,q})$ & 0 & $0$ & $\Z_2$ &$\Z_2$& $\ldots$\\
\hline
  \rule[-3mm]{0mm}{9mm}
 ${\rm Vec}_{\rr{Q}}^{2m}(\n{S}^{2,q})$ & 0 & $2\Z$ & $0$ & $\ldots$ &    \\
\cline{1-6}
 \rule[-3mm]{0mm}{9mm}
 ${\rm Vec}_{\rr{Q}}^{2m}(\n{S}^{3,q})$ & 0 & $0$ &$\ldots$  & & \\
\hline
\cline{1-6}
 \rule[-3mm]{0mm}{9mm}
 ${\rm Vec}_{\rr{Q}}^{2m}(\n{S}^{4,q})$ & 0 & $\ldots$ &  & &\\
 \hline
\end{tabular}\vspace{2mm}
 \caption{\footnotesize Complete classification of \virg{Quaternionic} vector bundles over involutive spheres of type $\n{S}^{p,q}$ in low dimension $d=p+q-1\leqslant 3$. The symbol $\text{\rm  \O}$ (empty set) denotes the impossibility of constructing \virg{Quaternionic} vector bundles. The symbol ${0}$ denotes the existence of a unique element which can be identified with the trivial product bundle in the even rank case. In the odd rank case, only possible over the free-involution spheres $\n{S}^{0,q}$, the notion of trivial bundle is not well defined, and even when there is only one representative (\eg the cases $q=1,2$) this is not given by a product bundle with a product \virg{Quaternionic} action. The classification by integers (even, odd) is attained by looking at the first Chern class of the underlying complex vector bundle. The $\Z_2$ classification is given by the FKMM-invariant.
  The results summarized in this table are discussed in detail in Section \ref{sect:class_invol_shper}.
 }
 \end{table}
 \end{center}

The case with free involution (the equivalence 
\eqref{eq:iso_sempl_A} allows to consider 
only the case $c=1$) is summarized in the Table \ref{tab:04}.3.

\medskip

\begin{remark}[Flip involution]{\upshape
Over $\n{T}^{2}=\n{S}^1\times\n{S}^1$ there is another interesting involution which is essentially different from the involutions described by \eqref{eq:intro_inv_tori}. Consider the \emph{flip} map $\gamma:(k,k')\mapsto(k',k)$ and 
the associated involutive space
$\n{T}^{2}_{\rm flip}:=(\n{T}^2,\gamma)$.
This space has a non-empty fixed point set $(\n{T}^{2}_{\rm flip})^\gamma\simeq\n{S}^1$ given by the \emph{diagonal} subset of points  $(k,k)$ and therefore it  admits only even rank \virg{Quaternionic} vector bundles. We anticipate that
\begin{equation}\label{eq:class_flip}
{\rm Vec}_{\rr{Q}}^{2m}\big(\n{T}^{2}_{\rm flip}\big)\;\stackrel{c_1}{\simeq}\;2\Z
\end{equation}
although we will study  this case in a {separated paper \cite{denittis-gomi-17??}} due to its relevance with the physics of a system of two identical one-dimensional particles.
{Anyway, the justification of \eqref{eq:class_flip}, which is quite laborious, requires the computation of $H^2_{\Z_2}(\n{T}^{2}_{\rm flip}|(\n{T}^{2}_{\rm flip})^\gamma,\Z(1))$ and the study of the morphism which connects this group with $H^2(\n{T}^{2}_{\rm flip},\Z)$ in a long exact sequence of type \eqref{eq:proof_aux_0}.}
{Let us point out that over $\n{T}^{2}$ there are only five inequivalent (non-trivial) involutions which are listed, for instance, in pg. 164 of \cite{sakuma-85}. In this paper the inequivalent involutions are denoted with $r_j$ and a simple inspection shows $r_1,r_2,r_3$ and $r_5$
correspond to the involutive tori $\n{T}^{1,0,1}, \n{T}^{0,2,0}, \n{T}^{1,1,0}$ and $\n{T}^{0,1,1}$, respectively. The map $r_4$ can be related with $\n{T}^{2}_{\rm flip}$ after some manipulation. In conclusion the content of Table 1.2 and Table 1.3 along with  \eqref{eq:class_flip} provide a classification of \virg{Quaternionic} vector bundles over all possible two-dimensional involutive tori. 
The inequivalent involutions for the three-dimensional torus are classified in \cite{kwun-tollefson-75}.}
}\hfill $\blacktriangleleft$
\end{remark}

\medskip

 \begin{center}
 \begin{table}[h]\label{tab:03}
 \begin{tabular}{|c||c|c|c|c|}
\hline
 \rule[-3mm]{0mm}{9mm}
 $a+b\leqslant 3$, $c=0$& $a=0$  & $a=1$ & $a=2$&$a=3$\\
\hline
 \hline
 \rule[-3mm]{0mm}{9mm}
  ${\rm Vec}_{\rr{Q}}^{2m}(\n{T}^{a,0,0})$& $\text{\rm  \O}$ & $0$ & $0$ & $0$   \\
\hline
 \rule[-3mm]{0mm}{9mm}
  ${\rm Vec}_{\rr{Q}}^{2m}(\n{T}^{a,1,0})$ & 0 & $2\Z$ & $(2\Z)^2$ &$\ldots$\\
\hline
  \rule[-3mm]{0mm}{9mm}
  ${\rm Vec}_{\rr{Q}}^{2m}(\n{T}^{a,2,0})$ & $\Z_2$ & $\Z_2\oplus (2\Z)^2$ & $\ldots$ &     \\
\hline
 \rule[-3mm]{0mm}{9mm}
  ${\rm Vec}_{\rr{Q}}^{2m}(\n{T}^{a,3,0})$ & ${\Z_2}^4$ & $\ldots$ &  &  \\
\hline
\end{tabular}\vspace{2mm}
 \caption{\footnotesize Complete classification of \virg{Quaternionic} vector bundles over involutive tori of type $\n{T}^{a,b,0}$ in low dimension $d=a+b\leqslant 3$. The existence of fixed points implies that only even rank \virg{Quaternionic} vector bundles are admissible.  The classification by (even) integers is attained by looking at the first Chern class of the underlying complex vector bundle. The $\Z_2$ classification is given by the FKMM-invariant. The results summarized in this table are discussed in detail in Section \ref{sect:class_invol_tori}.
 }
 \end{table}
 \end{center}

 \begin{center}
 \begin{table}[h]\label{tab:04}
 \begin{tabular}{|c||c|c|c|}
\hline
 \rule[-3mm]{0mm}{9mm}
 $a+b\leqslant 2$, $c=1$& $a=0$  & $a=1$ & $a=2$\\
\hline
 \hline
 \rule[-3mm]{0mm}{9mm}
  ${\rm Vec}_{\rr{Q}}^{m}(\n{T}^{a,0,1})$& $0$ & $\Z_2$ & ${\Z_2}^2$    \\
\hline
 \rule[-3mm]{0mm}{9mm}
  ${\rm Vec}_{\rr{Q}}^{m}(\n{T}^{a,1,1})$ & $2\Z$ & $\Z_2\oplus(2\Z)^2$ & $\ldots$ \\
\hline
  \rule[-3mm]{0mm}{9mm}
  ${\rm Vec}_{\rr{Q}}^{m}(\n{T}^{a,2,1})$ & $(2\Z)^2$ & $\ldots$ & $\ldots$      \\
\hline
 \end{tabular}\vspace{2mm}
 \caption{\footnotesize Complete classification of \virg{Quaternionic} vector bundles over involutive tori of type $\n{T}^{a,b,1}$ in low dimension $d=a+b+1\leqslant 3$. The absence of fixed points allows for   \virg{Quaternionic} vector bundles
of every rank (here $m\in\Z$).  The classification by (even) integers is given by the first Chern class of the underlying complex vector bundle. The $\Z_2$ classification is given by the FKMM-invariant. The results summarized in this table are discussed in detail in Section \ref{sect:class_invol_tori}.
 }
 \end{table}
 \end{center}

\section{FKMM-invariant: Definitions and properties}\label{sect:top_class}
At a topological level, isomorphism classes of
\virg{Quaternionic} vector bundles can be classified by the \emph{FKMM-invariant} \cite{furuta-kametani-matsue-minami-00,denittis-gomi-14-gen}.
In its original form the FKMM-invariant can be introduced only for a subfamily of \virg{Quaternionic} vector bundles provided that certain conditions are met. The aim of this section is to drop this unpleasant restriction by introducing a more general version of the FKMM-invariant.

\subsection{Basic definitions}
\label{sect:basic_def}
In this section we recall some basic facts about the category of \virg{Quaternionic} vector bundles and we refer to  \cite{denittis-gomi-14-gen} for a more systematic presentation. 

\medskip

An {involution} $\tau$ on a topological space $X$ is a homeomorphism of period 2, \ie  $\tau^2={\rm Id}_X$. The pair 
 $(X,\tau)$ will be called an  \emph{involutive space}. The \emph{fixed point} set of the   $(X,\tau)$
 is by definition $
X^{\tau}:= \{x\in X\ |\ \tau(x)=x\}
$. Henceforth, we will assume that: 
\begin{assumption}\label{ass:top}
{$X$ is a 
topological space which admits the structure of a $\Z_2$-CW-complex. The \emph{dimension} $d$ of $X$ is the maximal dimension of its cells.} 
\end{assumption}
\noindent
For the sake of completeness, let us recall that an involutive space $(X,\tau)$ has the structure of a $\Z_2$-CW-complex if it admits a skeleton decomposition given by gluing cells of different dimensions which carry a $\Z_2$-action. 
For a precise definition of the notion of 
$\Z_2$-CW-complex, the reader can refer to  \cite[Section 4.5]{denittis-gomi-14} or \cite{matumoto-71,allday-puppe-93}. {Assumption  \eqref{ass:top} allows the space $X$ to be made by several disconnected component. However, in the case of multiple components, we will tacitly assume that vector bundles built over $X$ possess fibers of constant rank on the whole $X$. Let us recall that a space with a CW-complex structure is automatically Hausdorff and paracompact and it is compact exactly when it is made by a finite number of cells \cite{hatcher-02}. Almost all the examples considered in this paper will concern with spaces with a finite CW-complex structure.}

\medskip

 We are now in position to introduce the principal object of interest in this work.

\begin{definition}[{\virg{Quaternionic}} vector bundles {\cite{dupont-69,denittis-gomi-14-gen}}]\label{defi:Q_VB}
A {\virg{Quaternionic}} vector bundle, or $\rr{Q}$-bundle, over  $(X,\tau)$
is a complex vector bundle $\pi:\bb{E}\to X$ endowed with a  homeomorphism $\Theta:\bb{E}\to \bb{E}$
 such that:
\begin{itemize}

\item[$(\rr{Q}_1)$] the projection $\pi$ is \emph{equivariant} in the sense that $\pi\circ \Theta=\tau\circ \pi$;
\vspace{1mm}
\item[$(\rr{Q}_2)$] $\Theta$ is \emph{anti-linear} on each fiber, \ie $\Theta(\lambda p)=\overline{\lambda}\ \Theta(p)$ for all $\lambda\in\C$ and $p\in\bb{E}$ where $\overline{\lambda}$ is the complex conjugate of $\lambda$;
\vspace{1mm}
\item[$(\rr{Q}_3)$] $\Theta^2$ acts fiberwise as the multiplication by $-1$, namely $\Theta^2|_{\bb{E}_x}=-\n{1}_{\bb{E}_x}$.
\end{itemize}
\end{definition}

\noindent
It
is always possible to endow
$\bb{E}$ with
a (essentially unique) Hermitian metric with respect to which $\Theta$ is an \emph{anti-unitary} map between conjugate fibers
\cite[Proposition 2.5]{denittis-gomi-14-gen}.
 
  \medskip
 
A vector bundle \emph{morphism} $f$ between two vector bundles  $\pi:\bb{E}\to X$ and $\pi':\bb{E}'\to X$ 
over the same base space
is a  continuous map $f:\bb{E}\to \bb{E}'$ which is \emph{fiber preserving} in the sense that  $\pi=\pi'\circ f$
 and that restricts to a {linear} map on each fiber $\left.f\right|_x:\bb{E}_x\to \bb{E}'_x$. Complex vector bundles over  $X$ together with vector bundle morphisms define a category and the symbol ${\rm Vec}^m_\C(X)$
 is used to denote the set of isomorphism classes  of vector bundles of rank $m$.
    Also 
 $\rr{Q}$-bundles define a category with respect to  \emph{$\rr{Q}$-morphisms}. A $\rr{Q}$-morphism $f$ between two  {\virg{Quaternionic}} vector bundles
 $(\bb{E},\Theta)$ and $(\bb{E}',\Theta')$ over the same involutive space $(X,\tau)$ 
  is a vector bundle morphism  commuting with the \virg{Quaternionic} structures, \ie $f\circ\Theta\;=\;\Theta'\circ f$. The set of isomorphisms classes of   $\rr{Q}$-bundles of  rank $m$ over $(X,\tau)$ 
  will be denoted by ${\rm Vec}_{\rr{Q}}^m(X,\tau)$.

\begin{remark}[{\virg{Real}} vector bundles]\label{rk_real}{\upshape
By changing  condition $(\rr{Q}_3)$
 of Definition \ref{defi:Q_VB}   with 
\begin{itemize}
\item[$(\rr{R})$] \emph{$\Theta^2$ acts fiberwise as the multiplication by $1$, namely $\Theta^2|_{\bb{E}_x}=\n{1}_{\bb{E}_x}$}\;
\end{itemize}
one ends in the category of \emph{\virg{Real}} (or $\rr{R}$)  \emph{vector bundles}. 
Isomorphism classes of rank $m$ $\rr{R}$-bundles  over the involutive space $(X,\tau)$ 
are denoted by ${\rm Vec}_{\rr{R}}^m(X,\tau)$. 
For more details we refer to \cite{denittis-gomi-14}.
}\hfill $\blacktriangleleft$
\end{remark}

\medskip

Consider the fiber $\bb{E}_x\simeq\C^m$ over a fixed point $x\in X^\tau$. In this case the restriction $\Theta|_{\bb{E}_x}\equiv J$ defines an  \emph{anti}-linear map $J : \bb{E}_x \to \bb{E}_x$ such that $J^2 = -\n{1}_{\bb{E}_x}$. Then,  fibers over fixed points are endowed with a \emph{quaternionic} structure in the sense of \cite[Remark 2.1]{denittis-gomi-14-gen}.
This fact has an important  consequence: if
$X^\tau\neq \text{\rm  \O}$ then every {\virg{Quaternionic}} vector bundle over $(X,\tau)$   has necessarily even rank \cite[Proposition 2.1]{denittis-gomi-14-gen}. 
However, odd rank $\rr{Q}$-bundles are possible in the case of a free involution, \ie when $X^\tau= \text{\rm  \O}$ (\cf Section \ref{sect:quat_lin_bund}).

\medskip

The set
 ${\rm Vec}_{\rr{Q}}^{2m}(X,\tau)$ is non-empty since it contains at least the \virg{Quaternionic} \emph{product  bundle} 
 $X\times\C^{2m}\to X$ endowed with the {product} $\rr{Q}$-structure
$\Theta_0 (x,{\rm v})=(\tau(x),Q\;\overline{{\rm v}})$ 
where the matrix $Q$ is given by
\beql{eq:Q-mat}
Q
\;:=\;\left(
\begin{array}{rr}
0 & -1    \\
1 &  0 
\end{array}
\right)\;\otimes\;\n{1}_m\;=\;
\left(
\begin{array}{rr|rr|rr}
0 & -1 &        &        &   &    \\
1 &  0 &        &        &   &    \\
\hline
  &    & \ddots &        &   &    \\
  &    &        & \ddots &   &    \\
\hline
  &    &        &        & 0 & -1 \\
  &    &        &        & 1 &  0
\end{array}
\right)\;.
\eeq
A $\rr{Q}$-bundle is said to be \emph{$\rr{Q}$-trivial}  if it is isomorphic  to the product $\rr{Q}$-bundle in the category of $\rr{Q}$-bundles.

\subsection{The determinant construction}
\label{subsec:det_construct}
Let $\bb{V}$ be a complex vector space of dimension $m$. The \emph{determinant} of  $\bb{V}$ is by definition ${\rm det}(\bb{V}):=\bigwedge^m\bb{V}$ where the symbol  $\bigwedge^m$ denotes the top exterior power of $\bb{V}$ (\ie the skew-symmetrized $m$-th tensor power of $\bb{V}$). This is a complex vector space of dimension one.
If $\bb{W}$ is a second vector space of the same dimension $m$
and $T:\bb{V}\to \bb{W}$ is a linear map then there is a naturally associated map ${\rm det}(T):{\rm det}(\bb{V})\to{\rm det}(\bb{W})$ which in the special case $\bb{V}= \bb{W}$
 coincides with the multiplication by the determinant of the endomorphism $T$.
 This determinant construction
is a functor from the category of vector spaces  to itself
and, by a standard argument \cite[Chapter 5, Section 6]{husemoller-94},  induces a functor on the category of complex vector bundles over an arbitrary space $X$. More precisely, for each rank $m$ complex vector bundle $\bb{E}\to X$, the associated \emph{determinant line bundle} ${\rm det}(\bb{E})\to X$ is the rank 1 complex vector bundle with fibers
\beql{eq:fib_descr}
 {\rm det}(\bb{E})_x\;=\; {\rm det}(\bb{E}_x)\qquad\quad x\in X \;.
\eeq
Each local  frame of sections $\{s_1,\ldots,s_m\}$ of $\bb{E}$ over the open set $\f{U}\subset X$ induces the  section $s_1\wedge\ldots\wedge s_m$ of ${\rm det}(\bb{E})$ which fixes a trivialization 
over  $\f{U}$. Given a  map $\varphi:X\to Y$ one can prove  the isomorphism ${\rm det}(\varphi^*(\bb{E}))\simeq \varphi^*({\rm det}(\bb{E}))$ which is a special case of the compatibility between pullback and tensor product.
Finally, if $\bb{E}=\bb{E}_1\oplus\bb{E}_2$ in the sense of the
Whitney sum then ${\rm det}(\bb{E})={\rm det}(\bb{E}_1)\otimes {\rm det}(\bb{E}_2)$.

\medskip

Let $(\bb{E},\Theta)$ be a rank $m$ \virg{Quaternionic} vector bundle over
$(X,\tau)$. The associated determinant line bundle ${\rm det}(\bb{E})$ inherits an involutive structure given by the map ${\rm det}(\Theta)$ which acts \emph{anti}-linearly between the fibers ${\rm det}(\bb{E})_x$ and ${\rm det}(\bb{E})_{\tau(x)}$ according to ${\rm det}(\Theta)(p_1\wedge\ldots\wedge p_m)= \Theta(p_1)\wedge\ldots\wedge \Theta(p_m)$. Clearly ${\rm det}(\Theta)^2$ is a fiber preserving map which coincides with the multiplication by $(-1)^{m}$. Hence, in view of  Remark \ref{rk_real} one has the following result.

\begin{lemma}\label{lemma:R_Q_det_bun}
Let $(\bb{E},\Theta)$ be a rank $m$ \virg{Quaternionic} vector bundle over
 $(X,\tau)$. The associated determinant line bundle  ${\rm det}(\bb{E})$ endowed with  the involutive structure ${\rm det}(\Theta)$ is
 a \virg{Real} line bundle  if $m$ is even and a \virg{Quaternionic} line bundle if $m$ is odd.
\end{lemma}

\medskip

Let $(\bb{E},\Theta)$ be a rank $m$ $\rr{Q}$-bundle over
 $(X,\tau)$ endowed with an equivariant Hermitian metric $\rr{m}$. These data  fix  a unique Hermitian metric $\rr{m}_{\rm det}$ on ${\rm det}(\bb{E})$ which is equivariant with respect to the  structure
induced by ${\rm det}(\Theta)$. More explicitly, if $(p_i,q_i)\in \bb{E}|_x\times\bb{E}|_x$, $i=1,\ldots,m$ then,
$$
\rr{m}_{\rm det}(p_1\wedge\ldots\wedge p_{m},q_1\wedge\ldots\wedge q_{m})\;:=\;\prod_{i=1}^{m}\rr{m}(p_i,q_i)\;.
$$
The line bundle $({\rm det}(\bb{E}),{\rm det}(\Theta))$ endowed with   $\rr{m}_{\rm det}$  is trivial if and only if there exists an isometric  equivariant isomorphism (in the appropriate category) with $X\times\C$. Equivalently, if and only if
 there exists a global equivariant section $s:X\to {\rm det}(\bb{E})$ of unit length (\cf \cite[Theorem 4.8]{denittis-gomi-14}). 
Let
 $$
 \n{S}({\rm det}(\bb{E}))\;:=\;\left\{p\in{\rm det}(\bb{E})\ |\  \rr{m}_{\rm det}(p,p)=1\right\}
 $$
 be the  \emph{circle bundle} underlying to  $({\rm det}(\bb{E}),{\rm det}(\Theta))$.
  Then the triviality of ${\rm det}(\bb{E})$ can be rephrased as the existence of a global equivariant section of $\n{S}({\rm det}(\bb{E}))\to X$.
The following result will play a crucial role in our construction.

\begin{proposition}[{\cite[Lemma 3.3]{denittis-gomi-14-gen}}]
\label{lemma:R_Q_det_bun2}
Let $(\bb{E},\Theta)$ be a \virg{Quaternionic} vector bundle over a space $X$ with trivial involution $\tau={\rm Id}_X$. Then, the associated  determinant line bundle ${\rm det}(\bb{E})$
endowed with the \virg{Real} structure ${\rm det}(\Theta)$
 is $\rr{R}$-trivial and admits a unique  \emph{canonical} $\rr{R}$-section $s:X\to\n{S}({\rm det}(\bb{E}))$.
 \end{proposition}

\medskip
\noindent
If $X^\tau\neq \text{\rm  \O}$   the restricted vector bundle $\bb{E}|_{X^\tau}\to{X^\tau}$ is a $\rr{Q}$-bundle over a space with trivial involution. Proposition \ref{lemma:R_Q_det_bun2} assures that the restricted line bundle ${\rm det}(\bb{E}|_{X^\tau})$
is $\rr{R}$-trivial with respect to the restricted \virg{Real} structure ${\rm det}(\Theta|_{X^\tau})$ and admits a distinguished $\rr{R}$-section
\beql{eq:triv_2}
s_{\bb{E}}\;:\;X^\tau\;\to\;\n{S}\big({\rm det}(\bb{E})|_{X^\tau}\big)\;.
\eeq
We will refer to $s_{\bb{E}}$ as the \emph{canonical section} associated to $(\bb{E},\Theta)$.

\subsection{A short reminder of the equivariant Borel cohomology}
\label{subsec:borel_cohom}

The proper cohomology theory for the study of vector bundles  in the category of spaces with involution is the {equivariant cohomolgy} introduced by  A.~Borel in \cite{borel-60}. This cohomology has been used for the topological classification of \virg{Real} vector bundles \cite{denittis-gomi-14} and plays also a role in the classification of \virg{Quaternionic} vector bundles \cite{denittis-gomi-14-gen}. A short   self-consistent summary of this cohomology theory can be found in \cite[Section 5.1]{denittis-gomi-14} and we refer to \cite[Chapter 3]{hsiang-75} and \cite[Chapter 1]{allday-puppe-93}
for a more complete  introduction to the subject.

\medskip

Since we need this tool we briefly recall the main steps of the
{Borel construction}. 
The \emph{homotopy quotient} of an involutive space   $(X,\tau)$ is the orbit space
\begin{equation}\label{eq:homot_quot}
{X}_{\sim\tau}\;:=\;X\times\ {\n{S}}^{\infty} /( \tau\times \theta_\infty)\;.
\end{equation}
Here $\theta_\infty$ is the {antipodal map} on the infinite sphere $\n{S}^\infty$ 
(\cf \cite[Example 4.1]{denittis-gomi-14}) and ${\n{S}}^{0,\infty}$ is used as short notation for the pair $(\n{S}^\infty,\theta_\infty)$.
The product space $X\times{\n{S}}^\infty$ (forgetting for a moment the $\Z_2$-action) has the \emph{same} homotopy type of $X$ 
since $\n{S}^\infty$ is contractible. Moreover, since $\theta_\infty$ is a free involution,  also the composed involution $\tau\times\theta_\infty$ is free, independently of $\tau$.
Let $\s{R}$ be any commutative ring (\eg, $\R,\Z,\Z_2,\ldots$). The \emph{equivariant} cohomology ring 
of $(X,\tau)$
with coefficients
in $\s{R}$ is defined as
$$
H^\bullet_{\Z_2}(X,\s{R})\;:=\; H^\bullet({X}_{\sim\tau},\s{R})\;.
$$
More precisely, each equivariant cohomology group $H^j_{\Z_2}(X,\s{R})$ is given by the
 singular cohomology group  $H^j({X}_{\sim\tau},\s{R})$ of the  homotopy quotient ${X}_{\sim\tau}$ with coefficients in $\s{R}$ and the ring structure is given, as usual, by the {cup product}.
As the coefficients of
the usual singular cohomology are generalized to \emph{local coefficients} (see \eg \cite[Section 3.H]{hatcher-02} or
\cite[Section 5]{davis-kirk-01}), the coefficients of the Borel  equivariant cohomology are also
generalized to local coefficients. Given an involutive space $(X,\tau)$ one can  consider the homotopy group $\pi_1({X}_{\sim\tau})$
and the associated  \emph{group ring} $\Z[\pi_1({X}_{\sim\tau})]$. Each module $\s{Z}$ over the group $\Z[\pi_1({X}_{\sim\tau})]$ is, by definition,
a \emph{local system} on $X_{\sim\tau}$.  Using this local system one defines, as usual, the equivariant cohomology with local coefficients in $\s{Z}$:
$$
H^\bullet_{\Z_2}(X,\s{Z})\;:=\; H^\bullet({X}_{\sim\tau},\s{Z})\;.
$$
We are particularly interested in modules $\s{Z}$ whose underlying groups are identifiable with $\Z$. 
For each involutive space  $(X,\tau)$, there always exists a particular family of local systems $\Z(j)$
labelled by $j\in\Z$. Here
 $\Z(j)\simeq X\times\Z$  denotes the $\Z_2$-equivariant local system on $(X,\tau)$  made equivariant  by the $\Z_2$-action $(x,l)\mapsto(\tau(x),(-1)^jl)$.
Given that the module structure depends only on the parity of $j$, one can consider only the $\Z_2$-modules ${\Z}(0)$ and ${\Z}(1)$. Noticing that ${\Z}(0)$ corresponds to the case of the trivial action of $\pi_1(X_{\sim\tau})$ on $\Z$ one has that $H^k_{\Z_2}(X,\Z(0))\simeq H^k_{\Z_2}(X,\Z)$ \cite[Section 5.2]{davis-kirk-01}.

\medskip

We recall the two important group isomorphisms 
\begin{equation}\label{eq:iso:eq_cohom}
\begin{aligned}
H^1_{\Z_2}\big(X,\Z(1)\big)\;&\simeq\;\big[X,\n{U}(1)\big]_{\Z_2}\;\simeq\;\big[X,\n{S}^{1,1}\big]_{\Z_2}\\
 H^2_{\Z_2}\big(X,\Z(1)\big)\;&\simeq\;{\rm Vec}_{\rr{R}}^1\big(X,\tau\big)\equiv {\rm Pic}_{\rr{R}}\big(X,\tau\big)\end{aligned}
\end{equation}
 involving the 
first two equivariant cohomology groups. 
The first isomorphism \cite[Proposition A.2]{gomi-13} says that the first equivariant cohomology group is isomorphic to the set of $\Z_2$-homotopy classes of equivariant maps $\varphi:X\to\n{U}(1)$ where the involution on $\n{U}(1)$ is induced by the complex conjugation, \ie $\varphi(\tau(x))=\overline{\varphi(x)}$.
Of course there is a natural identification of this target involutive space with the involutive sphere $\n{S}^{1,1}$.
 The second isomorphism is due to B.~Kahn \cite{kahn-59} and 
expresses the equivalence between the Picard group of \virg{Real} line bundles (in the sense of \cite{atiyah-66,denittis-gomi-14}) over  $(X,\tau)$ and the second equivariant cohomology group of this space.
A more modern proof of this isomorphism can be found in \cite[Corollary A.5]{gomi-13}.

\medskip

The fixed point subset $X^\tau\subset X$ is closed and $\tau$-invariant and the inclusion $\imath:X^\tau\hookrightarrow X$ extends to an inclusion $\imath:X^\tau_{\sim\tau}\hookrightarrow X_{\sim\tau}$ of the respective homotopy quotients. The \emph{relative} equivariant cohomology can be defined as usual by the identification
$$
H^\bullet_{\Z_2}(X|X^\tau,\s{Z})\;:=\; H^\bullet({X}_{\sim\tau}|X^\tau_{\sim\tau},\s{Z})\;.
$$
Consequently, one has the related long exact sequence in cohomology
\begin{equation}\label{eq:Long_seq}
\ldots\;\longrightarrow\;H^j_{\Z_2}(X|X^\tau,\s{Z})\;\stackrel{\delta_2}{\longrightarrow}\;H^j_{\Z_2}(X,\s{Z})\;\stackrel{r}{\longrightarrow}\;H^j_{\Z_2}(X^\tau,\s{Z})\;\stackrel{\delta_1}{\longrightarrow}\;H^{j+1}_{\Z_2}(X|X^\tau,\s{Z})\;\longrightarrow\;\ldots\;
\end{equation}
where the map $r:=\imath^*$ restricts  cochains on $X_{\sim\tau}$ to  cochains on $X^\tau_{\sim\tau}$. The $j$-th \emph{cokernel} of $r$ is by definition
\begin{equation}\label{eq:Long_seq_cokern}
{\rm Coker}^j(X|X^\tau,\s{Z})\;:=\;H^j_{\Z_2}(X^\tau,\s{Z})\;/\;r\big(H^j_{\Z_2}(X,\s{Z})\big)\;.
\end{equation}

Let us point out that with the same construction 
one can define {relative} cohomology theories $H^\bullet_{\Z_2}(X|Y,\s{Z})$
for each closed subset $Y\subseteq X$ which is $\tau$-invariant  $\tau(Y)=Y$. If $Y=\text{\rm  \O}$ then $H^j_{\Z_2}(X|\text{\rm  \O},\s{Z})\simeq H^j_{\Z_2}(X,\s{Z})$ by definition. Therefore it is 
suitable to declare $H^j_{\Z_2}(\text{\rm  \O},\s{Z})=0$ for consistency with the  exact sequence \eqref{eq:Long_seq}.

\subsection{A geometric model for the relative equivariant cohomology}
\label{subsec:geom_relat_cohom}
In this section we provide a geometric model for the equivariant relative cohomology group $H^2_{\Z_2}(X|Y,\n{Z}(1))$ which extends the second isomorphism in \eqref{eq:iso:eq_cohom}.  

\medskip

Let $(X,\tau)$ be an involutive space and $Y\subseteq X$ a closed $\tau$-invariant subspace $\tau(Y)=Y$ ({it is not required that $Y\subseteq X^\tau$}). Consider pairs $(\bb{L}, s)$ consisting of: (a) a \virg{Real} line bundle $\bb{L}\to X$ with a given \virg{Real} structure $\Theta$ and a Hermitian metric $\rr{m}$; (b) a (nowhere vanishing) \virg{Real} section $s:Y\to \n{S}(\bb{L}|_{Y})$ of the circle bundle associated to the restriction $\bb{L}|_{Y}\to Y$ (or equivalently a trivialization $h:\bb{L}|_{Y}\to Y\times\C$). 
Two pairs $(\bb{L}_1, s_1)$ and $(\bb{L}_2, s_2)$ built over the same involutive base space
$(X,\tau)$ and the same invariant subspace $Y$ are said to be \emph{isomorphic} if there is an $\rr{R}$-isomorphism of line bundles $f:\bb{L}_1\to \bb{L}_2$ (preserving the Hermitian structure) such that $f\circ s_1=s_2$.

\begin{definition}[Relative \virg{Real} Picard group]
For a given involutive space $(X, \tau)$ and a closed subspace $Y \subseteq X$  such that $\tau(Y)=Y$, we define ${\rm Pic}_{\rr{R}}(X|Y,\tau)$ to be the group of the isomorphism classes of pairs $(\bb{L}, s)$, with  group structure  given by the tensor product
$$
(\bb{L}_1, s_1)\; \otimes\; (\bb{L}_2, s_2)\; \simeq\; (\bb{L}_1\otimes \bb{L}_2, s_1  \otimes s_2).
$$
\end{definition}

\medskip

The main result of this section is the following characterization.

\begin{proposition}
\label{prop:nat_iso}
There is a natural isomorphism of groups
$$
\tilde{\kappa}\;:\;{\rm Pic}_{\rr{R}}(X|Y,\tau)\;\stackrel{\simeq}{\longrightarrow}\;H^2_{\Z_2}\big(X|Y,\Z(1)\big)\;
$$
{which reduces to the Kahn's isomorphism \eqref{eq:iso:eq_cohom} when $Y=\text{\rm  \O}$}.
\end{proposition}

\proof[{Proof} (sketch of).]
The Kahn's isomorphism $H^2_{\Z_2}(X,\Z(1))\simeq{\rm Pic}_{\rr{R}}(X,\tau)$
 can be proved roughly in two steps: The first step is to prove that ${\rm Vec}_{\rr{R}}^1(X,\tau)$ is isomorphic to the sheaf cohomology $H^1(\Z_2^\bullet \times X, \underline{\tilde{S}}^1_\bullet)$ of the simplicial space $\Z_2^\bullet \times X$ associated to the involutive space $(X, \tau)$ and this is carried out by realizing the sheaf cohomology as a \v{C}ech cohomology; The second step is to identify the sheaf cohomology with $H^2_{\Z_2}(X,\Z(1))$. For the full argument we refer to \cite[Appendix A]{gomi-13}.
   The same argument is replicable verbatim for the pair $Y\subseteq X$. Firstly, the use of the \v{C}ech cohomology allows to  prove the group isomorphism between ${\rm Pic}_{\rr{R}}(X|Y,\tau)$ and the relative sheaf cohomology $H^1(\Z_2^\bullet \times X|\Z_2^\bullet \times Y, \underline{\tilde{S}}^1_\bullet)$. Secondly,  this sheaf cohomology can be identified with the relative equivariant cohomology $H^2(X|Y,\Z(1))$ in the same way as in \cite[Appendix A]{gomi-13}.
\qed

\medskip

In the next sections we will exploit the result of Proposition  \ref{prop:nat_iso} in the special case of  $Y\subseteq X^\tau$. We recall that in this situation, as a consequence of Proposition \ref{lemma:R_Q_det_bun2}, 
the choice of $s$ is \emph{canonical} and corresponds to a  trivialization of the restricted \virg{Real} line bundle $\bb{L}|_{Y}\to Y$.

\subsection{Generalized FKMM-invariant}
\label{subsec:gen_FKMM_inv}
By combining the content of Proposition \ref{lemma:R_Q_det_bun2} with
that of Proposition \ref{prop:nat_iso}
we can introduce an \emph{intrinsic} invariant for the category of \virg{Quaternionic} vector bundles.
\begin{definition}[{FKMM-invariant for even rank $\rr{Q}$-bundles}]\label{def:gen_FKMM_inv}
Let $(\bb{E},\Theta)$ be an even rank \virg{Quaternionic} vector bundle over the involutive space $(X,\tau)$
and consider the pair $({\rm det}(\bb{E}), s_{\bb{E}})$ where ${\rm det}(\bb{E})$ is the determinant line bundle
associated to $\bb{E}$ endowed with the \virg{Real} structure induced by ${\rm det}(\Theta)$  and $s_{\bb{E}}$ is the canonical section described by \eqref{eq:triv_2}. The \emph{generalized FKMM-invariant} of  $(\bb{E},\Theta)$ is the cohomology class $\kappa(\bb{E},\Theta)\in H^2_{\Z_2}\big(X|X^\tau,\Z(1)\big)$ given by
$$
\kappa(\bb{E},\Theta)\;:=\;\tilde{\kappa}\big([({\rm det}(\bb{E}), s_{\bb{E}})]\big)
$$
where  $[({\rm det}(\bb{E}), s_{\bb{E}})]\in {\rm Pic}_{\rr{R}}(X|X^\tau,\tau)$ is the 
 isomorphism class of the pair $({\rm det}(\bb{E}), s_{\bb{E}})$ and $\tilde{\kappa}$ is the group isomorphism described in Proposition \ref{prop:nat_iso}. {The definition extends to the  case $X^\tau=\text{\rm  \O}$ by interpreting the class $\kappa(\bb{E},\Theta)$ as  the  (first) \virg{Real} Chern class of  ${\rm det}(\bb{E})\in {\rm Pic}_{\rr{R}}(X,\tau)$ according to the Kahn's isomorphism (\cf Corollary \ref{corol:II} (2)).}
\end{definition}

\medskip

\noindent
The following properties are immediate consequence of the last definition.
\begin{itemize}
\item[(1)] Isomorphic \virg{Quaternionic} vector bundles define the same FKMM-invariant;
\vspace{1mm}
\item[(2)] The FKMM-invariant is \emph{natural} under the pullback induced by equivariant maps;
\vspace{1mm}
\item[(3)] If $(\bb{E},\Theta)$ is $\rr{Q}$-trivial then $\kappa(\bb{E},\Theta)=0$; 
\vspace{1mm}
\item[(4)] The FKMM-invariant is \emph{additive} with respect to the Whitney sum and the abelian structure of $H^2_{\Z_2}\big(X|X^\tau,\Z(1)\big)$, namely
$$
\kappa(\bb{E}_1\oplus \bb{E}_2,\Theta_1\oplus\Theta_2)\;=\;\kappa(\bb{E}_1,\Theta_1)\;\cdot\;\kappa( \bb{E}_2,\Theta_2)
$$
for each pair of \virg{Quaternionic} vector bundles $(\bb{E}_1,\Theta_1)$ and $(\bb{E}_2,\Theta_2)$ over the same involutive space $(X,\tau)$.
\end{itemize}

\medskip

\begin{remark}[Comparison with the original definition of the FKMM-invariant]\label{rk:comp_origin}{\upshape
The original definition of the FKMM-invariant, firstly introduced in \cite{furuta-kametani-matsue-minami-00} and then extensively used in \cite{denittis-gomi-14-gen}, is strongly based on two crucial assumptions: (a)  $X^\tau\neq\text{\rm  \O}$ and (b) ${\rm det}(\bb{E})$ has to be $\rr{R}$-trivial (\cf \cite[Definition 3.1]{denittis-gomi-14-gen}).
Under this assumption the FKMM-invariant can be defined as an element in the cokernel of the restriction map 
$r: H^1_{\Z_2}(X,\Z(1))\to H^1_{\Z_2}(X^\tau,\Z(1))$. Moreover, the isomorphism described in \cite[Lemma 3.1]{denittis-gomi-14-gen} allows to describe the original FKMM-invariant as an equivariant map $\phi:X^\tau\to\n{U}(1)$ which 
 measures the difference between the canonical section $s_{\bb{E}}$ and a global \virg{Real} section of ${\rm det}(\bb{E})$ (\cf \cite[Definition 3.2 \& Remark 3.2]{denittis-gomi-14-gen}). On the other hand, the generalized invariant introduced in Definition \ref{def:gen_FKMM_inv} does not require  any extra restrictions on the nature of $(X,\tau)$ and $(\bb{E},\Theta)$ and fulfills all the structural properties of the original FKMM-invariant (compare the (1)-(4) above with the content of \cite[Theorem 3.1]{denittis-gomi-14-gen}). Moreover, as soon as assumptions (a) and (b) are met,  the generalized FKMM-invariant agrees with the original FKMM-invariant (see Proposition \ref{prop:genVSorig_inv} and Corollary \ref{corol:I} below). This allows  to state that  Definition \ref{def:gen_FKMM_inv} really \emph{extends} the notion of the FKMM-invariant.
}\hfill $\blacktriangleleft$
\end{remark}

\medskip

Let  ${\rm Coker}^1(X|X^\tau,\n{Z}(1))$ be the cokernel described in \eqref{eq:Long_seq_cokern} and $[X,\n{U}(1)]_{\Z_2}$ the set of  $\Z_2$-homotopy classes of equivariant maps between the involutive space $(X,\tau)$ and the group $\n{U}(1)$ endowed with the involution induced by the complex conjugation.
Let us  recall that  the isomorphism
$$
\big[X^\tau,\n{U}(1)\big]_{\Z_2}\;/\;\big[X,\n{U}(1)\big]_{\Z_2}\;\simeq\;{\rm Coker}^1(X|X^\tau,\n{Z}(1))
$$
holds under the assumption $X^\tau\neq\text{\rm  \O}$ \cite[Lemma 3.1]{denittis-gomi-14-gen}.

\medskip

\begin{proposition}\label{prop:genVSorig_inv}
Let $(\bb{E},\Theta)$ be an even rank \virg{Quaternionic} vector bundle over the involutive space $(X,\tau)$. Assume that: 
(a)  $X^\tau\neq\text{\rm  \O}$ and (b) ${\rm det}(\bb{E})$ is $\rr{R}$-trivial. Then generalized FKMM-invariant $\kappa(\bb{E},\Theta)$ introduced in Definition \ref{def:gen_FKMM_inv} is the \emph{injective} image of an element $[\phi]\in{\rm Coker}^1(X|X^\tau,\n{Z}(1))$.
\end{proposition}
\proof
The homomorphism $\delta_1$ and $\delta_2$  in  the long exact sequence 
\begin{equation}\label{eq:proof_aux_0}
\ldots\;\longrightarrow\;H^1_{\Z_2}\big(X,\Z(1)\big)\;\stackrel{r}{\longrightarrow}\;H^1_{\Z_2}\big(X^\tau,\Z(1)\big)\;\stackrel{\delta_1}{\longrightarrow}\;H^{2}_{\Z_2}\big(X|X^\tau,\Z(1)\big)\;\stackrel{\delta_2}{\longrightarrow}\;H^2_{\Z_2}\big(X,\Z(1)\big)\;\longrightarrow\;\ldots\;
\end{equation}
for the pair $X^\tau\hookrightarrow X$
can be interpreted as  homomorphisms 
$$
\delta_1\;:\; \big[X^\tau,\n{U}(1)\big]_{\Z_2}\;\longrightarrow\;{\rm Pic}_{\rr{R}}\big(X|X^\tau,\tau\big)\;,\qquad\quad \delta_2\;:\;{\rm Pic}_{\rr{R}}\big(X|X^\tau,\tau\big) \;\longrightarrow\; {\rm Pic}_{\rr{R}}\big(X,\tau\big)
$$
in view of the  isomorphism proved in Proposition \ref{prop:nat_iso}
and the
isomorphisms in \eqref{eq:iso:eq_cohom}, respectively.
Moreover,  the triviality of ${\rm det}(\bb{E})$ implies that the pair $({\rm det}(\bb{E}), s_{\bb{E}})$ which enters into the definition of the FKMM-invariant 
is an element of the  class $[(X\times\n{U}(1),\phi)]\in {\rm Pic}_{\rr{R}}(X|X^\tau,\tau)$ with $\phi:X^\tau\to\n{U}(1)$ is a suitable  equivariant map.
Therefore $\delta_1$ turns out to be  just the assignment of the class $[(X\times\n{U}(1),\phi)]$ to the map $[\phi]\in [X^\tau,\n{U}(1)]_{\Z_2}$ and  $\delta_2$ agrees with the map induced from $(\bb{L},s)\mapsto \bb{L}$. With this interpretation the result directly follows from the exact sequence. Finally, the injectivity is assured by
$$
0\longrightarrow\;{\rm Coker}^1(X|X^\tau,\n{Z}(1))\;\stackrel{\delta_1}{\longrightarrow}\;H^{2}_{\Z_2}\big(X|X^\tau,\Z(1)\big)\;\stackrel{\delta_2}{\longrightarrow}\;H^2_{\Z_2}\big(X,\Z(1)\big)\;\longrightarrow\;\ldots\;
$$
which follows from \eqref{eq:proof_aux_0} by general arguments.
\qed

\medskip

\noindent
As it emerges from the proof, the element $[\phi]$  in the statement of Proposition \ref{prop:genVSorig_inv} is represented by the canonical section $s_{\bb{E}}$ modulo the action (multiplication and restriction) of an equivariant map $s:X\to\n{U}(1)$ which can be interpreted as a global trivialization of the trivial line bundle 
${\rm det}(\bb{E})$. Therefore $[\phi]$ is just the original FKMM-invariant as introduced in  \cite{furuta-kametani-matsue-minami-00} and \cite[Definition 3.2]{denittis-gomi-14-gen}. As a consequence, we can rephrase Proposition \ref{prop:genVSorig_inv}
by saying that:

\medskip
\emph{\virg{
 The \emph{generalized} FKMM-invariant is just the image (under $\delta_1$) of the \emph{old} FKMM-invariant
whenever the latter can be defined.}}

\medskip

According to \cite[Definition 1.1]{denittis-gomi-14-gen} an \emph{FKMM-space} is an involutive space $(X,\tau)$ such that {$X^\tau$ is a non-empty finite set} and $H^2_{\Z_2}(X,\Z(1))=0$.

\begin{corollary}\label{corol:I}
Let $(\bb{E},\Theta)$ be an even rank \virg{Quaternionic} vector bundle over the FKMM-space $(X,\tau)$.  Then, the generalized FKMM-invariant $\kappa(\bb{E},\Theta)$ introduced in Definition \ref{def:gen_FKMM_inv} is represented by a map $\phi:X^\tau\to\{+1,-1\}$.
\end{corollary}
\proof
It has been proved in \cite[Lemma 3.1]{denittis-gomi-14-gen} that the condition $H^2_{\Z_2}(X,\Z(1))=0$ implies that $\delta_1$ defines an isomorphisms between ${\rm Coker}^1(X|X^\tau,\n{Z}(1))$ and $H^{2}_{\Z_2}\big(X|X^\tau,\Z(1)\big)$.
Hence, $\kappa(\bb{E},\Theta)$ can be view as an element in ${\rm Coker}^1(X|X^\tau,\n{Z}(1))$ which, modulo equivalences, is represented by an equivariant function $\phi:X^\tau\to\n{U}(1)$.  Since, by definition, $\tau$ acts trivially on $X^\tau$, $\phi$ maps into the  points of $\n{U}(1)$ fixed by the complex conjugation. \qed

\medskip

For the next result we need to recall that the Kahn's isomorphism $H^2_{\Z_2} (X,\Z(1) ) \simeq {\rm Pic}_{\rr{R}}\big(X,\tau\big)$ is realized by 
\emph{first \virg{Real} Chern class}
$c^{\rr{R}}_1$ (see  \cite{kahn-59} or \cite[Section 5.2]{denittis-gomi-14}).

\begin{corollary}\label{corol:II}
Let $(\bb{E},\Theta)$ be an even rank \virg{Quaternionic} vector bundle over the involutive space $(X,\tau)$
and
$$
\delta_2\;:\;H^{2}_{\Z_2}\big(X|X^\tau,\Z(1)\big)\;{\longrightarrow}\;H^2_{\Z_2}\big(X,\Z(1)\big)
$$
the homomorphism in the long exact sequence \eqref{eq:proof_aux_0} for  the pair $X^\tau\hookrightarrow X$. Then:
\begin{itemize}
\item[(1)] The homomorphism $\delta_2$ carries the FKMM-invariant $\kappa(\bb{E},\Theta)$ to the first \virg{Real} Chern class $c^{\rr{R}}_1({\rm det}(\bb{E}))$ of the \virg{Real} line bundle ${\rm det}(\bb{E})\to X$;
\vspace{1mm}
\item[(2)] If in addition $X^\tau={\text{\rm  \O}}$, then the FKMM-invariant $\kappa(\bb{E},\Theta)$ agrees with $c^{\rr{R}}_1({\rm det}(\bb{E}))$.
\end{itemize}
\end{corollary}
\proof From the proof of Proposition \ref{prop:genVSorig_inv} one knows that the   $\delta_2$ can be 
interpreted as the  homomorphism ${\rm Pic}_{\rr{R}}(X|X^\tau,\tau)\to{\rm Pic}_{\rr{R}}(X,\tau)$
induced by $(\bb{L},s)\mapsto \bb{L}$. Therefore (1) follows from the fact that the Kahn's isomorphism  is induced by $c^{\rr{R}}_1$. (2) follows just by observing that the condition $X^\tau=\text{\rm  \O}$, once inserted in the long exact sequence \eqref{eq:proof_aux_0},  forces $\delta_2$ to be an isomorphism.
\qed

\medskip

The original FKMM-invariant cannot be defined when $X^\tau = \text{\rm  \O}$ but, on the contrary, requires $X^\tau \neq \text{\rm  \O}$ {to be finite} and the triviality of ${\rm det}(\bb{E})$. On the other hand the generalized  FKMM-invariant 
introduced in Definition \ref{def:gen_FKMM_inv} bypasses these restrictions and provides a way to define a topological invariant  
also in the \emph{opposite} case when $X^\tau = \text{\rm  \O}$ and ${\rm det}(\bb{E})$ is non-trivial. In contrast to the definition of
{FKMM-space}
\cite[Definition 1.1]{denittis-gomi-14-gen} we can call \emph{\virg{anti-FKMM}} an involutive space $(X,\tau)$   such that   
$X^\tau=\text{\rm  \O}$ and $H^2_{\Z_2}(X,\Z(1))\neq0$. Corollary \ref{corol:I} says that the generalized FKMM-invariant agrees with the original invariant in the case of an FKMM-space and Corollary \ref{corol:II} (ii) shows that the 
generalized FKMM-invariant reduces to the first \virg{Real} Chern class in the anti-FKMM case. In  more general situations  the classification of even rank \virg{Quaternionic} vector bundles would require the generalized notion of  FKMM-invariant (in the form of Definition \ref{def:gen_FKMM_inv} )
 which can be considered a sort of  \emph{mixture} of the two mentioned extreme cases.

\subsection{Universal FKMM-invariant}
\label{subsec:univ_FKMM_inv}
Even rank \virg{Quaternionic} vector bundles can  be classified by maps with values in a classification space \cite[Theorem 2.4]{denittis-gomi-14-gen}. This is the basic fact which allows a description of the  generalized FKMM-invariant 
as a \emph{characteristic class}, namely as the image of a unique universal object under the pullback induced by the classification maps. This section provides a clarification and a simplification of some notions and results already discussed in \cite[Section 6]{denittis-gomi-14-gen}.

\medskip

Let us recall that the (even dimensional)  \emph{Grassmannian} is defined as
$$
G_{2m}(\C^\infty)\;:=\;\bigcup_{n=2m}^{\infty}\;G_{2m}(\C^n)\;,
$$
where,
for each pair $2m\leqslant n$, $G_{2m}(\C^n)\simeq\n{U}(n)/\big(\n{U}(2m)\times \n{U}(n-2m)\big)$ is the set of $2m$-dimensional (complex) subspaces of $\C^n$. 
The spaces $G_{2m}(\C^n)$ are finite CW-complexes and the space $G_{2m}(\C^\infty)$ inherits
the 
direct limit topology given by the inclusions $G_{2m}(\C^n)\hookrightarrow G_{2m}(\C^{n+1})\hookrightarrow\ldots$ induced by
$\C^n \subset \C^{n+1} \subset\ldots$.
Moreover, $G_{2m}(\C^\infty)$ can be endowed with an involution of quaternionic type in the following way:
Let
$\Sigma=\langle{\rm v}_1,{\rm v}_2,\ldots{\rm v}_{2m-1},{\rm v}_{2m}\rangle_\C$ be any
 $2m$-plane in $G_{2m}(\C^{2n})$  generated by the 
basis $\{{\rm v}_1,{\rm v}_2,\ldots{\rm v}_{2m-1},{\rm v}_{2m}\}$
and define a new point $\rho({\Sigma})\in G_{2m}(\C^{2n})$ as  the $2m$-plane spanned by $\langle Q\overline{\rm v}_1,Q\overline{\rm v}_2,\ldots,Q\overline{\rm v}_{2m-1},Q\overline{\rm v}_{2m}\rangle_{\C}$ where $\overline{\rm v}_j$ is the complex conjugate of  ${\rm v}_j$
and  $Q$ is the $2n\times 2n$ matrix \eqref{eq:Q-mat}. 
Notice that the definition of 
$\rho(\Sigma)$ does not depend on the choice of a particular basis and  the map $\rho:G_{2m}(\C^n)\to G_{2m}(\C^n)$  is an involution that makes the pair $(G_{2m}(\C^{2n}),\rho)$ into an involutive space. 
Since all the inclusions $G_{2m}(\C^{2n})\hookrightarrow G_{2m}(\C^{2(n+1)})\hookrightarrow\ldots$ are equivariant, the involution extends to 
 the infinite Grassmannian in such a way that $\hat{G}_{2m}(\C^\infty)\equiv(G_{2m}(\C^\infty),\rho)$ becomes an involutive space. 
Let $\Sigma=\rho(\Sigma)$ be a fixed point of $\hat{G}_{2m}(\C^\infty)$. Since $\rho$ acts on vectors as a quaternionic structure one has $\Sigma\simeq \n{H}^m$ (\cf \cite[Remark 2.1]{denittis-gomi-14-gen}).
This fact implies that the fixed point set of $\hat{G}_{2m}(\C^\infty)$ can be identified with the \emph{quaternionic} Grassmannian, namely $\hat{G}_{2m}(\C^\infty)^\rho\simeq G_{m}(\n{H}^\infty)$ (see \cite[Section 2.4]{denittis-gomi-14-gen} for more details).

\medskip

Each manifold $G_{m}(\C^{n})$ is the base space of a {canonical} rank $m$ complex vector bundle $\pi:\bb{T}_{m}^n\to G_m(\C^n)$ with total space $\bb{T}_m^n:=\{(\Sigma,{\rm v})\in (G_{m}(\C^{n})\times\C^n )\ |\ {\rm v}\in\Sigma\}$ and  bundle projection 
 $\pi(\Sigma,{\rm v})=\Sigma$. When  $n$ tends to infinity, the same construction leads to 
 the \emph{tautological} $m$-plane bundle $\pi:\bb{T}_m^\infty\to G_m(\C^\infty)$. 
The latter is the {universal} vector bundle  which classifies  complex vector bundles. In fact   any rank $m$ complex vector bundle $\bb{E}\to X$ is realized, up to isomorphisms, as a pullback $\bb{E}\simeq \varphi^\ast \bb{T}_m^\infty$
 with respect to a \emph{classifying map} $\varphi : X \to G_m(\C^\infty)$.
 Since pullbacks of homotopy equivalent maps yield isomorphic vector bundles one gets the well known  result ${\rm Vec}^m_\C(X) \simeq[X , G_m(\C^\infty)]
$.
This  result  extends to the category of even rank \virg{Quaternionic} vector bundles provided that the total space $\bb{T}_{2m}^\infty$ is endowed with a $\rr{Q}$-structure compatible with the involution $\rho$. This is done by  the 
{anti}-linear map $\Xi:\bb{T}_{2m}^\infty\to\bb{T}_{2m}^\infty$
defined by $\Xi:(\Sigma,{\rm v})\mapsto \big(\rho(\Sigma),Q\overline{\rm v}\big)$. The relation 
$\pi\circ\Xi=\rho\circ{\pi}$ assures that  $(\bb{T}_{2m}^\infty,\Xi)$ is a $\rr{Q}$-bundle over the involutive space  $\hat{G}_{2m}(\C^\infty)$. This     \emph{tautological} $\rr{Q}$-bundle classifies \virg{Quaternionic} vector bundles in the sense that ${\rm Vec}^{2m}_\rr{Q}(X,\tau)\simeq[X,\hat{G}_{2m}(\C^\infty)]_{\Z_2}$ where in the right-hand side  there is the set of $\Z_2$-homotopy  classes of  equivariant maps between $(X,\tau)$ and $\hat{G}_{2m}(\C^\infty)$
\cite[Theorem 2.4]{denittis-gomi-14-gen}.

\medskip

The construction of the generalized FKMM-invariant applies  to the  {universal} $\rr{Q}$-bundle $(\bb{T}_{2m}^\infty,\Xi)$. 
\begin{definition}[Universal FKMM-invariant]\label{def:univ_FKMM_inv}
The \emph{universal FKMM-invariant}
$$
\rr{h}_{\rm univ}\;\in\;H^2_{\Z_2}\big(\hat{G}_{2m}(\C^\infty)|\hat{G}_{2m}(\C^\infty)^\rho,\Z(1)\big)
$$
is the (generalized) FKMM-invariant of the \emph{tautological} $\rr{Q}$-bundle $(\bb{T}_{2m}^\infty,\Xi)$ as constructed in Definition \ref{def:gen_FKMM_inv}. More precisely
$\rr{h}_{\rm univ}:=\kappa(\bb{T}_{2m}^\infty,\Xi)$ is the image of 
$$
\big[\big({\rm det}(\bb{T}_{2m}^\infty), s_{\bb{T}_{2m}^\infty}\big)\big]\;\in\;{\rm Vec}_{\rr{R}}^1(\hat{G}_{2m}(\C^\infty)|\hat{G}_{2m}\big(\C^\infty)^\rho,\rho\big)
$$
under the  natural isomorphism $
\tilde{\kappa}$ described in Proposition \ref{prop:nat_iso}.
\end{definition}

\medskip

\noindent
The \emph{naturality} of the invariant $\kappa$   implies the following important result.
\begin{theorem}[Universality]
Let $(\bb{E},\Theta)$ be a rank $2m$ \virg{Quaternionic} vector bundle over the involutive space $(X,\tau)$ classified by the equivariant map 
$\varphi: X\to \hat{G}_{2m}(\C^\infty)$. The generalized FKMM-invariant of $(\bb{E},\Theta)$ verifies
$$
\kappa(\bb{E},\Theta)\;=\; \varphi^*\big(\rr{h}_{\rm univ}\big)
$$
where $\varphi^*$ is the homomorphism in cohomology induced by $\varphi$.
\end{theorem}

\medskip

The identification of $\rr{h}_{\rm univ}$ as an element of $H^2_{\Z_2}(\hat{G}_{2m}(\C^\infty)|\hat{G}_{2m}(\C^\infty)^\rho,\Z(1))$  requires a careful investigation  of the equivariant cohomology of $\hat{G}_{2m}(\C^\infty)$. Some aspects of this analysis are quite technical and are postponed in Appendix \ref{app:Cohom_comput}. The main results are summarized below.
\begin{proposition}[Identification]
The following facts hold:
\begin{itemize}
\item[(1)] The group $H^2_{\Z_2}(\hat{G}_{2m}(\C^\infty),\Z(1))\simeq\Z$ is generated by the (universal) first \virg{Real} Chern class 
$$
\rr{c}^{\rr{R}}_1\;:=\;c^{\rr{R}}_1\Big({\rm det}\big(\bb{T}_{2m}^\infty\big)\Big)
$$ 
of the \virg{Real} line bundle ${\rm det}(\bb{T}_{2m}^\infty)\to \hat{G}_{2m}(\C^\infty)$ associated to the 
{tautological} $\rr{Q}$-bundle
$(\bb{T}_{2m}^\infty,\Xi)$. Moreover, under the map $f:H^2_{\Z_2}(\hat{G}_{2m}(\C^\infty),\Z(1))\to
H^2({G}_{2m}(\C^\infty),\Z)$ which forgets the $\Z_2$- action the class $\rr{c}^{\rr{R}}_1$ is mapped in the  first universal Chern class $\rr{c}_1\in H^2(G_{2m}(\C^\infty),\Z)$;
\vspace{1mm}
\item[(2)] 
The homomorphism
$$
\delta_2\;:\;H^{2}_{\Z_2}\big(\hat{G}_{2m}(\C^\infty)|\hat{G}_{2m}(\C^\infty)^\rho,\Z(1)\big)\;\stackrel{\simeq}{\longrightarrow}\;H^2_{\Z_2}\big(\hat{G}_{2m}(\C^\infty),\Z(1)\big)
$$
 in the long exact sequence \eqref{eq:Long_seq} for  the pair $\hat{G}_{2m}(\C^\infty)^\rho\hookrightarrow\hat{G}_{2m}(\C^\infty)$ is, in fact, an isomorphism.
\vspace{1mm}
\item[(3)] The universal FKMM-invariant $\rr{h}_{\rm univ}$ is the generator of $H^{2}_{\Z_2}(\hat{G}_{2m}(\C^\infty)|\hat{G}_{2m}(\C^\infty)^\rho,\Z(1))\simeq\Z$ which is mapped  by $\delta_2$ in the first \virg{Real} Chern class $\rr{c}^{\rr{R}}_1$, \ie 
$$
\rr{h}_{\rm univ}\;=\; \delta_2^{-1}\big(\rr{c}^{\rr{R}}_1\big)\;.
$$
\end{itemize}
\end{proposition}
\proof The proof of (1) and (2) are given in Proposition \ref{propos:item(1)} and Proposition \ref{propos:item(2)}, respectively. Item (3) follows from Corollary \ref{corol:II} (1).
\qed

\medskip

\begin{remark}[Comparison with the \emph{old} definition of the universal FKMM-invariant]\label{rk:comp_origin_univ}{\upshape
The \emph{old} definition of the {universal} FKMM-invariant  \cite[Definition 6.1]{denittis-gomi-14-gen} is based on an intricate construction
that we briefly recall. The bundle map $\pi:{\rm det}(\bb{T}_{2m}^\infty)\to \hat{G}_{2m}(\C^\infty)$ defines an equivariant map between involutive spaces 
 $$
 \pi\;:\;\Big(\n{S}\big({\rm det}(\bb{T}_{2m}^\infty)\big),{\rm det}(\Xi)\Big)\;\longrightarrow\;\hat{G}_{2m}(\C^\infty)\;.
 $$
The pullback of the {tautological} $\rr{Q}$-bundle $(\bb{T}_{2m}^\infty,\Xi)$ induced by $\pi$ defines a 
$\rr{Q}$-bundle $(\pi^*\bb{T}_{2m}^\infty,\pi^*\Xi)$ over the involutive space $(\n{S}({\rm det}(\bb{T}_{2m}^\infty)),{\rm det}(\Xi))$. According to the  {old} definition, the  {universal} FKMM-invariant $\rr{K}_{\rm univ}$ is  the \emph{old} FKMM-invariant (here denoted with $\kappa_0$)    of  $(\pi^*\bb{T}_{2m}^\infty,\pi^*\Xi)$. More precisely
$$
\rr{K}_{\rm univ}\;:=\;\kappa_{0}(\pi^*\bb{T}_{2m}^\infty,\pi^*\Xi)\;\in\; {\rm Coker}^1\Big(\n{S}\big({\rm det}(\bb{T}_{2m}^\infty)\big)|\n{S}\big({\rm det}(\bb{T}_{2m}^\infty)\big)^\Xi,\n{Z}(1)\Big)\;\simeq\;\Z_2\;
$$
agrees with the non trivial element of  $\Z_2$  \cite[Theorem 6.2]{denittis-gomi-14-gen}.
It is interesting to notice that the $\rr{Q}$-line bundle $(\pi^*\bb{T}_{2m}^\infty,\pi^*\Xi)$ is of  FKMM-type \cite[Lemma 6.2]{denittis-gomi-14-gen} and consequently the description in Proposition \ref{prop:genVSorig_inv} applies. This means that the \emph{new}  FKMM-invariant 
$$
\kappa(\pi^*\bb{T}_{2m}^\infty,\pi^*\Xi)\;\in\;H^2_{\Z_2}\Big(\n{S}\big({\rm det}(\bb{T}_{2m}^\infty)\big)|\n{S}\big({\rm det}(\bb{T}_{2m}^\infty)\big)^\Xi,\n{Z}(1)\Big)
$$
agree with the (non trivial) image of $\kappa_{0}(\pi^*\bb{T}_{2m}^\infty,\pi^*\Xi)$
under the injective map $\delta_1$. Thus, one can replace in the definition of 
$\rr{K}_{\rm univ}$ above the {old} invariant $\kappa_{0}$ with the {new} invariant $\kappa$.  Since  $\kappa$ is \emph{natural} (by construction), it follows that
$$
\rr{K}_{\rm univ}\;:=\;\kappa(\pi^*\bb{T}_{2m}^\infty,\pi^*\Xi)\;=\;\pi^*\kappa(\bb{T}_{2m}^\infty,\Xi)
\;=\;\pi^*(\rr{h}_{\rm univ})\;,
$$
namely the {old} version of the {universal} FKMM-invariant is just the pullback of the {new} {universal} FKMM-invariant $\rr{h}_{\rm univ}$ under the projection $\pi$.
}\hfill $\blacktriangleleft$
\end{remark}

\section{\virg{Quaternionic} line bundles and FKMM-invariant}
\label{sect:quat_lin_bund}
As a matter of fact \virg{Quaternionic} vector bundles of odd rank can be defined only over involutive base spaces $(X,\tau)$  with a \emph{free} involution (meaning that $X^\tau=\text{\rm  \O}$) \cite[Proposition 2.1]{denittis-gomi-14-gen}. In this section we investigate the rank one case providing a classification scheme for 
\virg{Quaternionic} \emph{line} bundles. 
{As a byproduct we provides the definition of FKMM-invariant for \virg{Quaternionic} \emph{line} bundles (Definition \ref{rk:FKMMcomp_lin_bun}
) that will be used in Theorem \ref{theo:A_inject_free} to provide a FKMM-invariant for for odd-rank \virg{Quaternionic} vector bundles.}
As a matter of notational simplification we prefer to replace in the next the pompous notation $(\bb{L},\Theta)$ with lighter symbols like $\bb{L}_{\rr{Q}}$ and $\bb{L}_{\rr{R}}$ for \virg{Quaternionic} and \virg{Real} line bundles, respectively.

\subsection{The \virg{Quaternionic} Picard torsor}
We denote by ${\rm Pic}_{\rr{R}}(X,\tau)\equiv{\rm Vec}_{\rr{R}}^1(X,\tau)$ and
${\rm Pic}_{\rr{Q}}(X,\tau)\equiv{\rm Vec}_{\rr{Q}}^1(X,\tau)$ the sets of isomorphism classes of \virg{Real} and \virg{Quaternionic} line bundles
on $(X,\tau)$, respectively. 
The set ${\rm Pic}_{\rr{R}}(X,\tau)$ gives rise to an abelian group under the tensor
product of \virg{Real} line bundles and it is known as the \emph{\virg{Real} Picard group}.
On the other hand  ${\rm Pic}_{\rr{Q}}(X,\tau)$ does not possess a  group structure under the tensor
product.
 Instead, under the essential assumption that ${\rm Pic}_{\rr{Q}}(X,\tau)\neq\text{\rm  \O}$,
 we can use the tensor product to define a left group-action of
${\rm Pic}_{\rr{R}}(X,\tau)$  on the set ${\rm Pic}_{\rr{Q}}(X,\tau)$:
$$
\begin{aligned}
 &{\rm Pic}_{\rr{R}}(X,\tau)\;\times\; {\rm Pic}_{\rr{Q}}(X,\tau)&\;\longrightarrow\;&\ \ \ \ \ \ \ {\rm Pic}_{\rr{Q}}(X,\tau)&\\
 &\ \ \ \ \ \  \ \  \ ([\bb{L}_{\rr{R}}],[\bb{L}_{\rr{Q}}])&\;\longmapsto\;&\ \ \ \ \  [\bb{L}_{\rr{R}}\;\otimes\;\bb{L}_{\rr{Q}}]\;.&
\end{aligned}
$$
Let us  introduce some terminology. For a group $\n{G}$, a (left) \emph{$\n{G}$-torsor} , is a set $T$ with a simply transitive (left) $\n{G}$-action such that the map
$$
\begin{aligned}
 \phi\;:\;&\n{G}\;\times\; T&\;\longrightarrow\;&\ \ \ \ \ \ \ T\;\times\; T&\\
 &\  \ (g,t)&\;\longmapsto\;&\ \  \ \ \ \ \  (gt,t)\;.&
\end{aligned}
$$
is an isomorphism. 
In other words, the $\n{G}$-action on $T$ has to be free, and $T$ has to be identifiable with a single $\n{G}$-orbit. We notice that one may think of a $\n{G}$-torsor as a principal  $\n{G}$-bundle over a single point.
\begin{theorem}[\virg{Quaternionic} Picard torsor]\label{teo:torsor_pica}
Assume ${\rm Pic}_{\rr{Q}}(X,\tau)\neq\text{\rm  \O}$. Then
 ${\rm Pic}_{\rr{Q}}(X,\tau)$ is a torsor under the group-action induced by ${\rm Pic}_{\rr{R}}(X,\tau)$.
\end{theorem}
\proof
Given a \virg{Quaternionic} line bundle
$\bb{L}_{\rr{Q}}$ let $\bb{L}_{\rr{Q}}^*$ be
the associated \emph{dual} bundle. The 
tensor product $\bb{L}_{\rr{Q}}\otimes\bb{L}_{\rr{Q}}^*$ provides a representative for the trivial element in ${\rm Pic}_{\rr{R}}(X,\tau)$. This fact can be checked, for instance,  by looking at the transition functions.
Given any pair $\bb{L}_{\rr{R}},\bb{L}_{\rr{R}}'\in{\rm Pic}_{\rr{R}}(X,\tau)$ and a $\bb{L}_{\rr{Q}}\in{\rm Pic}_{\rr{Q}}(X)$ let assume that $\bb{L}_{\rr{R}}\otimes\bb{L}_{\rr{Q}}\simeq \bb{L}_{\rr{R}}'\otimes\bb{L}_{\rr{Q}}$. By tensorizing both sides with $\bb{L}_{\rr{Q}}^*$ one obtains that $\bb{L}_{\rr{R}}\simeq \bb{L}_{\rr{R}}'$, namely  ${\rm Pic}_{\rr{R}}(X,\tau)$ acts {freely} on
${\rm Pic}_{\rr{Q}}(X,\tau)$. On the other hand,
for each pair $\bb{L}_{\rr{Q}},\bb{L}_{\rr{Q}}'\in{\rm Pic}_{\rr{Q}}(X,\tau)$
there is a \virg{Real} line bundle 
$\bb{L}_{\rr{R}}:=\bb{L}_{\rr{Q}}'\otimes\bb{L}_{\rr{Q}}^*$ such that 
$\bb{L}_{\rr{R}}\otimes\bb{L}_{\rr{Q}}\simeq \bb{L}_{\rr{Q}}'$. Then  the action of ${\rm Pic}_{\rr{R}}(X,\tau)$ is also transitive.
\qed

\begin{corollary}[Classification of $\rr{Q}$-line bundles]\label{cor:picQ}
Let $(X,\tau)$ be an involutive space such that $X^\tau=\text{\rm  \O}$
and ${\rm Pic}_{\rr{Q}}(X,\tau)\neq\text{\rm  \O}$. Then, 
$$
{\rm Pic}_{\rr{Q}}(X,\tau)\;\simeq\;{\rm Pic}_{\rr{R}}(X,\tau)\;\simeq\;H^2_{\Z_2}\big(X,\Z(1)\big)
$$
is a bijection of sets.
\end{corollary}
\proof
The first bijection is just a consequence of Theorem \ref{teo:torsor_pica} which assures that ${\rm Pic}_{\rr{Q}}(X,\tau)$ can be realized as a single orbit under the left action of ${\rm Pic}_{\rr{R}}(X,\tau)$. However, the form of the  bijection is not natural but depends on the choice of an initial element in ${\rm Pic}_{\rr{Q}}(X,\tau)$. The second bijection (in fact a group isomorphism)
is proved in
\cite{kahn-59} or  \cite[Corollary A.5]{gomi-13}.
\qed

\medskip

{Theorem} \ref{teo:torsor_pica} provides a way to extend the notion of (generalized) FKMM-invariant for 
\virg{Quaternionic} line bundles.

\begin{definition}[FKMM-invariant for $\rr{Q}$-line bundles]\label{rk:FKMMcomp_lin_bun}
 Assume  that ${\rm Pic}_{\rr{Q}}(X,\tau)\neq\text{\rm  \O}$ and let us fix
arbitrarily a \emph{reference} element  $[\bb{L}_{\rm ref}]\in {\rm Pic}_{\rr{Q}}(X,\tau)$. Given a $\bb{L}_{\rr{Q}}\in {\rm Pic}_{\rr{Q}}(X,\tau)$ 
let $\bb{L}_{\rr{R}}\in {\rm Pic}_{\rr{R}}(X,\tau)$ be the unique (up to isomorphisms) element such that $[\bb{L}_{\rr{Q}}]=[\bb{L}_{\rr{R}}\otimes \bb{L}_{\rm ref}]$. The FKMM-invariant of  $\bb{L}_{\rr{Q}}$, normalized by $\bb{L}_{\rm ref}$, is by definition
$$
\kappa\big(\bb{L}_{\rr{Q}}\big)\;:=\;c_1^{\rr{R}}\big(\bb{L}_{\rr{R}}\big)\;.
$$
\end{definition}

\noindent
Evidently the definition of the $\kappa$-invariant in the line bundle case suffers of the ambiguity due to the choice of a {reference} \virg{Quaternionic} line bundle $\bb{L}_{\rm ref}$. However,  this ambiguity can be understood as the freedom to fix the normalization $\kappa(\bb{L}_{\rm ref})=0$, a fact which is common in the theory of characteristic classes. From Definition \ref{rk:FKMMcomp_lin_bun} it results that $\kappa$ is   a complete classifying invariant for ${\rm Pic}_{\rr{Q}}(X,\tau)$. Moreover, this definition is in accordance with the result in Corollary \ref{corol:II} (2).

\medskip

The structure of the \virg{Quaternionic} line bundles can be investigated in a more direct way under some additional assumptions. Consider a representative $\bb{L}$ in ${\rm Pic}_{\rr{Q}}(X,\tau)$ such that the underlying line bundle is trivial in the  complex category, \ie $\bb{L}\simeq X\times\C$. In this situation a \virg{Quaternionic} structure $\Theta$ is fixed by a continuous map $q:X\to\n{U}(1)$ such that $\Theta:(x,\lambda)\mapsto(\tau(x), q(x)\overline{\lambda})$. The \virg{Quaternionic} constraint requires $
q(\tau(x))\overline{q(x)}=-1$, or equivalently $
q(\tau(x))=-{q(x)}$, for all $x\in X$. Evidently, this condition cannot be verified  if the involution $\tau$ is not free.
The identification $\n{U}(1)\simeq\n{S}^1$ says that the 
\virg{Quaternionic} structure $\Theta$ is specified by a $\Z_2$-equivariant map $q$ from the involutive space $(X,\tau)$ into the  one-dimensional sphere endowed with the {antipodal} free involution   ${\n{S}}^{0,2}$. Since equivariant homotopy deformations produce isomorphic \virg{Quaternionic} vector bundles we can conclude that each equivariant homotopy class in $[X,{\n{S}}^{0,2}]_{\Z_2}$ identifies a \virg{Quaternionic} structure on the product line bundle $\bb{L}\simeq X\times\C$. Finally, we notice that the \virg{Quaternionic} structures induced by an equivariant map $q$ and its opposite $-q$ are $\rr{Q}$-isomorphic. In conclusion  we obtained that
\virg{Quaternionic} structures on a product line bundle are classified by $[X,{\n{S}}^{0,2}]_{\Z_2}/{\Z_{2}}$ where the $\Z_{2}$-action is induced by the multiplication by $-1$.
We are now in position to state the next result.
\begin{proposition}\label{prop:picQ_triv}
Let $(X,\tau)$ be an involutive space which verifies the following condition:
\begin{itemize}
\item[(a)] $X^\tau=\text{\rm  \O}$;
\vspace{1mm}
\item[(b)] $H^2(X,\Z)$ has no torsion;
\vspace{1mm}
\item[(c)] The pullback homomorphism $\tau^*:H^2(X,\Z)\to H^2(X,\Z)$ acts as the identity.
\end{itemize}
Then the following bijection
$$
{\rm Pic}_{\rr{Q}}(X,\tau)\;\simeq\; \big[X,{\n{S}}^{0,2}\big]_{\Z_2}/{\Z_{2}}
$$
holds true. The result is evidently valid under the stronger condition $H^2(X,\Z)=0$, which implies \emph{(b)} and \emph{(c)}.
\end{proposition}
\proof
Condition (a) is necessary for the existence of \virg{Quaternionic} line bundles. Let $\bb{L}$ be
an element in ${\rm Pic}_{\rr{Q}}(X,\tau)$. Conditions (b) and (c) are sufficient to ensure that $\bb{L}\simeq X\times\C$ in the complex category. In fact  the presence of a \virg{Quaternionic} structure induces an isomorphism of complex vector bundles 
$$
\tau^*\bb{L}\;\simeq\; \overline{\bb{L}}
$$
where $\overline{\bb{L}}$ is the \emph{conjugated} complex line bundle obtained by reversing the complex structure in the fibers of $\bb{L}$. This isomorphisms leads to the constraint
\begin{equation}
\tau^*c_1(\bb{L})\;=\;-c_1(\bb{L})
\end{equation}
for the first Chern class of $\bb{L}$. Condition (c) implies $2c_1(\bb{L})=0$ and condition (b) assures that $c_1(\bb{L})=0$. This fact forces $\bb{L}$ to be trivial (in the complex category) since complex line bundle are completely classified by the first Chern class.\qed

\medskip

One interesting aspect of the theory of \virg{Quaternionic} line bundles is that there is no natural candidate for the definition of a \emph{trivial} element in ${\rm Pic}_{\rr{Q}}(X,\tau)$!

\subsection{Classification over involutive spheres}\label{sect:inv_spher_lin_bund}
We discuss in this  section  the classification of \virg{Quaternionic} line bundles over the involutive spheres of type $\n{S}^{0,d}$.

\begin{example}[The Dupont $\rr{Q}$-line bundle over $\n{S}^{0,1}$]\label{ex:Dupont}
{\upshape
The zero-dimensional sphere $\n{S}^{0,1}$ coincides with the \emph{two-points} space $\{-1,+1\}$ endowed with the flip-action $\theta_{0,1}:\pm1\mapsto\mp 1$. We can construct a  \virg{Quaternionic} line bundle over  $\n{S}^{0,1}$ as follows: Let $\bb{L}:=\{-1,+1\}\times\C$ be the product line bundle and define   $\Theta_0: \bb{L}\to \bb{L}$ by
$$
\Theta_0\;:\;\big(\pm1,\lambda\big)\;\mapsto\;\big(\mp1,\pm\overline{\lambda}\big)\;,\qquad\quad \lambda\in\C\;.
$$
This example has been introduced for the fist time by J. L. Dupont in \cite{dupont-69}. In particular it proves that, ${\rm Pic}_{\rr{Q}}(\n{S}^{0,1})\neq\text{\rm  \O}$. The next natural question is whether there are other elements in ${\rm Pic}_{\rr{Q}}( \n{S}^{0,1})$ distinguished from the 
Dupont's example. A way to answer this question is to use the characterization of Corollary 
\ref{cor:picQ}. To compute the group $H^2_{\Z_2}( \n{S}^{0,1},\Z(1))$ let us start with the ordinary cohomology
$$
H^k\big(\{-1,+1\},\Z\big)\;\simeq\;
\left\{
\begin{aligned}
&\Z^2&& \ \text{if}\  \ \ k=0\\
&0&&\ \text{if}\  \ \ k\neq0\\
\end{aligned}
\right.
$$
and with the equivariant cohomology with fixed coefficients
$$
H^k_{\Z_2}\big( \n{S}^{0,1},\Z\big)\;\simeq\;H^k\big(\{\ast\},\Z\big)\;\simeq\;
\left\{
\begin{aligned}
&\Z&& \ \text{if}\  \ \ k=0\\
&0&&\ \text{if}\  \ \ k\neq0\\
\end{aligned}
\right.
$$
The application of the exact sequence in \cite[Proposition 2.3]{gomi-13}   produces
$H^2_{\Z_2}(\n{S}^{0,1},\Z\big)\simeq H^2_{\Z_2}(\n{S}^{0,1},\Z(1))$ and consequently
$$
{\rm Pic}_{\rr{Q}}\big(\n{S}^{0,1}\big)\;\simeq\; H^2_{\Z_2}\big(\n{S}^{0,1},\Z(1)\big)\;=\;0
$$
contains a single element that can be identified with the Dupont's example. The unicity (up to isomorphism) of this representative can be checked also in a more direct way. First of all we observe  that every complex vector bundle over $\{-1,+1\}$ is necessarily trivial. This means that every  \virg{Quaternionic} line bundle over  $\n{S}^{0,1}$ is built from the underlying complex line bundle  $\bb{L}:=\{-1,+1\}\times\C$. Any \virg{Quaternionic} structure on $\bb{L}$
can be specified by a $t\in[0,2\pi)$ according to
$$
\Theta_t\;:\;\big(\pm1,\lambda\big)\;\mapsto\;\big(\mp1,\pm\expo{\ii t}\overline{\lambda}\big)\;,\qquad\quad \lambda\in\C\;.
$$
On the other hand a $\rr{Q}$-isomorphism  between two \virg{Quaternionic} structures $\Theta_t$ and $\Theta_{t'}$ is provided by a map
$$
f_\phi\;:\;\big(\pm1,\lambda\big)\;\mapsto\;\big(\pm1,\phi(\pm1){\lambda}\big)\;,\qquad\quad \lambda\in\C\;
$$
subjected to the equivariant condition $f_\phi\circ\Theta_t =\Theta_{t'}\circ f_\phi$. 
For each pair $t,t'\in[0,2\pi)$ such a map is fixed by the prescription $\phi(+1):=+1$ and 
$\phi(-1):=\expo{\ii(t'-t)}$. This proves directly that each \virg{Quaternionic} line bundle over 
$\{-1,+1\}$ is isomorphic to the Dupont's example which corresponds to $t=0$. }\hfill $\blacktriangleleft$
\end{example}

\medskip

\begin{example}[$\rr{Q}$-line bundles over ${\n{S}}^{0,d+1}$ with $d\geqslant 3$]\label{Ex:lin_bun1}
{\upshape
Recall that ${\n{S}}^{0,d+1}$ agrees with $\n{S}^d$ as topological space and  $H^2(\n{S}^d,\Z)=0$ for $d\geqslant 3$. Then, we can use the result in Proposition \ref{prop:picQ_triv} to state that 
$$
{\rm Pic}_{\rr{Q}}\big({\n{S}}^{0,d+1}\big)\;\simeq\; \big[{\n{S}}^{0,d+1},{\n{S}}^{0,2}\big]_{\Z_2}/{\Z_{2}},\qquad\quad d\geqslant 3.
$$
However, an analysis of the equivariant homotopy shows that  
\begin{equation}\label{eq:homotZ2_sphere}
\big[{\n{S}}^{0,d+1},{\n{S}}^{0,2}\big]_{\Z_2}\;=\;\text{\rm  \O},\qquad\quad d\geqslant 2\;.
\end{equation}
and this leads to the conclusion that it is not possible to build \virg{Quaternionic} line bundle over involutive sphere ${\n{S}}^{0,d+1}$ of dimension bigger than two. As a matter of completeness, let us
justify equation \eqref{eq:homotZ2_sphere}. The involutive space ${\n{S}}^{0,d+1}$ can be seen as the total space of a $\Z_2$-sphere bundle ${\n{S}}^{0,d+1}\to \R P^d$ and it is well known that the Stiefel-Whitney class of this bundle $w_1({\n{S}}^{0,d+1})$ is given by the non-trivial element of $H^1(\R P^d,\Z_2)\simeq\Z_2$ for all $d\geqslant 1$.
Let us assume the existence of a $\Z_2$-equivariant map $q:{\n{S}}^{0,d+1}\to{\n{S}}^{0,2}$. Such a map  would induce a map $\tilde{q}: \R P^d\to \R P^1$ on the orbit spaces
and an isomorphism ${\n{S}}^{0,d+1}\simeq \tilde{q}^*{\n{S}}^{0,2}$ of bundles over $\R P^d$ where
$\tilde{q}^*{\n{S}}^{0,2}$ is the pullback of the $\Z_2$-sphere bundles ${\n{S}}^{0,2}\to \R P^1$. By functoriality, this would imply $w_1({\n{S}}^{0,d+1})=w_1(\tilde{q}^*{\n{S}}^{0,2})=\tilde{q}^*w_1({\n{S}}^{0,2})$ showing that $\tilde{q}^*w_1({\n{S}}^{0,2})$ is not trivial. On the other hand one has that
$$
\big[\R P^d, \R P^1\big]\;\simeq\;\big[\R P^d, K(\Z,1)\big]\;\simeq\;H^1\big(\R P^d,\Z\big)\;=\;0\;,\qquad\quad d\geqslant 2\;
$$
where  $\R P^1$ has been identified with the Eilenberg-MacLane space  $K(\Z,1)$. Therefore $\tilde{q}$ must be homotopy equivalent to the constant map and this contradicts the non-triviality of $\tilde{q}^*w_1({\n{S}}^{0,2})$. This contradiction shows that the set of $\Z_2$-equivariant map from ${\n{S}}^{0,d+1}$ into ${\n{S}}^{0,2}$ must be empty in agreement with \eqref{eq:homotZ2_sphere}.  }\hfill $\blacktriangleleft$
\end{example}

\medskip

\begin{example}[$\rr{Q}$-line bundles over ${\n{S}}^{0,2}$]\label{Ex:lin_bun2}
{\upshape
Let us start by proving that ${\rm Pic}_{\rr{Q}}({\n{S}}^{0,2})\neq\text{\rm  \O}$.
Since $H^2(\n{S}^1,\Z)=0$ we can use the same argument as in {Proposition} \ref{prop:picQ_triv}. 
The vanishing of the first Chern class implies that each (possible)
\virg{Quaternionic} line bundle over ${\n{S}}^{0,2}$ must be built over  the underlying product bundle  $\bb{L}_0:=\n{S}^1\times\C$. Then, a (possible) \virg{Quaternionic} structure $\Theta_q:\bb{L}_0\to \bb{L}_0$ must be specified   by a map $q:{\n{S}}^{1}\to\n{U}(1)$ which fulfills the constraint $q(-k)=-q(k)$ for all $k\in{\n{S}}^{1}$ according to the recipe
$$
\Theta_q\;:\;\big(k,\lambda\big)\;\mapsto\;\big(-k,q(k)\overline{\lambda}\big)\;,\qquad\quad (k,\lambda)\in\n{S}^1\times\C\;.
$$
A \emph{standard} choice for such a map is the following:
\begin{equation}\label{eq:mapq_0}
\begin{aligned}
 q_0\;:\;&\ \ \ \ {\n{S}}^1\;&\;\longrightarrow\;&\ \ \ \ \ \ \ \ \n{U}(1)&\\
 &(k_0,k_1)&\;\longmapsto\;&\ \  \ \ \ \ \  k_0+\ii k_1\;.&
\end{aligned}
\end{equation}
The existence of $q_0$ assures that  ${\rm Pic}_{\rr{Q}}({\n{S}}^{0,2})$ is nonempty and so the classification of Corollary \ref{cor:picQ} applies. The computation in 
Proposition \ref{prop:cohom_free_spher} gives 
$$
{\rm Pic}_{\rr{Q}}\big({\n{S}}^{0,2}\big)\;=\;0\;,
$$
namely there is only one isomorphism class of  \virg{Quaternionic} line bundles over ${\n{S}}^{0,2}$ with a representative  given by $(\bb{L}_0,q_0)$.
It is possible to verify this  claim in a more
  direct way.  A $\rr{Q}$-isomorphism  between two \virg{Quaternionic} structures $\Theta_q$ and $\Theta_{q_0}$ is specified by a map $\phi:\n{S}^1\to \n{U}(1)$ such that 
$$
f_\phi\;:\;\big(k,\lambda\big)\;\mapsto\;\big(k,\phi(k){\lambda}\big)\;,\qquad\quad (k,\lambda)\in\n{S}^1\times\C\;,
$$
and subjected to the equivariance condition $f_\phi\circ\Theta_q =\Theta_{q_0}\circ f_\phi$. This 
translates into
$$
q_0(k)\;=\;\phi(-k)\;\phi(k)\;q(k) \qquad\quad k\in\n{S}^1\;.
$$
In a general way
the map $q$ can be related to $q_0$ by a relation of the form
\begin{equation}\label{eq:Q_iso_Lin}
q(k)\;=\;q_0(k)^n\;\expo{\ii 2\pi\; \chi(k)}\qquad\quad k\in\n{S}^1\;
\end{equation}
for some $n\in\Z$ and $\chi:\n{S}^1\to\R$. The \virg{Quaternionic} constraints $-1= q_0(-k)\overline{q_0(k)}$ and $-1= q(-k)\overline{q(k)}$ impose the condition
$$
\expo{\ii 2\pi [\chi(-k)- \chi(k)]} \;=\;(-1)^{n+1}\qquad\quad k\in\n{S}^1\;.
$$
If $n$ were even  $\Delta\chi(k):=\chi(-k)- \chi(k)$ should take
 values in $\Z+\frac{1}{2}$. However,   $\Delta\chi$ is continuous and so $\Delta\chi(k)$ should take a constant value $N+\frac{1}{2}$ for some $N\in\Z$ and for all $k$. But this  is incompatible with the 
parity of 
$\Delta\chi$ which requires $\Delta\chi(-k)=-\Delta\chi(k)$. Then $n=2p+1$ must be odd and $\Delta\chi$ must take values in $\Z$. Again, the continuity and the parity of $\Delta\chi$ imply that 
$\Delta\chi(k)=0$, or equivalently $\chi(-k)= \chi(k)$ for all $k$. This allows to rewrite equation \eqref{eq:Q_iso_Lin} in the form 
$$
q_0(k)\;=\;q_0(k)^{-2p}\expo{-\ii 2\pi\; \chi(k)}q(k)\;=\;\phi(-k)\;\phi(k)\;q(k) \qquad\quad k\in\n{S}^1\;
$$
with  $\phi(k):=(-1)^{\frac{p}{2}}q_0(k)^{-p}\expo{-\ii \pi\; \chi(k)}$. The last equation shows that any  \virg{Quaternionic} structure $\Theta_q$ is $\rr{Q}$-isomorphic to the {standard} \virg{Quaternionic} structure $\Theta_{q_0}$.
  }\hfill $\blacktriangleleft$
\end{example}

\medskip

\begin{example}[$\rr{Q}$-line bundles over ${\n{S}}^{0,3}$]\label{Ex:lin_bun3}{\upshape
Again, let us start by showing that ${\rm Pic}_{\rr{Q}}({\n{S}}^{0,3})\neq\text{\rm  \O}$. This case is richer than those described in Example \ref{Ex:lin_bun1} and Example \ref{Ex:lin_bun2} since $H^2({\n{S}}^2,\Z)\simeq\Z$ which in turn implies that for each $\ell\in\Z$ there is an inequivalent complex line bundle $\bb{L}_\ell\to {\n{S}}^2$, specified by the Chern  class $\ell$. In principle each of these $\bb{L}_\ell$
can be used as the underlying line bundle for a \virg{Quaternionic} structure.
Equation \eqref{eq:homotZ2_sphere} says that it is not possible to endow the trivial line bundle $\bb{L}_0:={\n{S}}^2\times \C$ with a  \virg{Quaternionic} structure. Let us investigate the case $\ell=1$. A representative for 
$\bb{L}_1\to {\n{S}}^2$  can be constructed from the family of 
\emph{Hopf projections}  
\beql{eq:proj_sphe_Q_struct_line}
{P}_{\rm Hopf}(k_0,k_1,k_2)\;:=\;\frac{1}{2}\left(\n{1}_2\;+\;\sum_{j=0}^2k_j\; \sigma_{j+1}\right)\;,\qquad\quad k:=(k_0,k_1,k_2)\in\n{S}^2\;
\eeq
according to 
$$
\bb{L}_1\;:=\;\bigsqcup_{k\in \n{S}^2}{\rm Ran}\big({P}_{\rm Hopf}(k)\big)
$$
where $\sigma_1,\sigma_2$ and $\sigma_3$ are the  Pauli matrices and ${\rm Ran}\big({P}_{\rm Hopf}(k)\big)\subset\C^2$ is the one-dimensional subspace spanned by ${P}_{\rm Hopf}(k)$.
Moreover, $\bb{L}_1$ has a \virg{Quaternionic} structure induced by the anti-linear symmetry
$$
\Theta_1\; {P}_{\rm Hopf}(k_0,k_1,k_2)\; \Theta_1^*\;=\;{P}_{\rm Hopf}(-k_0,-k_1,-k_2)\;,\qquad\quad \Theta_1\;:=\;\sigma_2\circ C
$$
where $C$ is the operator which implements the complex conjugation on $\C^2$ (\cf Section \ref{sect:non-trivial_ex}). 
The existence of the \virg{Quaternionic} line bundle $(\bb{L}_1,\Theta_1)$
 assures that  ${\rm Pic}_{\rr{Q}}({\n{S}}^{0,3})$ is non-empty and so the 
classification of Corollary \ref{cor:picQ} applies.  
Proposition \ref{prop:cohom_free_spher} provides 
$$
{\rm Pic}_{\rr{Q}}\big({\n{S}}^{0,3}\big)\;\simeq\;H^2\big({\n{S}}^{0,3},\Z(1)\big)\;\simeq\;\Z\;,
$$
namely  there are $\Z$ distinct isomorphism classes of  $\rr{Q}$-line bundles over ${\n{S}}^{0,3}$. A look to the exact sequence \cite[Proposition 2.3.]{gomi-13}
$$
\begin{diagram}
&H^{1}_{\Z_2}\big({\n{S}}^{0,3},\Z\big)&\rTo^{}&        H^{2}_{\Z_2}\big({\n{S}}^{0,3},\Z(1)\big)&\rTo^{f}&   
H^{2}\big({\n{S}}^2,\Z\big)&\rTo^{}&  H^{2}_{\Z_2}\big({\n{S}}^{0,3},\Z\big) &\rTo^{}& H^{3}_{\Z_2}\big({\n{S}}^{0,3},\Z(1)\big)\vspace{-0mm}\\
&    \rotatebox{-90}{$=$}&&\rotatebox{-90}{$\simeq$} && \rotatebox{-90}{$\simeq$} && \rotatebox{-90}{$\simeq$} && \rotatebox{-90}{$=$}   \\
 &   0&&\Z && \Z &&\Z_2 &&0   \\                                    \end{diagram} 
$$
shows that the map $f$  which forgets the $\Z_2$-structure  acts as the multiplication by $2$.
By combining this fact with the isomorphism $c_1:{\rm Pic}_{\C}({\n{S}}^{2})\simeq H^{2}\big({\n{S}}^{2},\Z\big)$ induced by the first Chern class and the isomorphism $c_1^{\rr{R}}: {\rm Pic}_{\rr{R}}({\n{S}}^{0,3})\to H^{2}_{\Z_2}\big({\n{S}}^{0,3},\Z(1)\big)$ induced by the first \virg{Real} Chern class one concludes that there is an \emph{isomorphisms of groups}
$$
\begin{aligned}
 &{\rm Pic}_{\rr{R}}\big({\n{S}}^{0,3}\big)&\;\stackrel{\simeq}{\longrightarrow}\;&\ \ \ \ \ \ \ 2\Z&\\
 &  \ \ \  [\bb{L}_{\rr{R}}]&\;\longmapsto\;&\ \ \ \ \ c_1(\bb{L}_{\rr{R}})\;&
\end{aligned}
$$
where $\bb{L}_{\rr{R}}$ is any  \virg{Real} line bundle over ${\n{S}}^{0,3}$. The characterization of Theorem \ref{teo:torsor_pica} which describes ${\rm Pic}_{\rr{Q}}\big({\n{S}}^{0,3}\big)$ as a torsor over ${\rm Pic}_{\rr{R}}\big({\n{S}}^{0,3}\big)$ implies the \emph{bijection} of sets given by
$$
\begin{aligned}
 &\ \ \ \ \ \ \ \ \ {\rm Pic}_{\rr{Q}}\big({\n{S}}^{0,3}\big)&\;\stackrel{\simeq}{\longrightarrow}\;&\ \ \ \ \ \ \ \ \ \ \ \ \  2\Z\;+\;1&\\
 &  \ \ \  [\bb{L}_{\rr{Q}}]\;\simeq\;[\bb{L}_{\rr{R}}\otimes\bb{L}_1]&\;\longmapsto\;&\ \ \ \ \ c_1(\bb{L}_{\rr{Q}})\;=\;c_1(\bb{L}_{\rr{R}})\;+\;c_1(\bb{L}_{1})\;.&
\end{aligned}
$$
We point out that the \virg{Quaternionic} line bundle $\bb{L}_1$  has been used as the \emph{reference} element for the identification of  
${\rm Pic}_{\rr{Q}}({\n{S}}^{0,3})$ as a ${\rm Pic}_{\rr{R}}({\n{S}}^{0,3})$-orbit. Summarizing, one has:
\begin{itemize}
\item[(1)] \emph{A complex line bundle can support a unique (up to isomorphisms) \virg{Real} structure over the two-dimensional sphere endowed with the free antipodal action if and only if its first Chern class is even and different Chern classes distinguish between inequivalent  \virg{Real} line bundles;}
\vspace{1mm}
\item[(2)] 
\emph{A complex line bundle can support a unique (up to isomorphisms) \virg{Quaternionic} structure over the two-dimensional sphere endowed with the free antipodal action if and only if its first Chern class is odd and different Chern classes distinguish between inequivalent  \virg{Quaternionic} line bundles. Moreover, 
for each $\bb{L}_{\rr{Q}}\in {\rm Pic}_{\rr{Q}}({\n{S}}^{0,3})$ is specified by its
FKMM-invariant given by
$$
\kappa(\bb{L}_{\rr{Q}})\;=\;\frac{c_1(\bb{L}_{\rr{Q}})-1}{2}\;.
$$ 
constructed according to Definition \ref{rk:FKMMcomp_lin_bun}.
}
\end{itemize}
Let us mention that that results similar to (1) and (2)  have been recently derived in \cite{gat-robbins-15} with a different technique based on the analysis 
of obstructions to the construction of trivializing frames. It is interesting to note that our derivation of (1) and (2) turns out to be no more than an exercise once Theorem \ref{teo:torsor_pica} has been established.
 }\hfill $\blacktriangleleft$
\end{example}

\section{Topological classification of \virg{Quaternionic} vector bundles}\label{sect:top_class_low}
In this section we analyze the role of the FKMM-invariant in the classification of the \virg{Quaternionic} vector bundles. It turns out that the $\kappa$-invariant is an extremely efficient tool to solve the classification problem in \emph{low} dimensions.
Just to fix the terminology (for the present work) we say that:
\begin{definition}[Low dimension]\label{def:low_dim}
A topological space $X$ which verifies Assumption \ref{ass:top} is said to be \emph{low} dimensional if $0\leqslant d\leqslant 3$.
\end{definition}
%

\subsection{Stable rank condition}
\label{sect:stab_rang}
The {stable rank condition} for vector bundles expresses the pretty general fact that the non trivial topology can be concentrated in a sub-vector bundle of \emph{minimal} rank. This minimal value  depends on the dimensionality of the base space and on the category of vector bundles under consideration. For complex (as well as real or quaternionic) vector bundles the  stable rank condition is a well-known result (see \eg \cite[Chapter 9, Theorem 1.2]{husemoller-94}).
The proof of this result is based on an \virg{obstruction-type argument} which provides the explicit construction of a certain \emph{maximal} number of global sections \cite[Chapter 2, Theorem 7.1]{husemoller-94}. The  latter argument can be generalized to vector bundles over spaces with involution 
by means of the notion of  $\Z_2$-CW-complex \cite{matumoto-71,allday-puppe-93} (see also \cite[Section 4.5]{denittis-gomi-14}). A $\Z_2$-CW-complex is a  CW-complex  made by cells of various dimension that carry a $\Z_2$-action. 
These $\Z_2$-cells can be only of two types: They are \emph{fixed} if the action of $\Z_2$ is trivial or they are \emph{free} if they have no fixed points. 
Since this construction is modelled after the usual definition of CW-complex, just by replacing the \virg{point} by \virg{$\Z_2$-point}, (almost) all topological and homological properties valid for  CW-complexes have their \virg{natural} counterparts in the equivariant setting. The use of this technique is essential		for the determination of  the 
stable rank condition in the case of \virg{Real} vector bundles  \cite[Theorem 4.25]{denittis-gomi-14} and  even rank \virg{Quaternionic} vector bundles  \cite[Theorem 2.5]{denittis-gomi-14-gen}. In this section we discuss the   generalization of the  stable rank condition in the \virg{Quaternionic} category in situation not covered by in  \cite[Theorem 2.5]{denittis-gomi-14-gen} (\eg $X^\tau$ of any codimension)
 and for odd rank vector bundles. We start with the even dimensional case.
\begin{theorem}[Stable condition: even rank]
\label{theo:stab_ran_Q_even}
Let $(X,\tau)$ be an involutive space such that $X$ has a finite $\Z_2$-CW-complex decomposition of dimension $d$. 
Each rank $2m$ \virg{Quaternionic} vector bundle $(\bb{E},\Theta)$  over $(X,\tau)$ such that $d\leqslant 4m-3$
splits as
\beql{eq:stab_rank_Q}
\bb{E}\;\simeq\;\bb{E}_0\;\oplus\;(X\times\C^{2(m-\sigma)})
\eeq
where $\sigma:=[\frac{d+2}{4}]$ (here $[x]$ denotes the integer part of $x\in\R$), 
$\bb{E}_0$ is a (possible) non-trivial \virg{Quaternionic} vector bundle   of rank $2\sigma$ and the 	
remainder
 is the trivial  \virg{Quaternionic} vector bundle of rank $2(m-\sigma)$. As a consequence, one has
\beql{eq:stab_rank_Q_low_d}
{\rm Vec}^{2m}_{\rr{Q}}\big(X, \tau\big)\;\simeq\; {\rm Vec}^{2\sigma}_{\rr{Q}}\big(X, \tau\big)\qquad\quad \forall\ m \;\geqslant\; \frac{d+3}{4}\;
\eeq
and more specifically
\begin{align}
{\rm Vec}^{2m}_{\rr{Q}}\big(X, \tau\big)&\;=\; 0& \text{if}& \ \ d=0,1\ \ \ \ \ \ \forall\ m\in\N \label{eq:stab_rank_Q_low_d=1}\\
{\rm Vec}^{2m}_{\rr{Q}}\big(X, \tau\big)&\;\simeq\; {\rm Vec}^{2}_{\rr{Q}}\big(X, \tau\big)& \text{if}& \ \ 2\leqslant d\leqslant 5\ \ \ \forall\ m\in\N\;.\label{eq:stab_rank_Q_low_d>1}
\end{align}

\end{theorem}
\proof
The  claim above generalizes that of \cite[Theorem 2.5]{denittis-gomi-14-gen} 
because it includes also the cases: (a) $X^\tau=\text{\rm  \O}$ and, (b) $X^\tau$ a $\Z_2$-CW-complex of dimension bigger than zero. 
These extensions are subject to the possibility of generalizing in the same direction the key construction described in the proof of \cite[Proposition 2.7]{denittis-gomi-14-gen}. 
For the case (a) this generalization is trivial since we need only to assume that the number of fixed points is zero. At the first step of the inductive argument one must deal only with pairs of conjugated points $\{x_j,\tau(x_j)\}$ and one can construct $\rr{Q}$-pairs of independent sections $s_1,s_2$ over $\{x_j,\tau(x_j)\}$ just by choosing $s_1(x_j)={\rm v}_1$, $s_1(\tau{x_j})={\rm v}_2$, $s_2(x_j)=\Theta({\rm v}_2)$, $s_2(\tau{x_j})=\Theta({\rm v}_1)$ with  ${\rm v}_1,{\rm v}_2\in \C^{2m}\setminus\{0\}$
and ${\rm v}_2\neq \lambda \Theta({\rm v}_1)$ for some $\lambda\in\C$. At this point the rest of the 
argument follows exactly as in the proof of \cite[Proposition 2.7]{denittis-gomi-14-gen}. The validity of the case (b)
has been already anticipated and justified in \cite[Remark 2.4]{denittis-gomi-14-gen}.
\qed

\medskip

\begin{remark}[Stable rank condition for the  \virg{Real} case]\label{rk:stable_rank_Real}{\upshape
The same argument used in the proof of  Theorem \ref{theo:stab_ran_Q_even}  can be applied to extend the stable rank condition for  \virg{Real} vector bundles \cite[Theorem 4.25]{denittis-gomi-14} to the case  of a free action. In summary, under the conditions: (a)  $X^\tau=\text{\rm  \O}$ or (b') $X^\tau$ a $\Z_2$-CW-complex of dimension zero, one has that 
\beql{eq:stab_rank_R_low_d}
{\rm Vec}^{m}_{\rr{R}}\big(X, \tau\big)\;\simeq\; {\rm Vec}^{\sigma}_{\rr{R}}(X, \tau)\qquad\quad \forall\ m \;\geqslant\; \frac{d+1}{2}\;
\eeq
where $\sigma:=[\frac{d}{2}]$,
and more specifically
\begin{align}
{\rm Vec}^{m}_{\rr{R}}\big(X, \tau\big)&\;=\; 0& \text{if}& \ \ d=0,1\ \ \ \ \ \ \forall\ m\in\N \label{eq:stab_rank_R_low_d=1}\\
{\rm Vec}^{m}_{\rr{R}}\big(X, \tau\big)&\;\simeq\; {\rm Vec}^{1}_{\rr{R}}\big(X, \tau\big)& \text{if}& \ \ 2\leqslant d\leqslant 3\ \ \ \forall\ m\in\N\;.\label{eq:stab_rank_R_low_d>1}
\end{align}

The situation turns out to be quite different when $X^\tau$ is a $\Z_2$-CW-complex of dimension bigger than zero
as pointed out in \cite[Remark 4.24]{denittis-gomi-14}.
}\hfill $\blacktriangleleft$
\end{remark}

\medskip

The odd rank case requires a slight different analysis and strongly  depends on the existence
 of a \virg{Quaternionic}  line bundle. The following result provides the key argument.
\begin{lemma}\label{lemma:decmp_oddrankQbun}
Let $(X,\tau)$ be an involutive space such that $X^\tau=\text{\emph{\O}}$ and ${\rm Pic}_{\rr{Q}}(X,\tau)\neq\text{\emph{\O}}$. Then, one has a bijection
$$
{\rm Vec}^{m}_{\rr{Q}}\big(X, \tau\big)\;\simeq\;{\rm Vec}^{m}_{\rr{R}}\big(X, \tau\big)\;\qquad\quad \forall\ m\in\N\;.
$$
Moreover, 
$$
{\rm Vec}^{2m+1}_{\rr{Q}}\big(X, \tau\big)\;=\;\text{\emph{\O}}\ \ \ \ \ \Leftrightarrow \ \ \ \ \  {\rm Pic}_{\rr{Q}}\big(X,\tau\big)\;=\;\text{\emph{\O}}\;.
$$
\end{lemma}
\proof
The proof of the second claim is easy.
 If $[\bb{L}_\rr{Q}]\in {\rm Pic}_{\rr{Q}}\big(X,\tau\big)\neq\text{\rm  \O}$ one has that 
$\bb{L}_{\rr{Q}}\oplus(X\times\C^{2m})$ is a \virg{Quaternionic}  vector bundle of rank $2m+1$ and so   ${\rm Vec}^{2m+1}_{\rr{Q}}(X, \tau)\neq\text{\rm  \O}$.
Conversely, if $[(\bb{E}, \Theta)]\in {\rm Vec}^{2m+1}_{\rr{Q}}(X, \tau)\neq\text{\rm  \O}$, then  
the line bundle
${\rm det}(\bb{E})$ inherits a \virg{Quaternionic} structure given by ${\rm det}(\Theta)$  and so 
${\rm Pic}_{\rr{Q}}(X,\tau )\neq\text{\rm  \O}$. Let now  assume that ${\rm Pic}_{\rr{Q}}\big(X,\tau\big)\neq\text{\rm  \O}$ and choose a representative $\bb{L}_\rr{Q}$ of some element in ${\rm Pic}_{\rr{Q}}\big(X,\tau\big)$. The map
 $$
\begin{aligned}
 \bb{L}_\rr{Q}\otimes\; :&\ \ {\rm Vec}^{m}_{\rr{Q}}\big(X, \tau\big)&\;\stackrel{\simeq}{\longrightarrow}\;&\ \ {\rm Vec}^{m}_{\rr{R}}\big(X, \tau\big)&\\
 &\ \ \ \ \ \  \ \  \ [\bb{E}_{\rr{Q}}]&\;\longmapsto\;&\ \  [\bb{L}_{\rr{Q}}\;\otimes\;\bb{E}_{\rr{Q}}]\;&
\end{aligned}
$$ 
turns out to be a bijection of sets for each $m\in\N$. This fact can be proved with an argument similar to that in the proof of Theorem \ref{teo:torsor_pica} and is based on the observation that  $\bb{L}_\rr{Q}\otimes\bb{L}_\rr{Q}^*$ is equivalent to the trivial element in ${\rm Pic}_{\rr{R}}\big(X,\tau\big)$.
\qed

\medskip

\noindent
Let us point out that  the condition ${\rm Vec}^{2m}_{\rr{Q}}(X, \tau)\neq\text{\rm  \O}$ is 	always guaranteed by the existence of the trivial  
{\virg{Quaternionic} product  bundle} 
 $X\times\C^{2m}\to X$. The next result follows by combining Lemma \ref{lemma:decmp_oddrankQbun} and the stable rank condition for \virg{Real} vector bundles discussed in  Remark \ref{rk:stable_rank_Real}.

\begin{theorem}[Stable condition: odd rank]
\label{theo:stab_ran_Q_odd}
Let $(X,\tau)$ be an involutive space such that $X$ has a finite $\Z_2$-CW-complex decomposition of dimension $d$ and $X^\tau=\text{\emph{\O}}$. Assume also ${\rm Pic}_{\rr{Q}}(X,\tau)\neq\text{\emph{\O}}$. Then 
$$
{\rm Vec}^{2m+1}_{\rr{Q}}\big(X, \tau\big)
\;\simeq\;  {\rm Vec}^{\sigma}_{\rr{Q}}\big(X, \tau\big)\qquad\quad \forall\ m \;\geqslant\; \frac{d-1}{4}
$$
where $\sigma:=[\frac{d}{2}]$.
In low dimensions this provides
\begin{align}
{\rm Vec}^{2m+1}_{\rr{Q}}\big(X, \tau\big)&\;\simeq\; {\rm Pic}_{\rr{Q}}\big(X,\tau\big)\;=\; 0& \text{if}& \ \ d=0,1\ \ \ \ \ \ \forall\ m\in\N \label{eq:stab_rank_Q_low_d>1!!xxx}\\
{\rm Vec}^{2m+1}_{\rr{Q}}\big(X, \tau\big)&\;\simeq\; {\rm Pic}_{\rr{Q}}\big(X,\tau\big)& \text{if}& \ \ d=2,3\ \ \ \ \ \ \forall\ m\in\N \label{eq:stab_rank_Q_low_d=1_odd_case}\\
{\rm Vec}^{2m+1}_{\rr{Q}}\big(X, \tau\big)&\;\simeq\; {\rm Vec}^{2}_{\rr{Q}}\big(X, \tau\big)& \text{if}& \ \ d=4,5\ \ \ \ \ \ \forall\ m\in\N\;.\label{eq:stab_rank_Q_low_d>1!!}
\end{align}
\end{theorem}

\medskip

\noindent
We point out that the $0$ in the \eqref{eq:stab_rank_Q_low_d>1!!xxx} refers to the existence of a unique element which could also be different from a (trivial) product  
vector bundle.

\subsection{Injectivity in low dimension}
\label{sect:inject}
The FKMM-invariant introduced in Definition \ref{def:gen_FKMM_inv} provides a map
$$
\kappa\;:\;{\rm Vec}^{2m}_{\rr{Q}}\big(X, \tau\big)\;\longrightarrow\;H^2_{\Z_2}\big(X|X^\tau,\Z(1)\big)
$$
which can be used to classify \virg{Quaternionic} vector bundles over  $(X,\tau)$.
A remarkable property is that, under certain assumptions on $(X,\tau)$,
 $\kappa$ provides a  \emph{natural injection}.  A similar result  has been already proved in \cite[Theorem 1.1]{denittis-gomi-14-gen} under the hypothesis that: ($\alpha$) $(X,\tau)$ is a low dimensional space according to Definition \ref{def:low_dim}
 ; and ($\beta$) $(X,\tau)$ is an {FKMM-space}  
\cite[Definition 1.1]{denittis-gomi-14-gen}. While  ($\alpha$) is indispensable (when $d$ grows more invariants are needed to distinguish between inequivalent \virg{Quaternionic} vector bundles, \cf \cite[Theorem 1.3]{denittis-gomi-14-gen}), the assumption ($\beta$) can be weakened. First of all one can  renounce to the requirement $H^2_{\Z_2}(X,\Z(1))=0$ proper of an FKMM-space. Second, one can consider (separately) the case  $X^\tau\neq \text{\rm  \O}$ of \emph{any} codimension
and the case  $X^\tau= \text{\rm  \O}$. The first step for the study of the injectivity of  $\kappa$ consists  in the following generalization (and simplification) of \cite[Lemma 4.1]{denittis-gomi-14-gen}.
\begin{lemma}\label{lemma:inject_Q}
Let $(X,\tau)$ be a space with involution such that $X^\tau\neq \text{\rm  \O}$. Let $(\bb{E}_1,\Theta_1)$ and $(\bb{E}_2,\Theta_2)$ be rank 2 \virg{Quaternionic} vector bundles on $(X,\tau)$ such that:
\begin{itemize}
\item[(a)] There is an isomorphism $\Psi:\bb{E}_1{|_{X^\tau}}\to\bb{E}_2{|_{X^\tau}}$ 
of \virg{Quaternionic} vector bundles over $X^\tau$;
\vspace{1mm}
\item[(b)] $\kappa(\bb{E}_1,\Theta_1)=\kappa(\bb{E}_2,\Theta_2)$ in $H^2_{\Z_2}(X|X^\tau,\Z(1))$.
\end{itemize}
Then, there is an isomorphism of \virg{Real} line bundles $\psi:({\rm det}(\bb{E}_1),{\rm det}(\Theta_1))\to ({\rm det}(\bb{E}_2),{\rm det}(\Theta_2))$ on $(X,\tau)$ such that
${\rm det}(\Psi)=\psi|_{X^\tau}$.
\end{lemma}
\proof
Let $j=1,2$.
Recall that
the two determinant \virg{Real} line bundles $({\rm det}(\bb{E}_j),{\rm det}(\Theta_j))$
associated to  $(\bb{E}_j,\Theta_j)$ 
admit  unique  \emph{canonical} $\rr{R}$-sections
$s_{\bb{E}_j}: X^\tau \to \n{S}({\rm det}(\bb{E}_j)|_{X^\tau})$ for their restrictions   on $X^\tau$ (Proposition \ref{lemma:R_Q_det_bun2}). Assumption (a)
 and the uniqueness of the canonical section assure that the $\rr{R}$-isomorphism 
 ${\rm det}(\Psi):{\rm det}(\bb{E}_1)|_{X^\tau}\to {\rm det}(\bb{E}_2)|_{X^\tau}$ intertwines the canonical sections in the sense that
 $$
 s_{\bb{E}_2}\;=\;{\rm det}(\Psi)\;\circ\;  s_{\bb{E}_1}\;.
 $$
According to Definition \ref{def:gen_FKMM_inv}, the invariants $\kappa(\bb{E}_j,\Theta_j)$ agree with the isomorphism classes of the pairs $({\rm det}(\bb{E}_j), s_{\bb{E}_j})$.
Thus, by  assumption (b), there is also an $\rr{R}$-isomorphism $\psi: {\rm det}(\bb{E}_1)\to {\rm det}(\bb{E}_2)$  such that
$$
 s_{\bb{E}_2}\;=\;\psi|_{X^\tau}\;\circ\;  s_{\bb{E}_1}\;.
 $$
 The difference between the isomorphisms ${\rm det}(\Psi)$ and $\psi|_{X^\tau}$ is uniquely specified by a map $u:X^\tau\to\n{U}(1)$ according to the relation ${\rm det}(\Psi)= \psi|_{X^\tau}\cdot u$. This implies
 $$
 s_{\bb{E}_2}\;=\;{\rm det}(\Psi)\;\circ\;  s_{\bb{E}_1}
 \;=\;\big(\psi|_{X^\tau}\cdot u\big)\;\circ\;  s_{\bb{E}_1}
 \;=\;\big(\psi|_{X^\tau}\circ s_{\bb{E}_1}\big)\cdot u
 \;=\; s_{\bb{E}_2}\cdot u
 $$
namely $u$ is the \emph{constant map} $u\equiv 1$ on $X^\tau$ and so ${\rm det}(\Psi)=\psi|_{X^\tau}$. 
\qed

\medskip

Let us consider now a pair of rank 2 \virg{Quaternionic} vector bundles  $(\bb{E}_1,\Theta_1)$ and $(\bb{E}_2,\Theta_2)$ over $(X,\tau)$ 
related by an $\rr{R}$-isomorphism
 $\psi:({\rm det}(\bb{E}_1),{\rm det}(\Theta_1))\to ({\rm det}(\bb{E}_2),{\rm det}(\Theta_2))$ of the respective determinant line bundles.
The construction in \cite[Lemma 4.1]{denittis-gomi-14-gen} allows us to  define the 
locally trivial fiber bundle 
$$
\f{SU}\big(\bb{E}_1,\bb{E}_2,\psi\big)\;:=\;\bigsqcup_{x\in X}\;\left\{\Upsilon_x:\bb{E}_1|_x\to\bb{E}_2|_x\ \left|\ 
\begin{aligned}
&\text{vector bundle isomorphism}\\
&\text{such that}\ {\rm det}(\Upsilon_x)=\psi|_x\\
\end{aligned}
\right. \right\}
$$
whose typical fiber is modeled out of ${\rm S}\n{U}(2)$. It has been proved in \cite[Lemma 4.1]{denittis-gomi-14-gen} that:

\begin{itemize}
\item[(1)] There is a natural involution on $\f{SU}(\bb{E}_1,\bb{E}_2,\psi)$ covering $\tau$
which 
can be identified  with the standard involution on ${\rm S}\n{U}(2)$
given by
$\mu:u\mapsto -Q\overline{u}Q$ (see \cite[Remark 2.1]{denittis-gomi-14-gen} for more details) on the fixed point set
$X^\tau$;
\vspace{1.0mm}
\item[(2)] The set of equivariant sections of $\f{SU}(\bb{E}_1,\bb{E}_2,\psi)$ is in bijection with the set of $\rr{Q}$-isomorphisms
$\Upsilon:\bb{E}_1\to\bb{E}_2$  such that ${\rm det}(\Upsilon)=\psi$.
\end{itemize}

\medskip

\noindent
We are now in  position to prove the following crucial  result.

\begin{theorem}[Injectivity: even rank case]\label{theo:A_inject_fixed}
Let $(X,\tau)$ be a low dimensional involutive space in the sense of Definition \ref{def:low_dim}. Then:
\begin{itemize}
\item[(1)] If $X^\tau\neq\text{\rm  \O}$ the map
$$
\kappa\;:\;{\rm Vec}_{\rr{Q}}^{2m}\big(X,\tau\big)\;\longrightarrow\; H^{2}_{\Z_2}\big(X|X^\tau,\Z(1)\big)\;,\qquad\qquad m\in\N
$$
given by the 
 FKMM-invariant provides a natural injective group homomorphism;
\vspace{1.0mm}
\item[(2)] If $X^\tau=\text{\rm  \O}$ the map
$$
\kappa\;:\;{\rm Vec}_{\rr{Q}}^{2m}\big(X,\tau\big)\;\longrightarrow\; H^{2}_{\Z_2}\big(X,\Z(1)\big)\;,\qquad\qquad m\in\N
$$
given by the identification of $\kappa$ with the  first \virg{Real} Chern class of the associated 
determinant line bundle
(\cf Corollary \ref{corol:II} (2))
 provides a natural injective group homomorphism. In addition, this is an isomorphism if 
${\rm Pic}_{\rr{Q}}(X,\tau)\neq\text{\emph{\O}}$. 
\end{itemize}
\end{theorem}
\proof
For $d=0,1$ the claim is trivially true in both cases since ${\rm Vec}^{2m}_{\rr{Q}}(X, \tau)= 0$ in view \eqref{eq:stab_rank_Q_low_d=1}. For $d=2,3$ the bijection
${\rm Vec}^{2m}_{\rr{Q}}(X, \tau)\simeq{\rm Vec}^{2}_{\rr{Q}}(X, \tau)$ 
in equation \eqref{eq:stab_rank_Q_low_d>1} reduces the  proof to the fact that  
$\kappa$ is injective on ${\rm Vec}_{\rr{Q}}^{2}\big(X,\tau\big)$.\\
Let us start with the case (1). Let $(\bb{E}_1,\Theta_1)$ be a \virg{Quaternionic} vector bundle of rank 2 over 
$(X,\tau)$. The restriction $\bb{E}_1|_{X^\tau}$ is isomorphic to a 
bundle of   quaternionic lines over $X^\tau$ \cite[Proposition 2.2]{denittis-gomi-14-gen}. Since the dimension of 
$X^\tau$ is less than four one has that ${\rm Vec}_{\n{H}}^{1}(X^\tau)=0$ as a consequence of the stable rank condition for $\n{H}$-vector bundles \cite[Chapter 9, Theorem 1.2]{husemoller-94}. Therefore $\bb{E}_1|_{X^\tau}$ is $\rr{Q}$-trivial. Let $(\bb{E}_2,\Theta_2)$ be a second \virg{Quaternionic} vector bundle of rank 2 over 
$(X,\tau)$. Since also $\bb{E}_2|_{X^\tau}$ turns out to be  $\rr{Q}$-trivial for the same reason, there is an isomorphism of $\rr{Q}$-bundles $\Psi:\bb{E}_1|_{X^\tau}\to \bb{E}_2|_{X^\tau}$
given by the composition of the respective trivializations. If $\kappa(\bb{E}_1,\Theta_1)=\kappa(\bb{E}_2,\Theta_2)$  Lemma \ref{lemma:inject_Q} assures that 
there is an $\rr{R}$-isomorphism $\psi:{\rm det}(\bb{E}_1) \to {\rm det}(\bb{E}_2)$
such that ${\rm det}(\Psi)=\psi|_{X^\tau}$. This allows  to construct the
fiber bundle $
\f{SU}(\bb{E}_1,\bb{E}_2,\psi)$ and $\Psi$ provides an equivariant section of the restriction $
\f{SU}(\bb{E}_1,\bb{E}_2,\psi)|_{X^\tau}$. Now, the free $\Z_2$-celles of $X$ have dimensions less than $4$ and $\pi_j({\rm S}\n{U}(2))=0$ for $j=0,1,2$ (equivalently ${\rm S}\n{U}(2)$ is 2-connected). These facts are sufficient to build a global section $\Upsilon:X\to \f{SU}(\bb{E}_1,\bb{E}_2,\psi)$ which extends $\Psi$ in the sense that $\Upsilon|_{X^\tau}=\Psi$. The explicit construction is realized following the same arguments in the proof of \cite[Proposition 2.7]{denittis-gomi-14-gen}.
The existence of $\Upsilon$ implies the $\rr{Q}$-isomorphism $(\bb{E}_1,\Theta_1)\simeq (\bb{E}_2,\Theta_2)$ and this proves that $\kappa$ is injective. The group structure on ${\rm Vec}^{2m}_{\rr{Q}}(X, \tau)$, which makes $\kappa$ a group homomorphism, has been described in full detail 
in the proof of \cite[Theorem 1.1]{denittis-gomi-14-gen} and is based 
on the splitting \eqref{eq:stab_rank_Q_low_d>1} 
 and on the additivity of the FKMM-invariant with respect to the Whitney sum (\cf property (4) in Section \ref{subsec:gen_FKMM_inv}).
\\
The proof in the case (2) is quite similar. Let $(\bb{E}_1,\Theta_1)$ and $(\bb{E}_2,\Theta_2)$ be two \virg{Quaternionic} vector bundles of rank 2 over  $(X,\tau)$ and assume that $\kappa(\bb{E}_1,\Theta_1)=\kappa(\bb{E}_2,\Theta_2)$ in $H^{2}_{\Z_2}(X,\Z(1))$. This implies that there is an isomorphism of \virg{Real} line bundles
$\psi:{\rm det}(\bb{E}_1)\to{\rm det}(\bb{E}_2)$ which allows to build  the
fiber bundle $
\f{SU}(\bb{E}_1,\bb{E}_2,\psi)$. Since $X$ has free $\Z_2$-celles of  dimension less than $4$ 
and the typical fiber ${\rm S}\n{U}(2)$ is 2-connected it follows that  $
\f{SU}(\bb{E}_1,\bb{E}_2,\psi)$ admits a global section which provides a $\rr{Q}$-isomorphism between 
$(\bb{E}_1,\Theta_1)$ and $(\bb{E}_2,\Theta_2)$. The group structure can be introduced  as in the case (i). The last claim is a consequence of Lemma \ref{lemma:decmp_oddrankQbun}.
\qed

{\begin{corollary}\label{coro:sempl}
Let $(X,\tau)$ be a low dimensional involutive space in the sense of Definition \ref{def:low_dim}. If the involution is trivial,  $X=X^\tau$, it holds that
$$
{\rm Vec}_{\rr{Q}}^{2m}\big(X,\tau\big)\;=\; 0\;,\qquad\qquad m\in\N\;.
$$
\end{corollary}
\proof
It follows from the injectivity of $\kappa$ proved in 
Theorem
\ref{theo:A_inject_fixed} (1) and the observation that
$$
H^{2}_{\Z_2}\big(X|X,\Z(1)\big)\;=\;0
$$
is trivial.
\qed
}

\medskip

The odd rank case requires a base space with a 
 free involution $X^\tau=\text{\rm  \O}$ and it is completely determined by the nature of ${\rm Pic}_{\rr{Q}} (X,\tau )$. More precisely, when ${\rm Pic}_{\rr{Q}}(X,\tau\big) = \text{\rm  \O}$ is not possible to build \virg{Quaternionic} vector bundles (\cf Lemma \ref{lemma:decmp_oddrankQbun}) and the $\kappa$-invariant simply cannot be defined. The case ${\rm Pic}_{\rr{Q}}(X,\tau\big) \neq \text{\rm  \O}$ is described in the next result.

\begin{theorem}[Injectivity: odd rank case]\label{theo:A_inject_free}
Let $(X,\tau)$ be a low dimensional involutive space in the sense of Definition \ref{def:low_dim}. Assume $X^\tau=\text{\emph\O}$ and ${\rm Pic}_{\rr{Q}} (X,\tau )\neq \text{\emph\O}$.
Then, there is a   bijection
$$
\kappa\;:\;{\rm Vec}_{\rr{Q}}^{2m-1}\big(X,\tau\big)\;\stackrel{\simeq}{\longrightarrow}\; H^{2}_{\Z_2}\big(X,\Z(1)\big)\;,\qquad\qquad m\in\N
$$
and the form of $\kappa$ depends on a normalization given by the choice a {reference} element  $[\bb{L}_{\rm ref}]\in {\rm Pic}_{\rr{Q}}(X,\tau)$.
\end{theorem}
\proof
By combining equation \eqref{eq:stab_rank_Q_low_d=1_odd_case} with Corollary \ref{cor:picQ} one  obtains the  bijections
$$
{\rm Vec}_{\rr{Q}}^{2m-1}\big(X,\tau\big)\;\simeq\;{\rm Pic}_{\rr{Q}}\big(X,\tau\big)\;\simeq\;H^2_{\Z_2}\big(X,\Z(1)\big)\;,\qquad\quad m\in\N
$$
both depending  on the election of a reference element $[\bb{L}_{\rm ref}]\in {\rm Pic}_{\rr{Q}}(X,\tau)$. More precisely, one has the  isomorphisms
$$
[\bb{E}_\rr{Q}]
\;\simeq\;
\left\{
\begin{aligned}
&[\bb{L}_{\rm ref}]\;\otimes\;[X\times\C]^{\oplus(2m-1)}\;\simeq\;[\bb{L}_{\rm ref}]^{\oplus(2m-2)}\oplus[\bb{L}_{\rm ref}]&\ \ &d=0,1\\
&[\bb{L}_{\rm ref}]\;\otimes\;\Big([X\times\C]^{\oplus(2m-2)}\oplus \bb{L}_{\rr{R}}\Big)\;\simeq\;[\bb{L}_{\rm ref}]^{\oplus(2m-2)}\oplus[\bb{L}_{\rm ref}\otimes\bb{L}_{\rr{R}}]& \ \ &d=2,3\\
\end{aligned}
\right.
$$
which induce the one-to-one identifications
$$
[\bb{E}_\rr{Q}]\;\mapsto\;[\bb{L}_{\rm ref}\otimes\bb{L}_{\rr{R}}]\;\mapsto\;[\bb{L}_{\rr{R}}]\;\mapsto\;c_1^{\rr{R}}\big(\bb{L}_{\rr{R}}\big)
$$
with the convention that $\bb{L}_{\rr{R}}$ is automatically trivial when $d=0,1$.
The value of the   FKMM-invariant can be fixed in accordance to Definition \ref{rk:FKMMcomp_lin_bun}
by the prescription
$$
\kappa(\bb{E}_\rr{Q})\;=\; c_1^{\rr{R}}\big(\bb{L}_{\rr{R}}\big)
$$
which implies the normalization $\kappa(\bb{L}_{\rm ref}^{\oplus 2m-1})=0$.
\qed

\subsection{The question of the surjectivity of the FKMM-invariant}
\label{sect:surject_FKMM} 
Theorem \ref{theo:A_inject_fixed} and Theorem  \ref{theo:A_inject_free}
assure the injectivity of the FKMM-invariant $\kappa$ in the case of low dimensional base spaces ($d\leqslant3$). On the one hand,  this is enough to distinguish between different topological phases. On the other hand it could be 	
valuable to know whether the cohomology group where $\kappa$ maps is a complete set of invariants or if it is redundant. It turns out that, for  a big class  low dimensional spaces the FKMM-invariant provides bijections. For instance, this is the case for the involutive spheres $\n{S}^{p,q}$ and the involutive tori $\T^{a,b,c,}$ discussed 
 in Section \ref{sect:class_invol_shper} and Section \ref{sect:class_invol_tori} respectively, or for oriented
surfaces with  finitely many isolated  fixed points \cite[Corollary 4.2]{denittis-gomi-14-gen}. Therefore, one is legitimate  to ask whether the surjectivity in low dimension is  a general property of the FKMM-invariant or not!

 \medskip
 
Let us start with two observation: ($\alpha$) The FKMM-invariant fails to be surjective in dimension bigger than three where other invariants like the second Chern class start to play a role in the classification (\cf \cite[Theorem 1.4]{denittis-gomi-14-gen}); ($\beta$) The  odd rank case (which requires $X^\tau= \text{\rm  \O}$) is completely specified by the nature of ${\rm Pic}_{\rr{Q}}(X,\tau)$ as proved in Theorem  \ref{theo:A_inject_free}. Therefore, the question of the  surjectivity is relevant only for even rank \virg{Quaternionic} vector bundles over low dimensional involutive spaces.
We point out that the study of this question deserves a separated analysis 
for the cases $X^\tau= \text{\rm  \O}$ and   $X^\tau\neq \text{\rm  \O}$.

 \medskip	
 
The study of the surjectivity of $\kappa$ goes beyond the scope of this work. However, we can anticipate some results proved in  \cite{denittis-gomi-16}. Let $(X,\tau)$ be an involutive space of dimension $d$ which fulfills Assumption \ref{ass:top}. Then:
\begin{itemize}
\item[(1)] When  $d=0,1$ the $\kappa$-invariant provides a \virg{trivial} isomorphism.
\item[(2)] When  $d=2$ the $\kappa$-invariant provides  isomorphisms
$$
{\rm Vec}_{\rr{Q}}^{2m}\big(X,\tau\big)\;\stackrel{\kappa}{\simeq}\;
\left\{
\begin{aligned}
&H^2_{\Z_2}(X,\Z(1))&\ \ \ &\text{if}\ X^\tau= \text{\rm  \O}\\
&H^2_{\Z_2}(X|X^\tau,\Z(1))& \ \ \ &\text{if} \ X^\tau\neq \text{\rm  \O}\;.
\end{aligned} 
\right.
$$
\item[(3)] When  $d=3$ the $\kappa$-invariant \emph{fails} to be surjective.
\end{itemize}

\noindent
The proof of (1) amounts to show that the condition $d=0,1$ forces $H^2_{\Z_2}(X,\Z(1))=0$ for the free involution case $X^\tau= \text{\rm  \O}$ and $H^2_{\Z_2}(X|X^\tau,\Z(1))=0$
when $X^\tau\neq \text{\rm  \O}$. The proof of (2) is quite technical. The payoff of (1) and (2) is that one can see the $\kappa$-invariant as the
generalization of the isomorphisms \eqref{eq:intro_tqs3} and \eqref{eq:intro_tqs4} for the \virg{Quaternionic} category up to dimension two. The claim (3) marks the difference between the FKMM-invariant and the first (\virg{Real}) Chern class.

\medskip

The proof of (3) is based on the construction of an explicit counterexample 
 Let $\n{S}^3:=\{(z_0,z_1)\in\C^2\ |\ |z_0|^2+|z_1|^2=1  \}\subset\C^2$ be the three dimensional sphere  viewed as the unit sphere in $\C^2$. Consider the action of $\Z_4\simeq\{\pm1,\pm\ii\}$ on 
 $\n{S}^3$ given by $(z_0,z_1)\mapsto (\rho z_0,\rho z_1)$ for all $\rho\in\Z_4$. The quotient space
$$
 L(1;4)\;:=\;\n{S}^3/\Z_4
 $$
is called (3-dimensional) \emph{lens space} (\cf \cite[Example 18.5]{bott-tu-82} or \cite[Example 2.43]{hatcher-02} for more details). Since $\n{Z}_4$ 
 is preserved by the complex conjugation (with our convention)  the involution on $\n{S}^3\subset\C^2$ induced by 
$(z_0,z_1)\mapsto(\overline{z_0},\overline{z_1})$ descends to an involution $\tau$ on
 $L(1;4)$. The involutive space $(L(1;4),\tau)$ is a smooth (3-dimensional) manifold with a smooth involution, hence it admits a $\Z_2$-CW-complex structure \cite[Theorem 3.6]{may-96}. Moreover this space has a fixed point set of the type
 $
L(1;4)^\tau\simeq\;\n{S}^1\;\sqcup\;\n{S}^1
$. One can construct the \virg{Quaternionic} vector bundles over 
$(L(1;4),\tau)$ by  the \emph{clutching construction} providing 
\begin{equation}\label{eq:non_surj_alpha}
{\rm Vec}_{\rr{Q}}^{2m}\big(L(1;4),\tau\big)\;=\;\Z_4\;.
\end{equation}
On the other hand the computation of the equivariant cohomology provides
\begin{equation}\label{eq:non_surj_beta}
H^2_{\Z_2}\big(L(1;4)|L(1;4)^\tau,\Z(1)\big)\;=\;\Z_8\;.
\end{equation}
The proofs of \eqref{eq:non_surj_alpha} and \eqref{eq:non_surj_beta} are quite elaborate and technical and will be presented in \cite{denittis-gomi-16} for the slightly more general
 case of a lens space of type $L(1;2q)$ with $q\in\N$. In any case, the important message conveyed by \eqref{eq:non_surj_alpha} and \eqref{eq:non_surj_beta} is that the map
$$
\kappa\;:\;{\rm Vec}_{\rr{Q}}^{2m}\big(L(1;4),\tau\big)\;\longrightarrow\;H^2_{\Z_2}\big(L(1;4)|L(1;4)^\tau,\Z(1)\big)\;
$$
cannot be surjective, confirming the claim (3).

\subsection{Classification over low dimensional involutive spheres}
\label{sect:class_invol_shper}
In this section we apply the results of Section \ref{sect:stab_rang} and Section \ref{sect:inject} to provide the complete classification of \virg{Quaternionic} vector bundles over the involutive spheres of   type $\n{S}^{p,q}$ in   {low dimension } $d=p+q-1\leqslant 3$.
These results  have  been summarized in Table \ref{tab:01}.1. Since $\n{S}^{0,0}=\text{\rm  \O}$ we conventionally write
$$
{\rm Vec}^{k}_{\rr{Q}}\big(\n{S}^{0,0}\big)\;=\; \text{\rm  \O}\;.
$$

\subsubsection*{\bf $\bullet$ Spheres of type $\n{S}^{p,0}$}
These are the spheres with the trivial involution which fixes all points. 
{Corollary \ref{coro:sempl} immediately provides
$$
{\rm Vec}^{2m}_{\rr{Q}}\big(\n{S}^{p,0}\big)\;=\;0\;,\qquad\quad 1\leqslant p\leqslant 4\;,\qquad \forall\; m\in\N\;.
$$}

\subsubsection*{\bf $\bullet$ Spheres of type $\n{S}^{0,q}$} 
These are the spheres with free antipodal involution and are the only spheres which admit  \virg{Quaternionic} vector bundles of odd rank.
The line bundle case has been already discussed in Section \ref{sect:inv_spher_lin_bund}.

\medskip

\noindent
[q=1,2] -
 When $q=1$ the sphere $\n{S}^{0,1}$ is the two points set $\{-1,+1\}$
endowed with the flip involution $\pm1 \mapsto \mp1$ and
one has that
$$
{\rm Vec}^{k}_{\rr{Q}}\big(\n{S}^{0,1}\big)\;=\; 0\;,\qquad\quad \forall\ k\in\N.
$$
For $k=2m$   the unique element is given by the  (trivial) \virg{Quaternionic} product bundle over $\{-1,+1\}$ while for $k=2m+1$ the unique element is given by the direct sum of a rank $2m$  \virg{Quaternionic} trivial bundle  with the  Dupont \virg{Quaternionic} line bundle described in Example
\ref{ex:Dupont}. Also for $q=2$ one has 
$$
{\rm Vec}^{k}_{\rr{Q}}\big(\n{S}^{0,2}\big)\;=\; 0\;,\qquad\quad \forall\ k\in\N.
$$
When $k=2m$  the 
result is justified by equation \eqref{eq:stab_rank_Q_low_d=1} and the unique element represented   by the trivial  \virg{Quaternionic} product bundle.
When $k=2m+1$ the result is justified by \eqref{eq:stab_rank_Q_low_d=1_odd_case} and 
Example \ref{Ex:lin_bun2}. More precisely, in the odd rank case the unique element is represented by the direct sum of a  trivial  \virg{Quaternionic} vector bundle of rank $2m$ with the \virg{Quaternionic} line bundle $\bb{L}_0$ described in Example \ref{Ex:lin_bun2}.

\medskip

\noindent
[q=3] - In the odd rank case equation \eqref{eq:stab_rank_Q_low_d=1_odd_case} and the  Example \ref{Ex:lin_bun3} provide 
$$
{\rm Vec}^{2m-1}_{\rr{Q}}\big(\n{S}^{0,3}\big)\;\simeq\; {\rm Pic}_{\rr{Q}}\big(\n{S}^{0,3}\big)
\;\stackrel{}{\simeq}\;2\Z+1,\qquad\quad\ \ \forall\ m\in\N  \;
$$
where the (second) identification is given by the first Chern class $c_1$. This identification requires a reference element 
in ${\rm Pic}_{\rr{Q}}(\n{S}^{0,3})$ which can be chosen to be the
line bundle
$\bb{L}_1$  constructed in Example \ref{Ex:lin_bun3}. 
The  proof of Theorem \ref{theo:A_inject_free}  shows  that each $[\bb{E}_{\rr{Q}}]\in {\rm Vec}^{2m-1}_{\rr{Q}}(\n{S}^{0,3})$ can be uniquely represented
by an element $[\bb{L}_{\rr{R}}]\in  {\rm Pic}_{\rr{R}}(\n{S}^{0,3})$ according to the factorization $[\bb{E}_{\rr{Q}}]\simeq[\bb{L}_1]^{\oplus (2m-2)}\oplus [\bb{L}_1\otimes\bb{L}_{\rr{R}}]$.
Accordingly, one can fix the FKMM-invariant 
by the prescription
$$
\kappa\big(\bb{E}_{\rr{Q}}\big)\;:=\; c_1^{\rr{R}}\big(\bb{L}_{\rr{R}}\big)\;.
$$
The identification $2c_1^{\rr{R}}(\bb{L}_{\rr{R}})\simeq c_1(\bb{L}_{\rr{R}})$ between the \virg{Real} Chern class and the standard Chern class of $\bb{L}_{\rr{R}}$ discussed
 in Example \ref{Ex:lin_bun3}  provides 
$$
c_1\big(\bb{E}_{\rr{Q}}\big)\;\simeq\;(2m-2)\;c_1\big(\bb{L}_{1}\big)\;+\;c_1\big(\bb{L}_{1}\big)\;+\;c_1\big(\bb{L}_{\rr{R}}\big)\;\simeq\;(2m-1)\; c_1\big(\bb{L}_{1}\big)\;+\; 2\; c_1^{\rr{R}}(\bb{L}_{\rr{R}})
$$
Since $c_1\big(\bb{L}_{1}\big)=1$ and $c_1^{\rr{R}}(\bb{L}_{\rr{R}})\in\Z$ one obtains that 
the Chern class of any representative in ${\rm Vec}^{2m-1}_{\rr{Q}}(\n{S}^{0,3})$ is necessarily odd and it is sufficient for a complete characterization of the isomorphism class $[\bb{E}_{\rr{Q}}]$.
Moreover, the FKMM-invariant takes the form
$$
\kappa\big(\bb{E}_{\rr{Q}}\big)\;=\;\frac{c_1\big(\bb{E}_{\rr{Q}}\big)-(2m-1)}{2}\;,\qquad\quad [\bb{E}_{\rr{Q}}]\in  {\rm Vec}^{2m-1}_{\rr{Q}}\big(\n{S}^{0,3}\big)\;.
$$
In the even rank case one has
$$
{\rm Vec}^{2m}_{\rr{Q}}\big(\n{S}^{0,3}\big)\;\simeq\;{\rm Vec}^{2}_{\rr{Q}}\big(\n{S}^{0,3}\big)\;\stackrel{\kappa}{\longrightarrow}\;H^2_{\Z_2}\big(\n{S}^{0,3},\Z(1)\big)\;\simeq\;\Z\;
$$
where the first isomorphism is justified by \eqref{eq:stab_rank_Q_low_d>1} and the FKMM-invariant provides an injective map according to Theorem \ref{theo:A_inject_fixed} (2).
Moreover, the existence of  $\bb{L}_{1}$ assures  that $\kappa$ is indeed an isomorphism
More precisely, for each  $[\bb{E}_{\rr{Q}}]\in {\rm Vec}^{2m}_{\rr{Q}}(\n{S}^{0,3})$ 
 there exists a unique $[\bb{E}_{\rr{R}}]\in {\rm Vec}^{2m}_{\rr{R}}(\n{S}^{0,3})$ such that $[\bb{E}_{\rr{Q}}]\simeq [\bb{L}_{1}\otimes\bb{E}_{\rr{R}}]$. The determinant construction provides ${\rm det }(\bb{E}_{\rr{Q}})=\bb{L}_{1}^{\otimes 2m}\otimes {\rm det}(\bb{E}_{\rr{R}})$ and a straightforward computation shows that   
$$
c_1\big(\bb{E}_{\rr{Q}}\big)\;\simeq\;c_1\big({\rm det }(\bb{E}_{\rr{Q}})\big)\;\simeq\;2m\; c_1\big(\bb{L}_{1}\big)\;+\; c_1\big({\rm det }(\bb{E}_{\rr{R}})\big)\;\simeq\;2m\; c_1\big(\bb{L}_{1}\big)\;+\; 2\;c_1^{\rr{R}}\big({\rm det }(\bb{E}_{\rr{R}})\big)\;.
$$
Since $c_1(\bb{L}_{1})=1$ and $c_1^{\rr{R}}(\bb{L}_{\rr{R}})\in\Z$ one concludes that 
the Chern class of any representative in ${\rm Vec}^{2m}_{\rr{Q}}(\n{S}^{0,3})$ must be  even.
Moreover,  by definition one has
$$
\kappa\big(\bb{E}_{\rr{Q}}\big)\;:=\;c_1^{\rr{R}}\big({\rm det }(\bb{E}_{\rr{Q}})\big)\;=\;\frac{c_1\big(\bb{E}_{\rr{Q}}\big)}{2}\;,\qquad\quad [\bb{E}_{\rr{Q}}]\in  {\rm Vec}^{2m}_{\rr{Q}}\big(\n{S}^{0,3}\big)\;.
$$
Summarizing, we proved that 
\begin{equation}\label{eq:class_S2_antipodal}
\left\{
\begin{aligned}
{\rm Vec}^{2m}_{\rr{Q}}\big(\n{S}^{0,3}\big)&& \stackrel{c_1}{\simeq}\ \ \ &2\Z\\
{\rm Vec}^{2m-1}_{\rr{Q}}\big(\n{S}^{0,3}\big)&& \stackrel{c_1}{\simeq}\ \ \ &2\Z+1
\end{aligned}
\right.,\qquad\quad\ \ \forall\ m\in\N \;. 
\end{equation}
We point out that   the first Chern class $c_1$ 
turns out to be a complete invariant for 
the classification of $\rr{Q}$-bundles of any rank over $\n{S}^{0,3}$.
The \eqref{eq:class_S2_antipodal} recovers  the results recently presented in \cite{gat-robbins-15}.

\medskip

\noindent
[q=4] -  By combining Lemma \ref{lemma:decmp_oddrankQbun} and 
Example \ref{Ex:lin_bun1} one concludes
$$
{\rm Vec}^{2m-1}_{\rr{Q}}\big(\n{S}^{0,4}\big)\;=\;\text{\rm  \O}\;,\qquad\quad\ \ \forall\ m\in\N 
$$
for the odd rank case. For the even rank case 
one has
$$
{\rm Vec}^{2m}_{\rr{Q}}\big(\n{S}^{0,4}\big)\;\simeq\;{\rm Vec}^{2}_{\rr{Q}}\big(\n{S}^{0,4}\big)\;\stackrel{\kappa}{\longrightarrow}\;H^2_{\Z_2}\big(\n{S}^{0,4},\Z(1)\big)\;\simeq\;0\;
$$
where the first  isomorphism is given by \eqref{eq:stab_rank_Q_low_d>1}. The  injectivity of the $\kappa$-invariant proved in Theorem \ref{theo:A_inject_fixed} (2) implies
$$
{\rm Vec}^{2m}_{\rr{Q}}\big(\n{S}^{0,4}\big)\;=\;0\;,\qquad\quad\ \ \forall\ m\in\N 
$$
and the  unique element is represented by the   rank $2m$  trivial $\rr{Q}$-bundle.

\subsubsection*{\bf $\bullet$ Spheres of type $\n{S}^{1,q}$}
These are the spheres with TR-involution (two distinguished fixed points) which have been studied in   \cite{denittis-gomi-14-gen}.
We recall that:
$$
{\rm Vec}^{2m}_{\rr{Q}}\big(\n{S}^{1,0}\big)\;\simeq\;{\rm Vec}^{2m}_{\rr{Q}}\big(\n{S}^{1,1}\big)\;=\;0\;,\qquad\qquad 
{\rm Vec}^{2m}_{\rr{Q}}\big(\n{S}^{1,2}\big)\;\simeq\;{\rm Vec}^{2m}_{\rr{Q}}\big(\n{S}^{1,3}\big)\;\simeq\;\Z_2\;.
$$
In the non-trivial cases $\n{S}^{1,2}$ and $\n{S}^{1,3}$ the $\Z_2$-phase is discriminated by the FKMM-invariant.

\subsubsection*{\bf $\bullet$ Spheres of type $\n{S}^{2,q}$}
In this case the fixed point set $(\n{S}^{2,q})^\theta\simeq\n{S}^1$ is a circle and only even rank \virg{Quaternionic} vector bundles are possible. 
Equation 
\eqref{eq:stab_rank_Q_low_d>1} and Theorem \ref{theo:A_inject_fixed} (1) provide
$$
{\rm Vec}^{2m}_{\rr{Q}}\big(\n{S}^{2,q}\big)\;\simeq\;{\rm Vec}^{2}_{\rr{Q}}\big(\n{S}^{2,q}\big)\;\stackrel{\kappa}{\longrightarrow}\;H^2_{\Z_2}\big(\n{S}^{2,q}|\n{S}^1,\Z(1)\big)\;\qquad\quad q=1,2\ \ \ \ \forall\ m\in\N
$$
with $\kappa$ being an injection.

\medskip

\noindent
[q=1] -  Proposition \ref{lemma:surj_map_r_2Z} provides
$$
{\rm Vec}^{2}_{\rr{Q}}\big(\n{S}^{2,1}\big)\;\stackrel{\kappa}{\longrightarrow}\;H^2_{\Z_2}\big(\n{S}^{2,1}|\n{S}^1,\Z(1)\big)\;\simeq\;2\Z
$$
Moreover, one has  that the isomorphism $H^2_{\Z_2} (\n{S}^{2,1}|\n{S}^1,\Z(1))\simeq2\Z$ is induced by the inclusion 
$$
H^2_{\Z_2}\big(\n{S}^{2,1}|\n{S}^1,\Z(1)\big)\;\stackrel{\delta_2}{\hookrightarrow}\;H^2_{\Z_2}\big(\n{S}^{2,1},\Z(1)\big)\;\stackrel{f}{\simeq}\;H^2\big(\n{S}^{2},\Z\big)\;\stackrel{c_1}{\simeq}\;\Z
$$
by means of the first Chern class of the underlying complex vector bundle. 
The surjectivity of $\kappa$ can be proved as follow:
Let $\bb{E}_0$ be the trivial element in 
${\rm Vec}^{2}_{\rr{Q}}(\n{S}^{2,1})$ and $\bb{L}_1\in{\rm Pic}_{\rr{R}}(\n{S}^{2,1})$ such that $c_1(\bb{L}_1)$=1. An $\rr{R}$-line bundle with this property can be explicitly constructed as showed in the proof of Proposition \ref{lemma:surj_map_r_2Z}. Since $c_1(\bb{E}_0)=0$ the \virg{Quaternionic} vector bundle $\bb{E}_{2n}:=\bb{E}_0\otimes(\bb{L}_1^{\otimes n})$ has Chern class
$$
c_1\big(\bb{E}_{2n}\big)\;=\;{\rm rk}\big(\bb{E}_0\big)\; c_1\big(\bb{L}_1^{\otimes n}\big)\;=\;2n\;c_1\big(\bb{L}_1\big)\;=\;2n\;.
$$
This fact shows that $\kappa$ is also surjective. In particular, we proved the isomorphism
$$
{\rm Vec}^{2m}_{\rr{Q}}\big(\n{S}^{2,1}\big)\;\stackrel{c_1}\simeq\;2\Z
$$
which says that a vector bundle over $\n{S}^{2,1}$ can be endowed with a unique
$\rr{Q}$-structure if and only if its first Chern class is even.

\medskip

\noindent
[q=2] - 
Equation \eqref{eq:rel_cohom_S22} shows that $H^2_{\Z_2}(\n{S}^{2,2}|\n{S}^1,\Z(1))=0$, hence the  injectivity of $\kappa$ immediately implies that ${\rm Vec}^{2m}_{\rr{Q}}\big(\n{S}^{2,2}\big)=0$ for all $m\in\N$.

\subsubsection*{\bf $\bullet$ Spheres of type $\n{S}^{3,q}$} The only case of interest  in low dimension is $\n{S}^{3,1}$. This space contains 
a two-dimensional sphere as
 fixed point set $(\n{S}^{3,1})^\theta\simeq\n{S}^2$. Therefore,  only  the even rank case is relevant. 
Equation 
\eqref{eq:stab_rank_Q_low_d>1} and Theorem \ref{theo:A_inject_fixed} (1) provide
$$
{\rm Vec}^{2m}_{\rr{Q}}\big(\n{S}^{3,1}\big)\;\simeq\;{\rm Vec}^{2}_{\rr{Q}}\big(\n{S}^{3,1}\big)\;\stackrel{\kappa}{\longrightarrow}\;H^2_{\Z_2}\big(\n{S}^{3,1}|\n{S}^2,\Z(1)\big)\;
$$
with $\kappa$ being an injection. Since equation \eqref{eq:rel_cohom_S31} shows that 
$H^2_{\Z_2}(\n{S}^{3,1}|\n{S}^2,\Z(1))=0$ one concludes that ${\rm Vec}^{2m}_{\rr{Q}}(\n{S}^{3,1})=0$ for all $m\in\N$.

\subsection{Classification over low dimensional  involutive tori}
\label{sect:class_invol_tori}
This section is devoted to 
the complete classification of \virg{Quaternionic} vector bundles over involutive tori of the type $\n{T}^{a,b,c}$ in  low dimension $a+b+c\leqslant3$.
These results   have been summarized in Table \ref{tab:03}.2 and Table \ref{tab:04}.3. Since $\n{T}^{0,0,0}=\text{\rm  \O}$ we conventionally write
$$
{\rm Vec}^{k}_{\rr{Q}}\big(\n{T}^{0,0,0}\big)\;=\; \text{\rm  \O}\;.
$$

\medskip

\noindent
{\bf - Cases with fixed points -}\\
Involutive tori of type $\n{T}^{a,b,0}$ have non-empty fixed point sets. There are 
 three
different two-dimensional tori and four different three-dimensional tori
of this type. For all these cases only even rank $\rr{Q}$-bundles are possible due to the presence of fixed points.

\subsubsection*{\bf $\bullet$ The torus  $\T^{a,0,0}$}
Spaces of type $\T^{a,0,0}$ are $a$-dimensional tori
with the trivial involution which fixes all points. 
{Corollary \ref{coro:sempl} immediately provides
$$
{\rm Vec}^{2m}_{\rr{Q}}\big(\T^{a,0,0}\big)\;=\;0\;,\qquad\quad 1\leqslant a\leqslant 3\;,\qquad\forall\ m\in\N.
$$
}

\subsubsection*{\bf $\bullet$ The torus  $\T^{0,b,0}$}
The spaces $\T^{0,b,0}$ are $b$-dimensional tori with the  TR-involution ($2^b$ distinguished fixed points) which have been studied in   \cite{denittis-gomi-14-gen}. 
We recall that:
$$
{\rm Vec}^{2m}_{\rr{Q}}\big(\T^{0,2,0}\big)\;\simeq\;\Z_2,\qquad\qquad 
{\rm Vec}^{2m}_{\rr{Q}}\big(\T^{0,3,0}\big)\;\simeq\;{\Z_2}^4\qquad\qquad 
{\rm Vec}^{2m}_{\rr{Q}}\big(\T^{0,4,0}\big)\;\simeq\;{\Z_2}^{10}\oplus\Z\;.
$$
for all $m\in\N$. The $\Z_2$-phases are discriminated by the FKMM-invariant.

\subsubsection*{\bf $\bullet$ The torus  $\T^{1,1,0}$}
In this case the fixed point set $(\n{T}^{1,1,0})^\tau\simeq\n{S}^1\sqcup \n{S}^1$ is given by a disjoint union of  two circles. 
Equation 
\eqref{eq:stab_rank_Q_low_d>1} and Theorem \ref{theo:A_inject_fixed} (1) provide
$$
{\rm Vec}^{2m}_{\rr{Q}}\big(\T^{1,1,0}\big)\;\simeq\;{\rm Vec}^{2}_{\rr{Q}}\big(\T^{1,1,0}\big)\;\stackrel{\kappa}{\longrightarrow}\;H^2_{\Z_2}\big(\T^{1,1,0}|(\n{T}^{1,1,0})^\tau,\Z(1)\big)\;\simeq\;2\Z\;. 
$$
with $\kappa$ being an injection. The last isomorphism is proved in Proposition \ref{lemma:surj_map_r_torus_2Z}
and  is induced by the composition
$$
H^2_{\Z_2}\big(\T^{1,1,0}|(\n{T}^{1,1,0})^\tau,\Z(1)\big)\;\stackrel{\delta_2}{\hookrightarrow}\;H^2_{\Z_2}\big(\T^{1,1,0},\Z(1)\big)\;\stackrel{f}{\longrightarrow}\;H^2\big(\n{T}^{2},\Z\big)\;\stackrel{c_1}{\simeq}\;\Z\;
$$
where the forgetting map $f$ turns out to be surjective.
The surjectivity of  $\kappa$ can be verified as follow:
Let $\bb{E}_0$ be the trivial element in 
${\rm Vec}^{2}_{\rr{Q}}(\n{T}^{1,1,0})$ and $\bb{L}_{1,1}\in{\rm Pic}_{\rr{R}}(\n{T}^{1,1,0})$ the non trivial $\rr{R}$-line bundle
with  $c_1(\bb{L}_{1,1})=1$ constructed in the proof of Proposition \ref{lemma:surj_map_r_torus_2Z}.
Since $c_1(\bb{E}_0)=0$ the $\rr{Q}$-bundle $\bb{E}_{2n}:=\bb{E}_0\otimes(\bb{L}_{1,1}^{\otimes n})$ has Chern class
$$
c_1\big(\bb{E}_{2n}\big)\;=\;{\rm rk}\big(\bb{E}_0\big)\; c_1\big(\bb{L}_{1,1}^{\otimes n}\big)\;=\;2n\;c_1\big(\bb{L}_{1,1}\big)\;=\;2n\;.
$$
This fact proves that $\kappa$ is also surjective and in turn one has the isomorphism
$$
{\rm Vec}^{2m}_{\rr{Q}}\big(\n{T}^{1,1,0}\big)\;\stackrel{c_1}\simeq\;2\Z\;,\qquad\quad\ \ \forall\ m\in\N\;. 
$$
In particular a vector bundle over $\n{T}^{1,1,0}$ can be endowed with a unique
$\rr{Q}$-structure if and only if it has an even Chern class.
A similar result has been recently obtained in \cite{gat-robbins-15}.

\subsubsection*{\bf $\bullet$ The torus  $\T^{2,1,0}$}
The fixed point set $(\n{T}^{2,1,0})^\tau\simeq\n{T}^2\sqcup \n{T}^2$ consists of two disjoint copies of a two-dimensional torus. 
One has
$$
{\rm Vec}^{2m}_{\rr{Q}}\big(\T^{2,1,0}\big)\;\simeq\;{\rm Vec}^{2}_{\rr{Q}}\big(\T^{2,1,0}\big)\;\stackrel{\kappa}{\longrightarrow}\;H^2_{\Z_2}\big(\T^{2,1,0}|(\n{T}^{2,1,0})^\tau,\Z(1)\big)\;\simeq\;(2\Z)^2
$$
where the first isomorphism is justified in \eqref{eq:stab_rank_Q_low_d>1}
and Theorem \ref{theo:A_inject_fixed} (1) assures that
$\kappa$ is an injection.
 The last isomorphism is proved in Proposition \ref{lemma:surj_map_r_torus_2Z_210type}
and  is induced by the composition
$$
H^2_{\Z_2}\big(\T^{2,1,0}|(\n{T}^{2,1,0})^\tau,\Z(1)\big)\;\stackrel{\delta_2}{\hookrightarrow}\;H^2_{\Z_2}\big(\T^{2,1,0},\Z(1)\big)\;\stackrel{f}{\longrightarrow}\;H^2\big(\n{T}^{3},\Z\big)\;\stackrel{c_1}{\simeq}\;\Z^3\;
$$
where the forgetting map $f$ turns out to be bijective between the free part of $H^2_{\Z_2}(\T^{2,1,0},\Z(1))$ and a $\Z^2$-summand in $H^2(\n{T}^{3},\Z)$.
Let $\bb{E}_2\to\n{T}^{1,1,0}$ be the element  
of ${\rm Vec}^{2m}_{\rr{Q}}(\n{T}^{1,1,0})$ identified by the Chern class $c_1(\bb{E}_2)=2$
and consider the two pullbacks of $\bb{E}_2$ in $ {\rm Vec}^{2m}_{\rr{Q}}(\n{T}^{2,1,0})$ under the two distinct projections $\n{T}^{2,1,0}\to \n{T}^{1,1,0}$.
These two vector bundles provide a basis for the bijection
$$
{\rm Vec}^{2m}_{\rr{Q}}\big(\n{T}^{2,1,0}\big)\;\stackrel{c_1}\simeq\;(2\Z)^2\;,\qquad\quad\ \ \forall\ m\in\N\;
$$
showing in turn that $\kappa$ is bijective. In particular one has that a vector bundle over $\n{T}^{2,1,0}$ can be endowed with a unique
$\rr{Q}-$structure if and only if it has a Chern class
of type $(2n,2n',0)\in\Z^3$.

\subsubsection*{\bf $\bullet$ The torus  $\T^{1,2,0}$}
In this case the fixed point set $(\n{T}^{1,2,0})^\tau\simeq\n{S}^1\sqcup\n{S}^1\sqcup\n{S}^1\sqcup\n{S}^1$ is the disjoint union of four circles. 
From  
\eqref{eq:stab_rank_Q_low_d>1} and Theorem \ref{theo:A_inject_fixed} (1) one obtains
$$
{\rm Vec}^{2m}_{\rr{Q}}\big(\T^{1,2,0}\big)\;\simeq\;{\rm Vec}^{2}_{\rr{Q}}\big(\T^{1,2,0}\big)\;\stackrel{\kappa}{\longrightarrow}\;H^2_{\Z_2}\big(\T^{1,2,0}|(\n{T}^{1,2,0})^\tau,\Z(1)\big)\;\simeq\;\Z_2\;\oplus\;(2\Z)^2 
$$
with $\kappa$ being an injection. The last isomorphism, proved in Proposition \ref{lemma:surj_map_r_torus_2Z_120type}, enters in the  
composition of maps
$$
H^2_{\Z_2}\big(\T^{1,2,0}|(\n{T}^{1,2,0})^\tau,\Z(1)\big)\;\stackrel{\delta_2}{\longrightarrow}\;H^2_{\Z_2}\big(\T^{1,2,0},\Z(1)\big)\;\stackrel{f}{\longrightarrow}\;H^2\big(\n{T}^{3},\Z\big)\;\stackrel{c_1}{\simeq}\;\Z^3\;
$$
where $\delta_2$ acts injectively on the free part of $H^2_{\Z_2}(\T^{1,2,0}|(\n{T}^{1,2,0})^\tau,\Z(1))$ 
and the forgetting map $f$ is a bijection between the free part of $H^2_{\Z_2}(\T^{1,2,0},\Z(1))$  and a $\Z^2$-summand in $H^2(\n{T}^{3},\Z)$.
It turns out that the map $\kappa$ is also surjective. In fact,  the $\Z_2$-summand in 
${\rm Vec}^{2m}_{\rr{Q}}(\T^{1,2,0})$ is generated by the pullback of ${\rm Vec}^{2m}_{\rr{Q}}(\T^{0,2,0})\simeq\Z_2$ under the projection $\T^{1,2,0}\to \T^{0,2,0}$. The two $2\Z$-summands, instead, are generated by the pullback of 
${\rm Vec}^{2m}_{\rr{Q}}(\T^{1,1,0})\simeq2\Z$ under the two distinct projections $\T^{1,2,0}\to \T^{1,1,0}$. In conclusion one obtains that
$$
{\rm Vec}^{2m}_{\rr{Q}}\big(\n{T}^{1,2,0}\big)\;\stackrel{}\simeq\;\Z_2\;\oplus\;(2\Z)^2\;,\qquad\quad\ \ \forall\ m\in\N\;.
$$
This implies that one can build a
$\rr{Q}$-structure on $\T^{1,2,0}$ 
if and only if the underlying complex vector bundle 
has a first Chern class
of type $(2n,2n',0)\in\Z^3$. Moreover, if this is the case, it is possible to build to inequivalent $\rr{Q}$-structures.

\medskip
\medskip

\noindent
{\bf - Cases with free involution -}\\
As soon as $c\neq 0$ the involution over $\T^{a,b,c}$ turns out to be free. In this case the classification of  \virg{Quaternionic} vector bundles is simplified by two general results. The first of them concerns the possibility to reduce situations with $c\geqslant 2$ to the \emph{standard} situation $c=1$. 
\begin{proposition}\label{prop:cohom_Tab1-iso}
For each $a,b\in\N\cup\{0\}$ and $c\geqslant 2$ there is an identification of 
$\Z_2$-spaces 
$$
\n{T}^{a,b,c}\;\simeq\; \n{T}^{a+c-1,b,1}\;.
$$
\end{proposition}
\proof
Consider the identifications $\n{S}^1\equiv\n{U}(1)$ and $\n{T}^d\equiv\n{U}(1)\times\ldots\times \n{U}(1)$. By using the group structure of $\n{U}(1)$ we can define maps $f_c:\n{T}^d\to \n{T}^d$ of the type
$$
f_c\;:\;(z_1,\ldots, \underbrace{z_{d-c+1},\ldots,z_{d-1},z_d}_{\text{last}\; c\; \text{coord.}})\;\longmapsto\; (z_1,\ldots, \underbrace{z_{d-c+1}z_d,\ldots,z_{d-1}z_d}_{ c-1\; \text{coord.}},z_d)\;. 
$$
Topologically these maps are   homeomorphisms.
Since $\n{S}^{2,0}$ is identifiable with $\n{U}(1)$ endowed  with trivial involution $z\mapsto z$ and $\n{S}^{0,2}$ is identifiable with $\n{U}(1)$ with antipodal involution $z\mapsto -z$ one concludes that  $f_c$ is a $\Z_2$-homeomorphism between  
$\n{T}^{a,b,c}$ and $\n{T}^{a+c-1,b,1}$.
\qed

\medskip

\noindent
The second result concerns involutive spaces with the product structure  $X\times\n{S}^{0,2}$. 
\begin{proposition}\label{propos_S02}
For each involutive space $(X,\tau)$ the FKMM-invariant 
$$
\kappa\;:\; {\rm Vec}^{2}_{\rr{Q}}\big(X\times\n{S}^{0,2}\big)\;\stackrel{}{\longrightarrow}\;H^2_{\Z_2}\big(X\times\n{S}^{0,2},\Z(1)\big)
$$
is surjective. Moreover there is a bijection
$$
{\rm Pic}_{\rr{Q}}\big(X\times\n{S}^{0,2}\big)\;\simeq\;{\rm Pic}_{\rr{R}}\big(X\times\n{S}^{0,2}\big)\;\neq\;\text{\upshape\O}\;
$$
which preserves the first Chern class.
\end{proposition}
\proof
Since the involution on $X\times\n{S}^{0,2}$ is free the FKMM-invariant is given by the composition
$$
{\rm Vec}^{2}_{\rr{Q}}\big(X\times\n{S}^{0,2}\big)\;\stackrel{{\rm det}}{\longrightarrow}\;{\rm Pic}_{\rr{R}}\big(X\times\n{S}^{0,2}\big)\;\stackrel{c_1^{\rr{R}}}{\longrightarrow}\;H^2_{\Z_2}\big(X\times\n{S}^{0,2},\Z(1)\big)\;
$$
(\cf Corollary \ref{corol:II} (2)). In order to show that 
${\rm Pic}_{\rr{Q}}(X\times\n{S}^{0,2})\neq\text{\upshape\O}$ let us introduce a special $\rr{Q}$-line bundle by mimicking the construction in Example \ref{Ex:lin_bun2}. Consider the product line bundle $X\times\n{S}^{0,2}\times\C\to X\times\n{S}^{0,2}$ endowed with the  \virg{Quaternionic} structure
$$
\Theta_q\;:\;(x,k,\lambda)\;\mapsto\;(\tau(x),-k,q(x,k)\overline{\lambda})\;\qquad\quad (x,k,\lambda)\;\in\; X\times\n{S}^{0,2}\times\C
$$
where the map $q:X\times\n{S}^{0,2}\to\n{U}(1)$ must fulfill the constraint $q(\tau(x),-k)=-q(x,k)$. A possible choice is to  fix $q(x,k)=q_0(k)$ where $q_0$ is the map described by
\eqref{eq:mapq_0}. Let $\tilde{\bb{L}}_0\in {\rm Pic}_{\rr{Q}}(X\times\n{S}^{0,2})$ be this $\rr{Q}$-line bundle. By construction $c_1(\tilde{\bb{L}}_0)=0$ for the Chern class of the underlying complex line bundle. Let $\tilde{\bb{L}}^2_0:=\tilde{\bb{L}}_0\otimes\tilde{\bb{L}}_0$. This $\rr{R}$-line bundle  agrees with the trivial element in 
${\rm Pic}_{\rr{R}}(X\times\n{S}^{0,2})$. In fact the product structure of $\tilde{\bb{L}}^2_0$ is evident by construction. Moreover, the  $\rr{R}$-structure on $\tilde{\bb{L}}^2_0$ is induced by the map $q_0^2$ which can be factorized as $q_0(k)^2=\phi(-k)\phi(k)$ with $\phi:=\ii q_0$.
This shows that $q_0^2$ is isomorphic through $\phi$ to the constant map $1$.
For each $\bb{L}_\rr{R}\in{\rm Pic}_{\rr{R}}(X\times{\n{S}}^{0,2})$ 
consider the rank two  \virg{Quaternionic} vector bundle over $X\times\n{S}^{0,2}$ given by
$$
\bb{E}_\rr{Q}\;:=\;\big(\bb{L}_\rr{R}\;\oplus\;(X\times\n{S}^{0,2}\times\C)\big)\;\otimes\;\tilde{\bb{L}}_0\;.
$$
The associated determinant line bundle verifies
$$
{\rm det}\big(\bb{E}_\rr{Q}\big)\;\simeq\; \bb{L}_\rr{R}\;\otimes\;(X\times\n{S}^{0,2}\times\C)\otimes\;\tilde{\bb{L}}_0^2\;\simeq\;  \bb{L}_\rr{R}
$$
where we used the triviality of $\tilde{\bb{L}}_0^2$. This  proves the surjectivity of $\kappa$. The last claim is a consequence of Corollary \ref{cor:picQ}
and the existence of $\tilde{\bb{L}}_0$.
\qed

\medskip

\begin{corollary}\label{corol:classi_inv_tori}
Let $(X,\tau)$ be an involutive space that fulfills 
Assumption \ref{ass:top} and with dimension $d\leqslant 2$. Then the FKMM-invariant provides  a bijection
$$
{\rm Vec}^{m}_{\rr{Q}}\big(X\times\n{S}^{0,2}\big)\;\stackrel{\kappa}{\simeq}\;H^2_{\Z_2}\big(X\times\n{S}^{0,2},\Z(1)\big)\;\simeq\;  {\rm Pic}_{\rr{R}}(X\times\n{S}^{0,2})
$$
valid for each $m\in\N$ independently of its parity. In the odd rank case it is possible to normalize the FKMM-invariant $\kappa$ in such a way that the 
bijection
$$
{\rm Vec}^{2m-1}_{\rr{Q}}\big(X\times\n{S}^{0,2}\big)\;\simeq\; {\rm Pic}_{\rr{R}}(X\times\n{S}^{0,2})
$$
preserves the Chern class of the underlying complex vector bundle.
\end{corollary}
\proof
The condition about the dimension of $X$ assures the validity of the stable rank condition (\cf eq. \eqref{eq:stab_rank_Q_low_d>1} and eq. \eqref{eq:stab_rank_Q_low_d=1_odd_case}). 
In the even rank case the bijectivity of $\kappa$ is consequence of Theorem \ref{theo:A_inject_fixed} (2) and Proposition \ref{propos_S02}. In the odd rank case the bijection follows from Theorem \ref{theo:A_inject_free} and the existence of $\tilde{\bb{L}}_0\in {\rm Pic}_{\rr{Q}}(X\times\n{S}^{0,2})$. The use of $\tilde{\bb{L}}_0$ as reference element for the construction of the bijection and the fact of $c_1(\tilde{\bb{L}}_0)=0$ imply  the validity of the last claim. 
\qed

\subsubsection*{\bf $\bullet$ The torus  $\T^{0,1,1}$} 
In view of Corollary \ref{corol:classi_inv_tori} one has the bijection
$$
{\rm Vec}^{m}_{\rr{Q}}\big(\T^{0,1,1}\big)\;\stackrel{\kappa}{\simeq}\;H^2_{\Z_2}\big(\T^{0,1,1},\Z(1)\big)\;\simeq\;\Z\;\qquad\quad \forall\ m\in\N\:.
$$
The computation of the cohomology group is provided in Proposition \ref{prop:cohom_Tab1}.
However, we can add more information by using the  exact sequence \cite[Proposition 2.3]{gomi-13}

$$
\begin{diagram}
&H^2_{\Z_2}\big(\T^{0,1,1},\Z(1)\big)&\rTo^{f}&        H^2\big(\T^2,\Z\big)&\rTo^{}&   
H^{2}_{\Z_2}\big(\T^{0,1,1},\Z\big)&\rTo^{}&  H^3_{\Z_2}\big(\T^{0,1,1},\Z(1)\big) \vspace{-0mm}\\
&    \rotatebox{-90}{$\simeq$}&&\rotatebox{-90}{$\simeq$} && \rotatebox{-90}{$\simeq$} && \rotatebox{-90}{$\simeq$}   \\
 &   \Z&&\Z && \Z_2 &&0   \\                                    \end{diagram}\;. 
$$
Here we used
$$
\begin{aligned}
&H^{2}_{\Z_2}\big(\T^{0,1,1},\Z\big)\;\simeq\;H^2_{\Z_2}\big(\n{S}^{0,2},\Z\big)\;\oplus\; H^1_{\Z_2}\big(\n{S}^{0,2},\Z(1)\big)\;\simeq\;\Z_2\\
&H^{3}_{\Z_2}\big(\T^{0,1,1},\Z(1)\big)\;\simeq\;H^3_{\Z_2}\big(\n{S}^{0,2},\Z(1)\big)\;\oplus\; H^2_{\Z_2}\big(\n{S}^{0,2},\Z\big)\;\simeq\;0
\end{aligned}
$$
obtained by the application of the second isomorphism in \eqref{eq:appC_Gysin_exact}. 
The  exact sequence says that the forgetting map $f$ acts as the multiplication by two and the composition with the first Chern class produces the identification
$$
{\rm Vec}^{m}_{\rr{Q}}\big(\T^{0,1,1}\big)\;\stackrel{\kappa}{\simeq}\;H^2_{\Z_2}\big(\T^{0,1,1},\Z(1)\big)\stackrel{c_1}{\simeq}\;2\Z\;\qquad\quad \forall\ m\in\N\;.
$$
We point out that the validity of this result in the odd rank case is a consequence of the existence of a bijection which preserves the first Chern classes
as claimed by Corollary \ref{corol:classi_inv_tori}.

\subsubsection*{\bf $\bullet$ The tori  $\T^{1,0,1}\simeq\T^{0,0,2}$} Let us study the case $\T^{1,0,1}$.
Corollary \ref{corol:classi_inv_tori} and Proposition \ref{prop:cohom_Tab1} imply
$$
{\rm Vec}^{m}_{\rr{Q}}\big(\T^{1,0,1}\big)\;\stackrel{\kappa}{\simeq}\;H^2_{\Z_2}\big(\T^{1,0,1},\Z(1)\big)\;\simeq\;\Z_2\;\qquad\quad \forall\ m\in\N\;.
$$

\subsubsection*{\bf $\bullet$ The torus  $\T^{0,2,1}$} 
From Corollary \ref{corol:classi_inv_tori} and Proposition \ref{prop:cohom_Tab1} one obtains
$$
{\rm Vec}^{m}_{\rr{Q}}\big(\T^{0,2,1}\big)\;\stackrel{\kappa}{\simeq}\;H^2_{\Z_2}\big(\T^{0,2,1},\Z(1)\big)\;\simeq\;\Z^2\qquad\quad \forall\ m\in\N\;.
$$
A more precise description can be obtained by looking at  the exact sequence \cite[Proposition 2.3]{gomi-13}

$$
\begin{diagram}
&H^2_{\Z_2}\big(\T^{0,2,1},\Z(1)\big)&\rTo^{f}&        H^2\big(\T^3,\Z\big)&\rTo^{}&   
H^{2}_{\Z_2}\big(\T^{0,2,1},\Z\big)&\rTo^{}&  H^3_{\Z_2}\big(\T^{0,2,1},\Z(1)\big)
\vspace{-0mm} \\
&\rotatebox{-90}{$\simeq$}&&\rotatebox{-90}{$\simeq$} && \rotatebox{-90}{$\simeq$} && \rotatebox{-90}{$\simeq$}  \\
&\Z^2&&\Z^3 && {\Z_2}^2\oplus\Z &&\Z_2                                  \end{diagram}.
$$
The cohomology groups can be computed 
by using the second isomorphism in \eqref{eq:appC_Gysin_exact} which gives
$$
\begin{aligned}
&H^{2}_{\Z_2}\big(\T^{0,2,1},\Z\big)\;\simeq\;H^2_{\Z_2}\big(\T^{0,1,1},\Z\big)\;\oplus\; H^1_{\Z_2}\big(\n{S}^{0,2},\Z(1)\big)\;\oplus\; H^0_{\Z_2}\big(\n{S}^{0,2},\Z\big)\;\simeq\;{\Z_2}^2\;\oplus\;\Z\\
&H^{3}_{\Z_2}\big(\T^{0,2,1},\Z(1)\big)\;\simeq\;H^3_{\Z_2}\big(\T^{0,1,1},\Z(1)\big)\;\oplus\; H^2_{\Z_2}\big(\T^{0,1,1},\Z\big)\;\simeq\;\Z_2\;.
\end{aligned}
$$
The  exact sequence alone does not suffice for the determination of the forgetting map $f$. However, we can notice that $H^2_{\Z_2}(\T^{0,2,1},\Z(1))\simeq {\rm Vec}^{2}_{\rr{Q}}(\T^{0,2,1})$ contains two 
$2\Z$-summands generated by the pullbacks of 
${\rm Vec}^{2}_{\rr{Q}}(\T^{0,1,1})\simeq 2\Z$ under the two distinct projections $\T^{0,2,1}\to\T^{0,1,1}$. This fact implies that 
$f$ must act as $f:(n,n')\mapsto (2n,2n',0)$ with $n,n'\in\Z$.
The composition with the first Chern class finally produces the identification
$$
{\rm Vec}^{m}_{\rr{Q}}\big(\T^{0,2,1}\big)\;\stackrel{c_1}{\simeq}\;(2\Z)^2\;\qquad\quad \forall\ m\in\N\;.
$$

\subsubsection*{\bf $\bullet$ The tori  $\T^{1,1,1}\simeq\T^{0,1,2}$} Let us discuss the case $\T^{1,1,1}$. 
Corollary \ref{corol:classi_inv_tori} and Proposition \ref{prop:cohom_Tab1} imply
$$
{\rm Vec}^{m}_{\rr{Q}}\big(\T^{1,1,1}\big)\;\stackrel{\kappa}{\simeq}\;H^2_{\Z_2}\big(\T^{1,1,1},\Z(1)\big)\;\simeq\;{\Z_2}\;\oplus\;\Z^2\;\qquad\quad \forall\ m\in\N\;.
$$
However, a deeper analysis shows that 
$$
{\rm Vec}^{m}_{\rr{Q}}\big(\T^{1,1,1}\big)\;\simeq\;{\Z_2}\;\oplus\;(2\Z)^2\;\qquad\quad \forall\ m\in\N\;
$$
where the $\Z_2$-summand is the result of the pullback of ${\rm Vec}^{m}_{\rr{Q}}(\T^{1,0,1})\simeq\Z_2$ under the projection $\T^{1,1,1}\to\T^{1,0,1}$. In the same way, the two $2\Z$-summands are generated by the pullbacks of ${\rm Vec}^{m}_{\rr{Q}}(\T^{0,1,1})\simeq2\Z$ and ${\rm Vec}^{m}_{\rr{Q}}(\T^{1,1,0})\simeq2\Z$ under the projections 
$\T^{1,1,1}\to\T^{0,1,1}$ and $\T^{1,1,1}\to\T^{1,1,0}$, respectively. 

\subsubsection*{\bf $\bullet$ The tori  $\T^{2,0,1}\simeq\T^{1,0,2}\simeq\T^{0,0,3}$} 
Let us focus on the case $\T^{2,0,1}$. From Corollary \ref{corol:classi_inv_tori} and Proposition \ref{prop:cohom_Tab1} one obtains
$$
{\rm Vec}^{2}_{\rr{Q}}\big(\T^{2,0,1}\big)\;\stackrel{\kappa}{\simeq}\;H^2_{\Z_2}\big(\T^{2,0,1},\Z(1)\big)\;\simeq\;{\Z_2}^2\;\qquad\quad m\in\N\;.
$$
The two $\Z_2$-summands can be generated by the pullback of ${\rm Vec}^{m}_{\rr{Q}}(\T^{1,0,1})\simeq\Z_2$ under the two distinct projections $\T^{2,0,1}\to\T^{1,0,1}$.
%

\section{Application to prototype models of topological quantum systems}
\label{sect:non-trivial_ex}
Let $\{\Sigma_0,\ldots,\Sigma_4\}\in{\rm Mat}(\C^4)$ be an irreducible representation of the \emph{Clifford algebra} $C\ell_\C(5)$ generated by the rules $\Sigma_i\Sigma_j+\Sigma_j\Sigma_i=2\delta_{i,j}\n{1}_4$ for all $i,j=0,1,\ldots,4$. An explicit realization is given by
$$
\begin{aligned}
\Sigma_0\;:=\;\sigma_1\otimes\sigma_3&&\Sigma_1\;:=\;\sigma_2\otimes\sigma_3&&\Sigma_2\;:=\;\n{1}_2\otimes\sigma_1\\
&&\Sigma_3\;:=\;\n{1}_2\otimes\sigma_2&&\Sigma_4\;:=\;\sigma_3\otimes\sigma_3\\
\end{aligned}
$$
where
$$
\sigma_1\;:=\;\left(\begin{array}{cc}0 & 1 \\1 & 0\end{array}\right)\;,\qquad\sigma_2\;:=\;\left(\begin{array}{cc}0 & -\ii \\\ii & 0\end{array}\right)\;, \qquad\sigma_3\;:=\;\left(\begin{array}{cc}1 & 0 \\0 & -1\end{array}\right)
$$
are the Pauli matrices. With this special choice  the following relations hold true:  
$$
\Sigma_j^\ast=\Sigma_j\;,\qquad\quad\overline{\Sigma}_j:=C{\Sigma}_jC=(-1)^{j}\;{\Sigma}_j\;,\qquad\quad {\Sigma}_0\;{\Sigma}_1\;{\Sigma}_2\;{\Sigma}_3\;{\Sigma}_4\;=\;-\n{1}_4\;.
$$
where $C{\rm v}=\overline{\rm v}$ is the complex conjugation on $\C^4$. We need also the  relations for the trace
\beql{eq:trac_form}
{\rm Tr}_{\C^4}(\Sigma_j)\;=\;0\;,\qquad\quad {\rm Tr}_{\C^4}(\Sigma_i\Sigma_j)\;=\;4\delta_{i,j}\;.
\eeq
Let $J:=\sigma_2\otimes \n{1}_2$ and $\Theta:= C\circ J=-J\circ C$. Evidently $J^*=J=J^{-1}$ and  $\Theta^2=-\n{1}_4$ which in turn implies $\Theta^*=J\circ C$. Then
$$
\left\{
\begin{aligned}
&\Theta\;\Sigma_j\;\Theta^*\;=\;-\Sigma_j\;,\qquad\quad j=0,1,3,4\;.\\
&\Theta\;\Sigma_2\;\Theta^*\;=\;+\Sigma_2\;
\end{aligned}
\right.
$$

\medskip

Given an involutive space $(X,\tau)$ and five continuous functions $F_j:X\to \R$,  where $j=0,1,\ldots ,4,$ one can build a \emph{topological quantum system} (in the sense of \cite[Definition 1.1]{denittis-gomi-14}) $H:X\to{\rm Mat}(\C^4)$ given by
\begin{equation}\label{eq:TQS1}
H(x)\;:=\;\sum_{j=0}^4F_j(x)\;\Sigma_j\;.
\end{equation}
{It should be noted that \eqref{eq:TQS1} provides only a convenient and simple toy model. In principle one can consider more complicated Hamiltonians by adding terms proportional to the products $\Sigma_{j_1}\ldots\Sigma_{j_r}$ and relevant models of this type have been already discussed in the literature (\eg the Kane-Mele model \cite{kane-mele-05}). However, in order to simplify the presentation, only the simpler case \eqref{eq:TQS1} will be considered here.
}

\medskip

By imposing the conditions
\begin{equation}\label{eq:model0}
\left\{
\begin{aligned}
&F_j\big(\tau(x)\big)\;=\;-F_j(x)\;,\qquad\quad j=0,1,3,4\;,\\
&F_2\big(\tau(x)\big)\;=\;+F_2(x)
\end{aligned}
\right.\qquad\qquad x\in\ X
\end{equation}
one can verify that
\begin{equation}\label{eq:model1}
\Theta\;H(x)\;\Theta^*\;=\; H\big(\tau(x)\big)\;,\qquad\qquad x\in\ X\;.
\end{equation}
Equation \eqref{eq:model1} says that $x\mapsto H(x)$ defines a {topological quantum system} with an \emph{odd} time reversal symmetry in the sense of \cite[Definition 1.2]{denittis-gomi-14}. According to a commonly used notation, the class of these systems is denoted by the label \emph{class AII} (see \eg \cite{altland-zirnbauer-97,schnyder-ryu-furusaki-ludwig-08}).

\medskip

A simple computation provides
$$
{H}(x)^2\;=\;{Q}(x)\;\n{1}_4\;,\qquad\quad {Q}(x)\;:=\;\sum_{j=0}^4F_j(x)^2
$$
and, after assuming the \emph{zero gap condition}   ${Q}>0$, one can define two projection-valued maps $P_\pm:X\to {\rm Mat}(\C^4)$ given by 
\begin{equation}\label{eq:model0_2}
{P}_\pm(x)\;:=\;\frac{1}{2}\left({\n{1}_4}\;\pm\;\sum_{j=0}^4\frac{F_j(x)}{\sqrt{Q(x)}}\; \Sigma_j\right)\;.
\end{equation}
The projection relations 
 ${P}_\pm(x)^2={P}_\pm(x)$, ${P}_\pm(x) {P}_\mp(x)=0$ and ${P}_+(x)\oplus{P}_-(x)={\n{1}_4}$ 
 and the spectral relations $[{P}_\pm(x),{H}(x)]=0$ and ${P}_\pm(x){H}(x){P}_\pm(x)=\pm\sqrt{Q(x)}{P}_\pm(x)$
can be directly checked. The conditions  $
{\rm Tr}_{\C^4}{P}_\pm(x)=2$ shows that each  map $P_\pm:X\to {\rm Mat}(\C^4)$ defines a continuous family of two dimensional planes in $\C^4$, namely a rank two complex vector bundle over $X$. 
Moreover, under the condition \eqref{eq:model0} which assures  
\begin{equation}\label{eq:model2}
\Theta\;P_\pm(x)\;\Theta^*\;=\; P_\pm\big(\tau(x)\big)\;,\qquad\qquad x\in\ X\;,
\end{equation}
one obtains that the vector bundles $\bb{E}_\pm\to X$ associated to $P_\pm(x)$ are indeed \virg{Quaternionic}.
For more details about the construction of the vector bundles $\bb{E}_\pm$ we refer to  \cite[Section 2 \& Section 6]{denittis-gomi-14}. 
Define $f_j(x):=F_j(x)/\sqrt{Q(x)}$. The family of these functions $f_j:X\to\R$ is subjected to the pointwise constraint  $\sum_{j=0}^4f_j(x)^2=1$. Therefore, the vector valued map $\phi:=(f_0,f_1,\ldots,f_4)$ provides a continuous map $\phi:X\to\n{S}^4$. This map identifies uniquely the projections $P_\pm(x)$ through the formula \eqref{eq:model0_2}. 

\medskip

The last observation suggests the introduction of the  \emph{Hopf vector bundle}  $\bb{E}_{\rm Hopf}\to \n{S}^4$ defined by the   projection-valued   map ${P}_{\rm Hopf}:\n{S}^4 \to{\rm Mat}(\C^4)$ given by
\beql{eq:proj_sphe2}
{P}_{\rm Hopf}(k_0,k_1,\ldots,k_4)\;:=\;\frac{1}{2}\left(\n{1}_4\;+\;\sum_{j=0}^4k_j\; \Sigma_j\right)\;.
\eeq
The involution $\Theta$ induces a \virg{Quaternionic} structure on $\bb{E}_{\rm Hopf}$ provided that the base space $\n{S}^4$   is endowed with the  involution  $(k_0,k_1,k_2,k_3,k_4)\mapsto(-k_0,-k_1,k_2,-k_3,-k_4)$. This space, up to a reordering of the indices, agrees with the \emph{TR sphere} $\n{S}^{1,4}$ described in Section \ref{sect:intro} (see also \cite[Example 4.2]{denittis-gomi-14}). Then,  $(\bb{E}_{\rm Hopf}, \Theta)$ is an element of ${\rm Vec}_{\rr{Q}}^m(\n{S}^{1,4})$. This set has been classified in \cite[Theorem 1.4 (i)]{denittis-gomi-14-gen} and we know that ${\rm Vec}_{\rr{Q}}^2(\n{S}^{1,4})\simeq \Z$ is completely classified
by the second Chern class $c_2$. Moreover, the FKMM-invariant $\kappa$ of  elements in ${\rm Vec}_{\rr{Q}}^m(\n{S}^{1,4})$  takes values in $\Z_2$ and is fixed by the parity of $c_2$ . The second Chern class of $\bb{E}_{\rm Hopf}$ is $c_2(\bb{E}_{\rm Hopf})=1$ (see \eg \cite[Remark 6.1]{denittis-gomi-14}) and consequently one has 
$$
\kappa(\bb{E}_{\rm Hopf})\;=\;(-1)^{c_2(\bb{E}_{\rm Hopf})}\;=\;-1.
$$
Therefore, the {Hopf vector bundle}  $\bb{E}_{\rm Hopf}$ provides an example of non-trivial $\rr{Q}$-bundle over $\n{S}^{1,4}$ with a non-trivial FKMM-invariant.

\medskip

A comparison between \eqref{eq:proj_sphe2} and \eqref{eq:model0_2} shows that  the vector bundles $\bb{E}_\pm\to X$ can be obtained from $\bb{E}_{\rm Hopf}\to \n{S}^4$ by a pullback with respect a map $\phi:X\to\n{S}^4$ which amounts to the replacement of $\pm f_j(x)$ with $k_j$ in \eqref{eq:proj_sphe2}. This means that all  possible vector bundle generated by a topological quantum system of type \eqref{eq:TQS1} can be labelled by  continuous maps $\phi:X\to\n{S}^4$. Moreover, since homotopy equivalent maps generate (via pullback)  isomorphic vector bundles we obtain that the possible topological phases of the  system  \eqref{eq:TQS1} can be labelled by the set  $[X,\n{S}^4]$ of classes of homotopy equivalence functions. However, if we want to respect the odd time reversal symmetry we need to take into account the constraints 
\eqref{eq:model0}. These are compatible with the involution $\theta_{1,4}$ since
$$
\begin{aligned}
\theta_{1,4}\big(f_0(x),f_1(x),f_2(x),f_3(x),f_4(x)\big)\;&=\;\big(-f_0(x),-f_1(x),f_2(x),-f_3(x),-f_4(x)\big)\\
&=\;\big(f_0(\tau(x)),f_1(\tau(x)),f_2(\tau(x)),f_3(\tau(x)),f_4(\tau(x))\big)
\end{aligned}
$$
and one deduces that the map $\phi$ is indeed $\Z_2$-equivariant from $(X,\tau)$ to ${\n{S}}^{1,4}$. Therefore, the topology of 
a  systems \eqref{eq:TQS1} with an odd time reversal symmetry imposed by \eqref{eq:model0}
is encoded in   the set  $[X,{\n{S}}^{1,4}]_{\Z_2}$ of classes of  $\Z_2$-homotopy equivalence  of equivariant functions. In summary  we showed that:
\begin{proposition}[Sufficient condition for non-trivial phases]
Each class AII topological quantum system given by a \emph{gapped} Hamiltonian of type \eqref{eq:TQS1} subjected to the constraints 
\eqref{eq:model0} is obtained as the pullback of the 
{Hopf $\rr{Q}$-bundle}  $\bb{E}_{\rm Hopf}$ with respect to a map $[\phi]\in[X,{\n{S}}^{1,4}]_{\Z_2}$. In this sense $[X,{\n{S}}^{1,4}]_{\Z_2}$ provides a (generally non injective)
complete set of labels for the topological phases of these systems. Finally, by naturality, the FKMM-invariant for these systems is obtained as
$$
\kappa\big(\phi^*\bb{E}_{\rm Hopf}\big)\;=\;\phi^*\big(\kappa(\bb{E}_{\rm Hopf})\big)\;=\;\phi^*\big(-1)
$$
and is non-trivial whenever the equivariant  map $\phi:X\to {\n{S}}^{1,4}$ induces a homomorphism
$$
\phi^*\;:\;H^{2}_{\Z_2}\big({\n{S}}^{1,4}|({\n{S}}^{1,4})^\theta,\Z(1)\big)\;{\longrightarrow}\;H^{2}_{\Z_2}\big(X|X^\tau,\Z(1)\big)
$$
which is \emph{injective}.
\end{proposition}

\medskip

In Appendix \ref{sect:cohom_sper} we computed
\begin{equation}
H^{2}_{\Z_2}\big({\n{S}}^{1,d}|({\n{S}}^{1,d})^\theta,\Z(1)\big)\;\simeq\;
\left\{
\begin{aligned}
&0&\ \ \text{if}\ \ &\ d=0,1 \\
&\Z_2&\ \ \text{if}\ \ &\ d\geqslant 2\;.
\end{aligned}
\right.
\end{equation}
Since the condition $f_j\equiv 0$ is compatible with the constraints 
\eqref{eq:model0} we can try to extract information about this \virg{simplified models}. First of all let us notice that if some of the 
$f_j$'s vanishes then only a sub-vector bundle of the {Hopf $\rr{Q}$-bundle}  $\bb{E}_{\rm Hopf}$  is responsible for the pullback. To be more precise, let us introduce the \emph{embedding} maps $\imath:{\n{S}}^{1,d-1}\to{\n{S}}^{1,d}$
which identify ${\n{S}}^{1,d-1}$ with the subset of ${\n{S}}^{1,d}$ fixed by the condition
 $k_j=0$ on one of the coordinates 
which behaves non-trivially under the involution $\theta_{1,d}$ (according to the above notation this means $j\neq 2$). Then, we have a sequence of equivariant maps 
$$
{\n{S}}^{1,0}\;\stackrel{\imath}{\longrightarrow}\;{\n{S}}^{1,1}\;\stackrel{\imath}{\longrightarrow}\;\ldots\;\stackrel{\imath}{\longrightarrow}\;{\n{S}}^{1,d}\;\stackrel{\imath}{\longrightarrow}\;{\n{S}}^{1,d+1}\;\ldots
$$
and a related sequence of \emph{restricted} {Hopf $\rr{Q}$-bundle}
$$
\bb{E}^{(iv-n)}_{\rm Hopf}\;:=\;{\imath^n}^*\bb{E}_{\rm Hopf}\;\in\;{\rm Vec}_{\rr{Q}}^2\big({\n{S}}^{1,4-n}\big)
$$
where we used the short notation $\imath^n:=\imath\circ\ldots\circ \imath$ and the convention $\bb{E}^{(iv)}_{\rm Hopf}\equiv \bb{E}_{\rm Hopf}$. We know from the classification established in \cite{denittis-gomi-14-gen} that 
$$
{\rm Vec}_{\rr{Q}}^2\big({\n{S}}^{1,0}\big)\;\simeq\; {\rm Vec}_{\rr{Q}}^2\big({\n{S}}^{1,1}\big)\;=\;0
$$
which implies that $\bb{E}^{(i)}_{\rm Hopf}$ and $\bb{E}^{(0)}_{\rm Hopf}$ are both trivial. On the other hand we know that 
$$
{\rm Vec}_{\rr{Q}}^2\big({\n{S}}^{1,3}\big)\;\simeq\; {\rm Vec}_{\rr{Q}}^2\big({\n{S}}^{1,2}\big)\;=\;\Z_2
$$
are completely classified by the FKMM-invariant and by naturality
$$
\kappa\big(\bb{E}^{(iii)}_{\rm Hopf}\big)\;=\;\imath^*\kappa\big(\bb{E}_{\rm Hopf}\big)\;=\;\imath^*(-1)\;,\qquad\quad
\kappa\big(\bb{E}^{(ii)}_{\rm Hopf}\big)\;=\;{\imath^2}^*\kappa\big(\bb{E}_{\rm Hopf}\big)\;=\;{\imath^2}^*(-1)\;.
$$
Since the maps $\imath:{\n{S}}^{1,d-1}\to {\n{S}}^{1,d}$ induce isomorphisms (see Lemma \ref{lemma:isomaps_restr})
$$
\imath^*\;:\;H^{2}_{\Z_2}\big({\n{S}}^{1,d}|({\n{S}}^{1,d})^\theta,\Z(1)\big)\;\stackrel{\simeq}{\longrightarrow}H^{2}_{\Z_2}\big({\n{S}}^{1,d-1}|({\n{S}}^{1,d-1})^\theta,\Z(1)\big)\;,\qquad\qquad d\geqslant 2
$$
we can infer that 
$$
\bb{E}^{(iii)}_{\rm Hopf}\;\in\;{\rm Vec}_{\rr{Q}}^2({\n{S}}^{1,3})\;,\qquad\qquad \bb{E}^{(ii)}_{\rm Hopf}\;\in\;{\rm Vec}_{\rr{Q}}^2({\n{S}}^{1,2})
$$
are non-trivial representatives of the respective classes. These considerations have implications in the analysis of 
topological quantum systems  of type \eqref{eq:TQS1}.
\begin{proposition}[The case $F_2\not\equiv 0$]
\label{propo:nec1}
Consider class AII topological quantum systems given by  \emph{gapped} Hamiltonians of type \eqref{eq:TQS1} subjected to the constraints 
\eqref{eq:model0} and $F_2\not\equiv0$. A necessary condition to produce a non-trivial topological phases is that at least two of the functions $\{F_0,F_1,F_3,F_4\}$ have  to be  not identically zero. If this is the case, the set $[X,{\n{S}}^{1,4-n}]_{\Z_2}$ provides a (generally non injective) 
complete set of labels for the topological phases on these systems. Here  $n=0,1,2$ represents the number of functions in  $\{F_0,F_1,F_3,F_4\}$ which are identically zero.
  Finally, by naturality, the FKMM-invariant of these systems is given by
$$
\kappa\big(\phi^*\bb{E}^{(iv-n)}_{\rm Hopf}\big)\;=\;\phi^*\big(\kappa(\bb{E}^{(iv-n)}_{\rm Hopf})\big)\;=\;\phi^*\big(-1)\;,\qquad\quad n=0,1,2
$$
and is non-trivial whenever the equivariant  map $\phi:X\to {\n{S}}^{1,4-n}$ induces a homomorphism
$$
\phi^*\;:\;H^{2}_{\Z_2}\big({\n{S}}^{1,4-n}|({\n{S}}^{1,4-n})^\theta,\Z(1)\big)\;{\longrightarrow}\;H^{2}_{\Z_2}\big(X|X^\tau,\Z(1)\big)
$$
which is \emph{injective}.
\end{proposition}

\medskip

\noindent
It should be emphasized that the calculation of $H^{2}_{\Z_2}({\n{S}}^{1,4-n}|({\n{S}}^{1,4-n})^\theta,\Z(1))$ is quite easy since the cohomology classes can be represented in terms of sign maps defined on the fixed-point set $({\n{S}}^{1,4-n})^\theta$.
At this point the verification of the injectivity of $\phi^*$ is not a difficult problem provided that the equivariant cohomology of $X$ can be computed. The latter, however,  in general may be a difficult problem! 

\medskip

\noindent 
Let us now discuss the implications of the condition $F_2\equiv0$. In this case the  vector bundles $\bb{E}_\pm\to X$ are generated by the pullback with respect to a map $\phi:X\to {\n{S}}^{0,4}\subset {\n{S}}^{1,4}$ where ${\n{S}}^{0,4}$ is the subspace  fixed by the constraint $k_2=0$. In particular ${\n{S}}^{0,4}$ is the three dimensional sphere with \emph{antipodal} free involution. The 
 equivariant {embedding}  $\eta:{\n{S}}^{0,4}\to{\n{S}}^{1,4}$ produces by pullback
the \emph{restricted} {Hopf $\rr{Q}$-bundle}
$$
\hat{\bb{E}}_{\rm Hopf}\;:=\;{\eta}^*\bb{E}_{\rm Hopf}\;\in\;{\rm Vec}_{\rr{Q}}^2\big({\n{S}}^{0,4}\big)\;=\;0\;.
$$
Since the  vector bundles $\bb{E}_\pm\to X$ subjected to the condition $F_2\equiv0$ are obtained as pullbacks
of $\hat{\bb{E}}_{\rm Hopf}$ which is  trivial
one obtains:
\begin{proposition}[The case $F_2\equiv 0$]\label{propo:nec2}
Consider class AII topological quantum systems given by  \emph{gapped} Hamiltonians of type \eqref{eq:TQS1} subjected to the constraints 
\eqref{eq:model0} and $F_2\equiv0$. These
systems possesses only the trivial phase.
\end{proposition}

\appendix

\section{About the equivariant cohomology of the Classifying space}\label{app:Cohom_comput}
The cohomology of ${G}_{2m}(\C^\infty)$ is the polynomial ring 
\begin{equation}\label{eq:univ_chern_class!}
H^\bullet\big(G_{2m}(\C^\infty),\Z\big)\;\simeq\;\Z[\rr{c}_1,\ldots,\rr{c}_{2m}]
\end{equation}
generated by the \emph{universal} Chern classes $\rr{c}_k\in H^{2k}\big(G_{2m}(\C^\infty),\Z\big)$
 \cite[Theorem 14.5]{milnor-stasheff-74}. The involution $\rho$ on the space $G_{2m}(\C^\infty)$ described in Section \ref{subsec:univ_FKMM_inv} induces a homomorphism
$\rho^*$ on each cohomology group $H^{k}(G_{2m}(\C^\infty),\Z)$.
$\rho^*$ acts on the {universal} Chern classes as follows:
\begin{lemma}\label{lemma:inv_cohom_gras}
It holds that
$$
\rho^*(\rr{c}_k)=(-1)^k\rr{c}_k\;,\qquad\quad k=1,2,\ldots,2m\;.
$$
\end{lemma}
\proof
Since $\rr{c}_k$ is the $k$-th Chern class of the {tautological} vector bundle $\bb{T}_{2m}^\infty\to G_{2m}(\C^\infty)$
it follows by naturality  that $\rho^*(\rr{c}_k)$ is the $k$-th Chern class of the pullback vector bundle $\rho^*(\bb{T}_{2m}^\infty)$. The presence of the $\rr{Q}$-structure implies that the fibers of $\bb{T}_{2m}^\infty$ and $\rho^*(\bb{T}_{2m}^\infty)$ relative to a given point
of the base space are related by an anti-linear transform. The result follows from the same argument as in \cite[Lemma 14.9]{milnor-stasheff-74}.
\qed

\medskip
The equivariant cohomology of a fixed point $\{\ast\}$ plays the role of the coefficient system for the  equivariant cohomology of an equivariant space. We recall that (\cf \cite[Section 2.3]{gomi-13} and \cite[Section 5.1]{denittis-gomi-14})
\begin{equation}\label{eq:gen_HPR}
H_{\Z_2}^\bullet\big(\{\ast\},\Z\big)\;=\;H^\bullet\big(\R P^\infty,\Z\big)\;\simeq\;\Z[\rr{t}]/(2\rr{t})
\end{equation}
where the generator $\rr{t}\in H^2\big(\R P^\infty,\Z\big)\simeq\Z_2$ obeys $2\rr{t}=0$ and
\begin{equation}\label{eq:gen_HPR1}
H_{\Z_2}^\bullet\big(\{\ast\},\Z(1)\big)\;=\;H^\bullet\big(\R P^\infty,\Z(1)\big)\;\simeq\;\rr{t}^{{1}/{2}}\; \Z[\rr{t}]/(2\rr{t}^{{1}/{2}},2\rr{t})
\end{equation}
where  also $\rr{t}^{{1}/{2}}\in H^1\big(\R P^\infty,\Z(1)\big)\simeq\Z_2$ is subjected to  $2\rr{t}^{{1}/{2}}=0$. The relation $\rr{t}^{{1}/{2}}\cup\rr{t}^{{1}/{2}}=\rr{t}$ given by the cup  product provides the ring structure
\begin{equation}\label{eq:gen_HPR2}
H_{\Z_2}^\bullet\big(\{\ast\},\Z\big)\;\oplus\;H_{\Z_2}^\bullet\big(\{\ast\},\Z(1)\big)\;\simeq\;\Z[\rr{t}^{1/2}]/(2\rr{t}^{1/2})\;.
\end{equation}
The generators $\rr{t}^{{1}/{2}}$, $\rr{t}$ and $\rr{t}^{3/2}:=\rr{t}^{{1}/{2}}\cup\rr{t}\in H_{\Z_2}^3\big(\{\ast\},\Z(1)\big)$  play an important role in the description of the
(low dimensional) equivariant cohomology of the involutive space $\hat{G}_{2m}(\C^\infty)\equiv ({G}_{2m}(\C^\infty),\rho)$.
\begin{lemma}\label{theo:cohom_G}
The 
cohomology groups of the involutive space $\hat{G}_{2m}(\C^\infty)$ up to the degree  $4$  are summarized in the Table A.1. Moreover, the generators of $H^1_{\Z_2}(\hat{G}_{2m}(\C^\infty),\Z(1))$, 
$H^2_{\Z_2}(\hat{G}_{2m}(\C^\infty),\Z)$ and $H^3_{\Z_2}(\hat{G}_{2m}(\C^\infty),\Z(1))$ can be identified with $\rr{t}^{1/2}$, $\rr{t}$ and $\rr{t}^{3/2}$, respectively.
 Similarly, the generator of the $\Z_2$-summand in the group $H^4_{\Z_2}(\hat{G}_{2m}(\C^\infty),\Z)$ can be identified with $\rr{t}^{2}:=\rr{t}\cup\rr{t}$.

 \begin{center}
 \vspace{-3mm}
 \begin{table}[h]\label{tab:AAA}
 \begin{tabular}{|r||c|c|c|c|c|}
\hline
  & $k=0$  & $k=1$ & $k=2$  &  $k=3$  &  $k=4$\\
\hline
 \hline
 \rule[-3mm]{0mm}{9mm}
 $H^k\big({G}_{2m}(\C^\infty),\Z\big)$& $\Z$ & $0$ & $\Z$ & $0$ &  $\Z^2$   \\
\cline{1-6}
 \rule[-3mm]{0mm}{9mm}
$H^k_{\Z_2}\big(\hat{G}_{2m}(\C^\infty),\Z\big)$ & $\Z$ & $0$ & $\Z_2$ & $\Z_2$  &  $\Z^2\;\oplus\;\Z_2$  \\
\hline
  \rule[-3mm]{0mm}{9mm}
$H^k_{\Z_2}\big(\hat{G}_{2m}(\C^\infty),\Z(1)\big)$ & $0$ & $\Z_2$ & $\Z$ & $\Z_2$  &  $\Z_2$   \\
\hline
\end{tabular}
\vspace{1mm}
\caption{}
 \end{table}
 \end{center}
%
\end{lemma}
\proof
The ordinary (non-equivariant) cohomology is described by \eqref{eq:univ_chern_class!}. The computation of the equivariant cohomology requires the use of the Leray-Serre spectral sequence. For more details we refer to \cite[Appendix D]{denittis-gomi-15} and references therein. We know that there is a converging spectral sequence 
$$
E_2^{p,q}\;\Rightarrow\;H^{p+q}\big(\hat{G}_{2m}(\C^\infty),\Z(j)\big)\;
$$
which in page 2 is given by the \emph{group cohomology} of $\Z_2$,
$$
E_2^{p,q}\;:=\;H_{\rm group}^p\big(\Z_2;H^q({G}_{2m}(\C^\infty),\Z)\otimes_\Z\Z(j)\big)\;,
$$
where $j=0,1$ and the coefficient systems  $H^q({G}_{2m}(\C^\infty),\Z)\otimes_\Z\Z(j)$ are regarded as  $\Z_2$-modules with respect to the involution $\rho^*$ on $H^q({G}_{2m}(\C^\infty),\Z)$ described in Lemma \ref{lemma:inv_cohom_gras} and the action $n\mapsto (-1)^j n$ on $\Z(j)$. The page 2 is concentrated in the first quadrant, meaning that $E_2^{p,q}=0$
 if $p < 0$ or $q < 0$. In the case $j=1$ the  values of the 2-page  $E_2^{p,q}$ are showed in the following table:\\

\begin{center}
\vspace{-3mm}
 \begin{table}[h]\label{tab:E_2_d=4}
 \begin{tabular}{c||c|c|c|c|c|c|c|}
 \cline{1-7}
 \rule[-3mm]{0mm}{9mm}
$q=5$ & $0$ & $0$ & $0$ &$0$&$0$&$0$\\
\cline{1-7}
 \rule[-3mm]{0mm}{9mm}
$q=4$ & $0$ & ${\Z_2}^2$ & $0$ &${\Z_2}^2$&$0$&${\Z_2}^2$ \\
\cline{1-7}
 \rule[-3mm]{0mm}{9mm}
$q=3$ & $0$ & $0$ & $0$ &$0$&$0$&$0$\\
 \hline
 \rule[-3mm]{0mm}{9mm}
 $q=2$& $\Z$ & $0$ & $\Z_2$ & $0$ & $\Z_2$& $0$  \\
\cline{1-7}
 \rule[-3mm]{0mm}{9mm}
$q=1$ & $0$ & $0$ & $0$ &$0$&$0$&$0$\\
\cline{1-7}
 \rule[-3mm]{0mm}{9mm}
$q=0$ & $0$ & $\Z_2$ & $0$ &$\Z_2$&$0$&$\Z_2$\\
\hline
 \hline
 \rule[-3mm]{0mm}{9mm}
$E^{p,q}_2\  (j=1)$& $p=0$ & $p=1$ &$p=2$&$p=3$&$p=4$&$p=5$\\ 
\end{tabular}
 \end{table}
 \end{center}
\vspace{-4mm}

\noindent
Let us first notice that $E_2^{p,2i+1}=0$ as a consequence of the fact that the cohomology  $H^q({G}_{2m}(\C^\infty),\Z)$ is concentrated in even degree. This implies that the differentials $d_2:E^{p,q}_2\to E^{p+2,q-1}_2$ are trivial so that $E^{p,q}_3\simeq E^{p,q}_2$. Moreover, the differentials
$$
d_3\;:\;E^{p,2}_3\;\to\; E^{p+3,0}_3\;,\qquad\quad p=0,2,4,\ldots
$$ 
have to be trivial since  $E^{p,0}_3$ must survive inside the direct summand $E^{p,0}_2\simeq H_{\Z_2}^p(\{\ast\},\Z(1))$ of the decomposition 
\begin{equation}\label{EQ:SPEC_SEQ_GN1}
H^{p}\big(\hat{G}_{2m}(\C^\infty),\Z(1)\big)\;\simeq\;H_{\Z_2}^p\big(\{\ast\},\Z(1)\big)\;\oplus\;\tilde{H}^{p}\big(\hat{G}_{2m}(\C^\infty),\Z(1)\big)
\end{equation}
where $\tilde{H}^{\bullet}$ denotes the relative cohomology with respect to a fixed point. In fact, if $d_3$
were non trivial, some of the $E^{p+3,0}_3$ would be killed producing a contradiction. For instance, suppose that 
$d_3:E^{2,2}_3 \to  E^{5,0}_3$ is non trivial. Then, $E^{5,0}_4=0$ and so $E^{5,0}_4\simeq E^{5,0}_5\simeq\ldots\simeq E^{5,0}_\infty=0$. However, from a general property of the spectral sequence $E^{5,0}_\infty$ injects into $H^{5}(\hat{G}_{2m}(\C^\infty),\Z(1))$. Consider the following commutative diagram
\beql{eq:diag00_CW}
\begin{diagram}
          &                & E_5^{p,0}(\{\ast\})\simeq H_{\Z_2}^5\big(\{\ast\},\Z(1)\big)\simeq \Z_2&        \\
          & \ldTo^{\jmath_1^*} &                                        &\rdTo^{\jmath_2^*}& \\
E^{5,0}_5&         &                       \rTo^{s}                 &                &H^{5}\big(\hat{G}_{2m}(\C^\infty),\Z(1)\big)
\end{diagram}
\eeq
where $s$ is the injection induced by the spectral sequence and $\jmath_2^*$ is the homomorphism induced by the collapsing (equivariant) map $\jmath:\hat{G}_{2m}(\C^\infty)\to\{\ast\}$. Also $\jmath_2^*$ is injective due to the decomposition 
\eqref{EQ:SPEC_SEQ_GN1}. The map $\jmath_1^*$ can be seen as the homomorphism induced by $\jmath$ between the spectral sequence associated to the equivariant cohomology of the fixed point $E_2^{p,0}(\{\ast\})\simeq E_\infty^{p,0}(\{\ast\})\simeq H_{\Z_2}^\bullet(\{\ast\},\Z(1))$ and the spectral sequence for the cohomology of $\hat{G}_{2m}(\C^\infty)$.
Since $s\circ \jmath_1^*= \jmath_2^*$ is injective also $\jmath_1^*$ has to be injective, proving that $E^{5,0}_\infty\neq 0$. This contradiction shows that $d_3$ is trivial.
The triviality of $d_3$ implies that $E_4^{p,q}\simeq E_2^{p,q}$ for all $0\leqslant p+q\leqslant 4$. Moreover, all the differentials  $d_r:E^{p,q}_r \to  E^{p+r,q-r+1}_r$ are automatically trivial for $0\leqslant p+q\leqslant 4$ and $r\geqslant 5$, so that one concludes $E_4^{p,q}\simeq E_\infty^{p,q}$ in the range $0\leqslant p+q\leqslant 4$. In this situation the extension problem
$$
0\;\longrightarrow\;F^{p+1}H^{p+q}\big(\hat{G}_{2m}(\C^\infty),\Z(1)\big)\;\longrightarrow\;F^{p}H^{p+q}\big(\hat{G}_{2m}(\C^\infty),\Z(1)\big)\;\longrightarrow\; E_\infty^{p,q}\;\longrightarrow\;0
$$
can be (easily) solved. Notice that, among the pairs   $0\leqslant p+q\leqslant 4$ such that $p+q=n$ there is at least one pair which gives a non-trivial $E_4^{p,q}$. 
In conclusion one finds that 
$$
\begin{aligned}
&H^{0}\big(\hat{G}_{2m}(\C^\infty),\Z(1)\big)&\simeq\ \ &E^{0,0}_\infty&\simeq\ \ &0\\
&H^{1}\big(\hat{G}_{2m}(\C^\infty),\Z(1)\big)&\simeq\ \ &E^{1,0}_\infty&\simeq\ \ &\Z_2\\
&H^{2}\big(\hat{G}_{2m}(\C^\infty),\Z(1)\big)&\simeq\ \ &E^{0,2}_\infty&\simeq\ \ &\Z\\
&H^{3}\big(\hat{G}_{2m}(\C^\infty),\Z(1)\big)&\simeq\ \ &E^{3,0}_\infty&\simeq\ \ &\Z_2\\
&H^{4}\big(\hat{G}_{2m}(\C^\infty),\Z(1)\big)&\simeq\ \ &E^{2,2}_\infty&\simeq\ \ &\Z_2\;.\\
\end{aligned}
$$

\medskip
The case $j=0$ can be proved  in a similar way. In this case the  values of the 2-page  are summarized in the following table:\\

\begin{center}
\vspace{-3mm}
 \begin{table}[h]\label{tab:E_2_d=4_B}
 \begin{tabular}{c||c|c|c|c|c|c|c|}
 \cline{1-7}
 \rule[-3mm]{0mm}{9mm}
$q=5$ & $0$ & $0$ & $0$ &$0$&$0$&$0$\\
\cline{1-7}
 \rule[-3mm]{0mm}{9mm}
$q=4$ & $\Z^2$ & $0$ & ${\Z_2}^2$ &$0$&${\Z_2}^2$&$0$ \\
\cline{1-7}
 \rule[-3mm]{0mm}{9mm}
$q=3$ & $0$ & $0$ & $0$ &$0$&$0$&$0$\\
 \hline
 \rule[-3mm]{0mm}{9mm}
 $q=2$& $0$ & $\Z_2$ & $0$ & $\Z_2$ & $0$& $\Z_2$  \\
\cline{1-7}
 \rule[-3mm]{0mm}{9mm}
$q=1$ & $0$ & $0$ & $0$ &$0$&$0$&$0$\\
\cline{1-7}
 \rule[-3mm]{0mm}{9mm}
$q=0$ & $\Z$ & $0$ & $\Z_2$ &$0$&$\Z_2$&$0$\\
\hline
 \hline
 \rule[-3mm]{0mm}{9mm}
$E^{p,q}_2\  (j=0)$& $p=0$ & $p=1$ &$p=2$&$p=3$&$p=4$&$p=5$\\ 
\end{tabular}
 \end{table}
 \end{center}
\vspace{-3mm}

\noindent
 As above we have that $E^{p,q}_3\simeq E^{p,q}_2$ and one can show that the maps $d_3:E_3^{p,2}\to E^{p+3,0}_3$ are trivial
for all $p$ since $E^{p,0}_2\simeq H^p_{\Z_2}(\{\ast\},\Z)$ has to survive to contribute as a direct summand  of $H^p_{\Z_2}(\hat{G}_{2m}(\C^\infty),\Z)$ due to the presence of a fixed point. Therefore, one can determine  $E^{p,q}_\infty$ for all $p+q\leqslant 3$. For $p+q=4$ one does not have enough information to determine whether $d_3:E_3^{0,4}\to E^{3,2}_3$
is trivial or not. However, in any event,
the kernel of this map is $\Z^2$ and this is enough to compute $E^{p,q}_\infty$ also for all $p+q=4$. Since $E_3^{4,0}$ must survive as the contribution from $H^4_{\Z_2}(\{\ast\},\Z)$,  the extension problem for $H^4_{\Z_2}(\hat{G}_{2m}(\C^\infty),\Z)$ is trivial, and the computation is completed.

Finally we notice that the identification with the generators $\rr{t}^{1/2}$, $\rr{t}$,  $\rr{t}^{3/2}$ and $\rr{t}^2$ is clear from the 
direct summand argument.
\qed

\medskip

\noindent
The last result has     important consequences on the description of the cohomology of $\hat{G}_{2m}(\C^\infty)$.
\begin{proposition}\label{propos:item(1)}
The group $H^2_{\Z_2}(\hat{G}_{2m}(\C^\infty),\Z(1))\simeq\Z$ is generated by the (universal) first \virg{Real} Chern class 
$$
\rr{c}^{\rr{R}}_1\;:=\;c^{\rr{R}}_1\Big({\rm det}\big(\bb{T}_{2m}^\infty\big)\Big)
$$ 
of the \virg{Real} line bundle ${\rm det}(\bb{T}_{2m}^\infty)\to \hat{G}_{2m}(\C^\infty)$ associated to the 
{tautological} $\rr{Q}$-bundle. Moreover, under the map $f:H^2_{\Z_2}(\hat{G}_{2m}(\C^\infty),\Z(1))\to
H^2({G}_{2m}(\C^\infty),\Z)$ which forgets the $\Z_2$-action the class $\rr{c}^{\rr{R}}_1$ is mapped in the  first universal Chern class $\rr{c}_1$.
\end{proposition}
\proof
The exact sequence \cite[Proposition 2.3]{gomi-13}
$$
H^{1}_{\Z_2}\big(\hat{G}_{2m}(\C^\infty),\Z\big)\;\stackrel{\delta}{\longrightarrow}\;H^2_{\Z_2}\big(\hat{G}_{2m}(\C^\infty),\Z(1)\big)\;\stackrel{f}{\longrightarrow}\;H^2\big({G}_{2m}(\C^\infty),\Z\big)\;\stackrel{g}{\longrightarrow}\;H^{2}_{\Z_2}\big(\hat{G}_{2m}(\C^\infty),\Z\big)\;\ldots\;
$$
and $H^{1}_{\Z_2}\big(\hat{G}_{2m}(\C^\infty),\Z\big)=0$ imply that $f$ is injective. Moreover,  
$$
\stackrel{g}{\longrightarrow}\;H^{2}_{\Z_2}\big(\hat{G}_{2m}(\C^\infty),\Z\big)\;\stackrel{\delta'}{\longrightarrow}\;H^3_{\Z_2}\big(\hat{G}_{2m}(\C^\infty),\Z(1)\big)\;{\longrightarrow}\;H^3\big({G}_{2m}(\C^\infty),\Z\big)\;\ldots\;,
$$
along with $H^3\big({G}_{2m}(\C^\infty),\Z\big)=0$ and $H^{2}_{\Z_2}\big(\hat{G}_{2m}(\C^\infty),\Z\big)\simeq\Z_2\simeq H^3_{\Z_2}\big(\hat{G}_{2m}(\C^\infty),\Z(1)\big)$,
implies that $\delta'$ is an isomorphism.  Therefore $g=0$ and 
$$
f\;:\; H^2_{\Z_2}\big(\hat{G}_{2m}(\C^\infty),\Z(1)\big)\;\stackrel{\simeq}{\longrightarrow}\;H^2\big({G}_{2m}(\C^\infty),\Z\big)
$$
is an isomorphism. The general behavior of the \virg{Real} Chern classes under the forgetting map is
$$
f\;:\; c^{\rr{R}}_1\Big({\rm det}\big(\bb{T}_{2m}^\infty\big)\Big)\;\longmapsto\; c_1\Big({\rm det}\big(\bb{T}_{2m}^\infty\big)\Big)\;.
$$
Finally, the equality $c_1({\rm det}(\bb{T}_{2m}^\infty))=c_1(\bb{T}_{2m}^\infty)$ and the definition of universal Chern class  $\rr{c}_1=c_1(\bb{T}_{2m}^\infty)$ conclude the proof.
\qed

\medskip

Since the fixed point set $\hat{G}_{2m}(\C^\infty)^\rho$ can be identified with  $G_{m}(\n{H}^\infty)$ one has that
\begin{equation}\label{eq:isograsm_quat}
H^\bullet\big(\hat{G}_{2m}(\C^\infty)^\rho,\Z\big)\;\simeq\;H^\bullet\big(G_{m}(\n{H}^\infty),\Z\big)\;.
\end{equation}
 The latter is given by
 the polynomial ring 
\begin{equation}\label{eq:univ_chern_class}
H^\bullet\big(G_{m}(\n{H}^\infty),\Z\big)\;\simeq\;\Z[\rr{q}_1,\ldots,\rr{q}_{m}]
\end{equation}
generated by the \emph{universal (symplectic) Pontryagin classes} $\rr{q}_k\in H^{4k}( G_{m}(\n{H}^\infty),\Z)$
 \cite[Chapter 17]{husemoller-94}.

\begin{lemma}\label{lemma:ecos}
It holds that
$$
H^\bullet_{\Z_2}\big(\hat{G}_{2m}(\C^\infty)^\rho,\Z\big)\;\oplus\; H^\bullet_{\Z_2}\big(\hat{G}_{2m}(\C^\infty)^\rho,\Z(1)\big)\;\simeq\;\Z[\rr{t}^{1/2},\rr{q}_1,\ldots,\rr{q}_{m}]/(2\rr{t}^{1/2})
$$
where the class $\rr{t}^{1/2}$ agrees with the generator of $H^1_{\Z_2}\big(\{\ast\},\Z(1)\big)\simeq\Z_2$ and $\rr{q}_k\in H^{4k}(\hat{G}_{2m}(\C^\infty)^\rho,\Z)$ are the symplectic Pontryagin classes (up to the identification \eqref{eq:isograsm_quat}).
\end{lemma}
\proof
The involution on $\hat{G}_{2m}(\C^\infty)^\rho\simeq G_{m}(\n{H}^\infty)$ is trivial, hence 
$$
H^\bullet_{\Z_2}\big(\hat{G}_{2m}(\C^\infty)^\rho,\Z(j)\big)\;\simeq\; H^\bullet\big({G}_{m}(\n{H}^\infty)\times\R P^\infty,\Z(j)\big)\;,\qquad\qquad j=0,1\;
$$
by definition.
The K\"unneth formula and the absence of torsion in $H^\bullet({G}_{m}(\n{H}^\infty))$ imply that
$$
H^\bullet_{\Z_2}\big(\hat{G}_{2m}(\C^\infty)^\rho,\Z(j)\big)\;\simeq\;H^\bullet_{\Z_2}\big(\{\ast\},\Z(j)\big)\;\otimes\; H^\bullet\big({G}_{m}(\n{H}^\infty),\Z\big)
$$
More precisely, a comparison with  \eqref{eq:gen_HPR} and \eqref{eq:gen_HPR1} provides that
$$
H^\bullet_{\Z_2}\big(\hat{G}_{2m}(\C^\infty)^\rho,\Z(j)\big)\;\simeq\;\left\{
\begin{aligned}
&\Z[\rr{t},\rr{q}_1,\ldots,\rr{q}_{m}]/(2\rr{t})&\ \ \text{if}\ \ &\ j=0\\
&\rr{t}^{{1}/{2}}\; \Z[\rr{t},\rr{q}_1,\ldots,\rr{q}_{m}]/(2\rr{t}^{{1}/{2}},2\rr{t})&\ \ \text{if}\ \ &\ j=1
\end{aligned}
\right.
$$
where  $\rr{t}^{{1}/{2}}\in H^1_{\Z_2}\big(\{\ast\},\Z(1)\big)$,  $\rr{t}\in H^2\big(\{\ast\},\Z\big)\simeq\Z_2$ and 
$\rr{q}_k\in H^{4k}(\hat{G}_{2m}(\C^\infty)^\rho,\Z)$ are the symplectic Pontryagin classes. The final result is a consequence of  \eqref{eq:gen_HPR2}.
\qed

\medskip

\noindent
The  low order cohomology  of $\hat{G}_{2m}(\C^\infty)^\rho$ is summarized in the following table:

 \begin{center}
 \vspace{-0mm}
 \begin{table}[h]
 \begin{tabular}{|r||c|c|c|c|c|}
\hline
  & $k=0$  & $k=1$ & $k=2$  &  $k=3$  &  $k=4$\\
\hline
 \hline
 \rule[-3mm]{0mm}{9mm}
 $H^k\big(\hat{G}_{2m}(\C^\infty)^\rho,\Z\big)$& $\Z$ & $0$ & $0$ & $0$ &  $\Z$   \\
\cline{1-6}
 \rule[-3mm]{0mm}{9mm}
$H^k_{\Z_2}\big(\hat{G}_{2m}(\C^\infty)^\rho,\Z\big)$ & $\Z$ & $0$ & $\Z_2$ & $0$  &  $\Z\;\oplus\;\Z_2$  \\
\hline
  \rule[-3mm]{0mm}{9mm}
$H^k_{\Z_2}\big(\hat{G}_{2m}(\C^\infty)^\rho,\Z(1)\big)$ & $0$ & $\Z_2$ & $0$ & $\Z_2$  &  $0$   \\
\hline
\end{tabular} 
 \end{table}
 \end{center}
\vspace{-3mm}

The relative equivariant cohomology of the pair $\hat{G}_{2m}(\C^\infty)^\rho\hookrightarrow\hat{G}_{2m}(\C^\infty)$ can be conputed with the help of the long exact sequence \eqref{eq:Long_seq}.

\begin{proposition}\label{propos:item(2)}
It holds that
 \begin{center}
 \vspace{-3mm}
 \begin{table}[h]
 \begin{tabular}{|c||c|c|c|c|c|}
\hline
  & $k=0$  & $k=1$ & $k=2$  &  $k=3$  &  $k=4$\\
\hline
\hline
  \rule[-3mm]{0mm}{9mm}
$H^k_{\Z_2}\big(\hat{G}_{2m}(\C^\infty)|\hat{G}_{2m}(\C^\infty)^\rho,\Z(1)\big)$ & $0$ & $0$ & $\Z$ & $0$  &  $\Z_2$   \\
\hline
\end{tabular}
 \end{table}
 \end{center}
\vspace{-5mm}
and the map 
$$
\delta_2\;:\; H^2_{\Z_2}\big(\hat{G}_{2m}(\C^\infty)|\hat{G}_{2m}(\C^\infty)^\rho,\Z(1)\big)\;\stackrel{\simeq}{\longrightarrow}\;H^2_{\Z_2}\big(\hat{G}_{2m}(\C^\infty),\Z(1)\big)
$$
is an isomorphism.
\end{proposition}
\proof
The restriction map $\hat{G}_{2m}(\C^\infty)^\rho\hookrightarrow\hat{G}_{2m}(\C^\infty)$ induces  homomorphisms
$$
r\;:\; H^k_{\Z_2}\big(\hat{G}_{2m}(\C^\infty),\Z(1)\big)\;{\longrightarrow}\;H^k_{\Z_2}\big(\hat{G}_{2m}(\C^\infty)^\rho,\Z(1)\big).
$$
The analysis in Lemma \ref{theo:cohom_G} and Lemma \ref{lemma:ecos} shows that the homomorphisms in degree $k=1$ and $k=3$ are in fact the isomorphisms  which fix the generators $\rr{t}^{{1}/{2}}$ and $\rr{t}^{{3}/{2}}$, respectively.
To conclude the proof one needs to inspect  the long exact sequence \eqref{eq:Long_seq} for the relative equivariant cohomology $H^k_{\Z_2}(\hat{G}_{2m}(\C^\infty)|\hat{G}_{2m}(\C^\infty)^\rho,\Z(1))$. The degree $k=0$ is trivial. In  degree $k=1$ the exact sequence reads
$$
\begin{aligned}
0\;&\stackrel{\delta_1}{\longrightarrow}\;H^1_{\Z_2}\big(\hat{G}_{2m}(\C^\infty)|\hat{G}_{2m}(\C^\infty)^\rho,\Z(1)\big)\;\stackrel{\delta_2}{\longrightarrow}\;\Z_2\;\stackrel[\simeq]{r}{\longrightarrow}\;\Z_2\;\stackrel{\delta_1}{\longrightarrow}\;
\end{aligned}
$$
and this shows that $H^1_{\Z_2}(\hat{G}_{2m}(\C^\infty)|\hat{G}_{2m}(\C^\infty)^\rho,\Z(1))=0$.
In degree $k=2$ this implies 
$$
\begin{aligned}
0\;&\stackrel{\delta_1}{\longrightarrow}\;H^2_{\Z_2}\big(\hat{G}_{2m}(\C^\infty)|\hat{G}_{2m}(\C^\infty)^\rho,\Z(1)\big)\;\stackrel{\delta_2}{\longrightarrow}\;H^2_{\Z_2}\big(\hat{G}_{2m}(\C^\infty),\Z(1)\big)\;\stackrel{r}{\longrightarrow}\;0
\end{aligned}
$$
and in turn the isomorphism between $H^2_{\Z_2}(\hat{G}_{2m}(\C^\infty)|\hat{G}_{2m}(\C^\infty)^\rho,\Z(1))$
and $H^2_{\Z_2}(\hat{G}_{2m}(\C^\infty),\Z(1))\simeq\Z$.
In degree $k=3$ one has
$$
\begin{aligned}
0\;&\stackrel{\delta_1}{\longrightarrow}\;H^3_{\Z_2}\big(\hat{G}_{2m}(\C^\infty)|\hat{G}_{2m}(\C^\infty)^\rho,\Z(1)\big)\;\stackrel{\delta_2}{\longrightarrow}\;\Z_2\;\stackrel[\simeq]{r}{\longrightarrow}\;\Z_2\;\stackrel{\delta_1}{\longrightarrow}\;
\end{aligned}
$$
which implies $H^3_{\Z_2}(\hat{G}_{2m}(\C^\infty)|\hat{G}_{2m}(\C^\infty)^\rho,\Z(1))=0$. Finally, in degree $k=4$ the exact sequence reads
$$
\begin{aligned}
0\;&\stackrel{\delta_1}{\longrightarrow}\;H^4_{\Z_2}\big(\hat{G}_{2m}(\C^\infty)|\hat{G}_{2m}(\C^\infty)^\rho,\Z(1)\big)\;\stackrel{\delta_2}{\longrightarrow}\;\Z_2\;\stackrel{r}{\longrightarrow}\;0\;
\end{aligned}
$$
and so $H^4_{\Z_2}(\hat{G}_{2m}(\C^\infty)|\hat{G}_{2m}(\C^\infty)^\rho,\Z(1))\simeq\Z_2$.\qed


\section{About the equivariant cohomology of involutive spheres}
\label{sect:cohom_sper}

\subsection
{TR-involution}
Let ${\n{S}}^{1,d}:=({\n{S}}^d,\theta_{1,d})$ be the $d$-dimensional
{sphere} with \emph{TR-involution} 
$\theta_{1,d}:(k_0,k_1,\ldots,k_d)\mapsto(k_0,-k_1,\ldots,-k_d)$ (\cf  \cite[Example 4.2]{denittis-gomi-14}).
The equivariant cohomology groups
$H^{k}
_{\Z_2}({\n{S}}^{1,d},\Z(1))$
 have been computed in  \cite[Section 5.4]{denittis-gomi-14}:
\beql{eq:equi_cohom_spher_d_loc}
H^{\rm even}
_{\Z_2}\big({\n{S}}^{1,d},\Z(1)\big)\;\simeq\;0\qquad\quad H^{\rm odd}_{\Z_2}\big({\n{S}}^{1,d},\Z(1)\big)\;\simeq\;
\left\{
\begin{aligned}
&\Z_2\oplus\Z&&\text{if}\ \ k=d\\
&\Z_2&&\text{if}\ \  k< d\\
&\Z_2\oplus \Z_2&&\text{if}\ \ k>d\;.
\end{aligned}
\right.
 \eeq
Moreover, since  $({\n{S}}^{1,d})^\theta=\{-1,+1\}_{\rm fixed}$ consists of two fixed points (in each dimension), one has that
$$
H^k_{\Z_2}\big(({\n{S}}^{1,d})^\theta,\Z(1)\big)\;\simeq\;H_{\Z_2}^k\big(\{\ast\},\Z(1)\big)\;\oplus\; H_{\Z_2}^k\big(\{\ast\},\Z(1)\big)\;\simeq\;
\left\{
\begin{aligned}
&\Z_2\oplus\Z_2&&\text{if}\ \  k\ \text{is odd}\\
&0&&\text{if}\ \  k\ \text{is even}\;.
\end{aligned}
\right.\;
$$
The groups $H^{2}_{\Z_2}\big({\n{S}}^{1,d}|({\n{S}}^{1,d})^\theta,\Z(1)\big)$ for $d\geqslant 2$ have been computed in \cite[Proposition A.1]{denittis-gomi-14-gen}:
$$
H^{2}_{\Z_2}\big({\n{S}}^{1,d}|({\n{S}}^{1,d})^\theta,\Z(1)\big)\;\simeq\;\Z_2\;,\qquad\qquad d\geqslant 2\;.
$$
When $d=0$ the equality ${\n{S}}^{1,0}=\{-1,+1\}_{\rm fixed}=({\n{S}}^{1,d})^\theta$ immediately implies
$$
H^{2}_{\Z_2}\big({\n{S}}^{1,0}|({\n{S}}^{1,0})^\theta,\Z(1)\big)\;=\;0\;.
$$
The case $d=1$ can be studied with the exact sequence \eqref{eq:Long_seq} which in this case reads
$$
0\;\stackrel{}{\longrightarrow}\;H^{1}_{\Z_2}\big({\n{S}}^{1,1}|({\n{S}}^{1,1})^\theta,\Z(1)\big)
\;\stackrel{\delta_2}{\longrightarrow}\;\Z_2\oplus\Z\;\stackrel{r}{\longrightarrow}\;\Z_2\oplus\Z_2\;\stackrel{\delta_1}{\longrightarrow}\;H^{2}_{\Z_2}\big({\n{S}}^{1,1}|({\n{S}}^{1,1})^\theta,\Z(1)\big)\;\stackrel{}{\longrightarrow}\;0\;.
$$
The same argument as in the proof of \cite[Proposition A.1]{denittis-gomi-14-gen} shows that $r$ acts bijectively on the $\Z_2$ summand. On the other hand each homomorphism $\Z\to\Z_2$ has $\Z$ as kernel. Thus
$$
H^{1}_{\Z_2}\big({\n{S}}^{1,1}|({\n{S}}^{1,1})^\theta,\Z(1)\big)\;=\;\Z\;.
$$
Finally, since the $\Z_2$-summand   in $\Z_2 \oplus \Z$ is diagonally embedded into $\Z_2 \oplus \Z_2$
and the image of the $\Z$-summand  is $\{ (1, -1), (-1, 1) \}$,
one concludes that $r$ is surjective, and so
$$
H^{2}_{\Z_2}\big({\n{S}}^{1,1}|({\n{S}}^{1,1})^\theta,\Z(1)\big)\;=\;0\;.
$$
Let  $\imath:{\n{S}}^{1,d-1}\to {\n{S}}^{1,d}$ be  the \emph{embedding} maps
which identify ${\n{S}}^{1,d-1}$ with the subset of ${\n{S}}^{1,d}$ defined by the constraint
 $k_d=0$.
\begin{lemma}\label{lemma:isomaps_restr}
The homomorphisms
$$
\imath^*\;:\;H^{2}_{\Z_2}\big({\n{S}}^{1,d}|({\n{S}}^{1,d})^\theta,\Z(1)\big)\;\longrightarrow\; H^{2}_{\Z_2}\big({\n{S}}^{1,d-1}|({\n{S}}^{1,d-1})^\theta,\Z(1)\big)
$$
induced by $\imath:{\n{S}}^{1,d-1}\to{\n{S}}^{1,d}$ are isomorphisms for all $d\geqslant 3$.
\end{lemma}
\proof
For  $d\geqslant 2$ the contribution to $H^{1}_{\Z_2}({\n{S}}^{1,d},\Z(1))\simeq\Z_2$ is given by the summand
$H^{1}_{\Z_2}\big(\{\ast\},\Z(1)\big)$ where the fixed point $\ast\in {\n{S}}^{1,d}$ is preserved by the {embedding}. Therefore, $\imath^*:H^{1}_{\Z_2}({\n{S}}^{1,d},\Z(1))\to H^{1}_{\Z_2}({\n{S}}^{1,d-1},\Z(1))$ is an isomorphism. Also  the maps $\imath^*:H^{1}_{\Z_2}(({\n{S}}^{1,d})^\theta,\Z(1))\to H^{2}_{\Z_2}(({\n{S}}^{1,d-1})^\theta,\Z(1))$ are isomorphisms since the fixed point set $\{-1,+1\}_{\rm fixed}$ is preserved by the {embedding}. The 
 naturality of the exact sequence for pairs
 concludes the proof.
\qed

\subsection
{Antipodal involution}\label{set:antipodal}
Let ${\n{S}}^{0,d+1}:=({\n{S}}^d,\theta_{0,d+1})$ be the $d$-dimensional sphere with (free) \emph{antipodal involution} 
$\theta_{0,d+1}:(k_0,k_1,\ldots,k_d)\mapsto(-k_0,-k_1,\ldots,-k_d)$ (\cf  \cite[Example 4.1]{denittis-gomi-14}).
We want to compute the equivariant cohomology of ${\n{S}}^{0,d+1}$. 
The ordinary integer cohomology of the sphere is given by
$$
H^k\big(\n{S}^d,\Z\big)\;\simeq\;
\left\{
\begin{aligned}
&\Z&& \ \text{if}\  \ \ k=0,d\\
&0&&\ \text{if}\  \ \ k\neq0,d\;.\\
\end{aligned}
\right.
$$
The equivariant cohomology with integer coefficients  can be computed by observing that $\theta_{0,d+1}$ acts freely with orbit space ${\n{S}}^{0,d+1}/\theta_{0,d+1}\simeq \R P^d$. Therefore
$$
H^k_{\Z_2}\big({\n{S}}^{0,d+1},\Z\big)\;\simeq\; H^k\big( \R P^d,\Z\big)\;\simeq\;
\left\{
\begin{aligned}
&\Z&& \ \text{if}\  \ \ k=0\\
&\Z_2&& \ \text{if}\  \ \ k\ \text{even}\ \ 0<k<d \\
&\Z&& \ \text{if}\  \ \ k=d\ \text{odd}\\
&\Z_2&& \ \text{if}\  \ \ k=d\ \text{even} \\
&0&&\  \text{otherwise}\;\\
\end{aligned}
\right.
$$
by construction.
The equivariant cohomology with $\Z(1)$ coefficients can be   investigated by the help of the 
\emph{Gysin exact sequence}.
\begin{proposition}\label{prop:cohom_free_spher}
Let ${\n{S}}^{0,d+1}$ be the {sphere} endowed  with the free antipodal involution. Then:
\begin{itemize}
\item[(1)] In odd dimension, $d=2n+1$:
$$
H^k_{\Z_2}\big({\n{S}}^{0,2n+2},\Z(1)\big)\;\simeq\; 
\left\{
\begin{aligned}
&\Z_2&& \ \text{if}\  \ \ k\ \text{odd}\ \ 1\leqslant k\leqslant 2n+1 \\
&0&&\  \text{otherwise}\;.\\
\end{aligned}
\right.
$$
\vspace{1mm}
\item[(2)] In even dimension, $d=2n$:
$$
H^k_{\Z_2}\big( {\n{S}}^{0,2n+1},\Z(1)\big)\;\simeq\; 
\left\{
\begin{aligned}
&\Z_2&& \ \text{if}\  \ \ k\ \text{odd}\ \ 1\leqslant k<2n \\
&\Z&& \ \text{if}\  \ \ k=2n\\
&0&&\  \text{otherwise}\;.\\
\end{aligned}
\right.
$$
\end{itemize}
\end{proposition}
\proof (1)
Let  $\pi:\underline{\R}^{d+1}\to \{\ast\}$ be the $\Z_2$-equivariant real vector bundle with non-trivial $\Z_2$-action given by $\theta_{0,d+1}:(k_0,\ldots,k_d)\mapsto (-k_0,\ldots,-k_d)$. The $d$-dimensional  {antipodal sphere} ${\n{S}}^{0,d+1}$ coincides with  the sphere-bundle 
of  $\pi:\underline{\R}^{d+1}\to \{\ast\}$, \ie ${\n{S}}^{0,d+1}=\n{S}(\underline{\R}^{d+1})$. The use the Gysin exact sequence requires the specification of the \emph{equivariant orientability} and
the \emph{equivariant Euler class} of  $\pi:\underline{\R}^{d+1}\to \{\ast\}$.
If $d$ is odd, $\pi:\underline{\R}^{d+1}\to \{\ast\}$ is equivariant orientable since its first $\Z_2$-equivariant Stiefel-Whiteny class vanishes, \ie $w_1^{\Z_2}(\underline{\R}^{d+1})=0$. This can be proved by observing that
$$
w_1^{\Z_2}(\underline{\R}^{d+1})\;\in\; H^1_{\Z_2}(\{\ast\},\Z_2)\;\simeq\;H^1(\R P^\infty,\Z_2)\simeq\;{\rm Hom}_\Z(\Z_2,\Z_2)\;\simeq\;\Z_2
$$
agrees with the homomorphism induced by ${\rm det}(\theta_{0,d+1}):\Z_2\to\Z_2$ and   ${\rm det}(\theta_{0,d+1})=+1$ when  $d$ is odd. According to the proof of \cite[Proposition 2.6]{gomi-13}, the $\Z_2$-equivariant Euler class $\chi^{\Z_2}(\underline{\R}^{1})$ of $\pi:\underline{\R}^{1}\to \{\ast\}$ can be identified with the class $\rr{t}^{1/2}$
which generates the ring \eqref{eq:gen_HPR2}, so that
$$
\chi^{\Z_2}(\underline{\R}^{d+1})\;=\;\chi^{\Z_2}(\underline{\R}^{1})^{d+1}\;=\;\rr{t}^{(d+1)/2}\;\in\;H^{d+1}_{\Z_2}(\{\ast\},\Z)\;.
$$
Now, we are ready to use the Gysin sequence for the 
$\pi: {\n{S}}^{0,d+1}=\n{S}(\underline{\R}^{d+1})\to \{\ast\}$:
$$
\;H^{k-d-1}_{\Z_2}\big(\{\ast\},\Z(1)\big)\;\stackrel{\chi\;\cup}{\longrightarrow}\;H^{k}_{\Z_2}\big(\{\ast\},\Z(1)\big)
\;\stackrel{\pi^*}{\longrightarrow}\;H^{k}_{\Z_2}\big({\n{S}}^{0,d+1},\Z(1)\big)\;\stackrel{\pi_*}{\longrightarrow}\;H^{k-d}_{\Z_2}\big(\{\ast\},\Z(1)\big)\;\stackrel{\chi\;\cup}{\longrightarrow}\;\ldots
$$
where $\chi\cup$ denotes the cup product with $\chi^{\Z_2}(\underline{\R}^{d+1})$. The map $\pi^*$ turns out to be an isomorphism when $k\leqslant d$. Moreover, $\chi\cup$ is an isomorphism for  
$k\geqslant d+1$ and this shows that $H^{k}_{\Z_2}({\n{S}}^{0,d+1},\Z(1))=0$ for $k\geqslant d+1$.\\
(2) When $d$ is even  the main difference
is that the $\Z_2$-vector bundle $\pi:\underline{\R}^{d+1}\to \{\ast\}$ is \emph{not} equivariantly orientable since $w_1^{\Z_2}(\underline{\R}^{d+1})\simeq {\rm det}(\theta_{0,d+1})=-1$. This implies that the relevant 
Gysin sequence reads 
$$
\;H^{k-d-1}_{\Z_2}\big(\{\ast\},\Z\big)\;\stackrel{\chi\;\cup}{\longrightarrow}\;H^{k}_{\Z_2}\big(\{\ast\},\Z(1)\big)
\;\stackrel{\pi^*}{\longrightarrow}\;H^{k}_{\Z_2}\big({\n{S}}^{0,d+1},\Z(1)\big)\;\stackrel{\pi_*}{\longrightarrow}\;H^{k-d}_{\Z_2}\big(\{\ast\},\Z\big)\;\stackrel{\chi\;\cup}{\longrightarrow}\;\ldots
$$
and, as in the odd case,  one deduces that    $\pi^*$ is an isomorphism for $k\leqslant d$ and $H^{k}_{\Z_2}( {\n{S}}^{0,d+1},\Z(1) )=0$ for $k\geqslant d+1$.
\qed

\medskip
\medskip

\noindent
Proposition \ref{prop:cohom_free_spher} produces the following table in low dimension:  
 \begin{center}
 \vspace{-3mm}
 \begin{table}[h]
 \begin{tabular}{|c||c|c|c|c|c|c|c|c|}
\hline
 \rule[-3mm]{0mm}{9mm}
 $H^k_{\Z_2}\big({\n{S}}^{0,d+1},\Z(1)\big)$ & $k=0$  & $k=1$ & $k=2$  &  $k=3$  &  $k=4$&  $k=5$&  $k=6$& $\ldots$\\
\hline
 \hline
 \rule[-3mm]{0mm}{9mm}
 $d=1$& $0$ & ${\Z_2}$ & $0$ & $0$ &  $0$ &  $0$ &  $0$ &$\ldots$  \\
\hline
 \rule[-3mm]{0mm}{9mm}
$d=2$ & $0$ & $\Z_2$ & $\Z$ & $0$  &  $0$ &  $0$ &  $0$ & $\ldots$  \\
\hline
  \rule[-3mm]{0mm}{9mm}
$d=3$ & $0$ & $\Z_2$ & $0$ & $\Z_2$  &  $0$ &  $0$ &  $0$  & $\ldots$ \\
\hline
  \rule[-3mm]{0mm}{9mm}
$d=4$ & $0$ & $\Z_2$ & $0$ & $\Z_2$  &  $\Z$ &  $0$ &  $0$  & $\ldots$  \\
\hline
\rule[-3mm]{0mm}{9mm}
$d=5$ & $0$ & $\Z_2$ & $0$ & $\Z_2$  &  $0$  & $\Z_2$  &  $0$  & $\ldots$\\
\hline
\end{tabular}
 \end{table}
 \end{center}
 \vspace{-5mm}

\subsection{More general involutions}

Let $\n{S}^{p,q}$ be an involutive sphere of type $(p,q)$ according to the notation introduced in Section \ref{sect:intro}. There is  an identification of $\Z_2$-spaces 
$$
\n{S}^{p,q}\;\simeq\; \underbrace{\n{S}^{2,0}\;\wedge\;\ldots\;\wedge\;\n{S}^{2,0}}_{(p-1)\text{-times}}\;\wedge\;\underbrace{\n{S}^{1,1}\;\wedge\;\ldots\;\wedge\;\n{S}^{1,1}}_{q\text{-times}}\;,\qquad\quad p\geqslant 1
$$
where $\wedge$ denotes the \emph{smash product} of topological spaces \cite[Chapter 0]{hatcher-02}. Moreover, the equivariant \emph{suspension} isomorphism \cite[Section 2.4]{gomi-13} provides 
$$
\tilde{H}^k_{\Z_2}\big(\n{S}^{p,q},\Z(j)\big)\;\simeq\;\tilde{H}^{k-q}_{\Z_2}\big(\n{S}^{p,0},\Z(j-q)\big)\;\simeq\;{H}^{k-q-(p-1)}_{\Z_2}\big(\{\ast\},\Z(j-q)\big)
$$
where $\tilde{H}^k_{\Z_2}$ is the \emph{reduced} cohomology. Finally, let us recall that in presence of a fixed point $\{\ast\}\in X^\tau$ the exact sequence \eqref{eq:Long_seq} can be extended to the 
reduced cohomology theory and one obtains
\begin{equation}\label{eq:Long_seq_red}
\ldots\;\longrightarrow\;H^k_{\Z_2}\big(X|X^\tau,\s{Z}\big)\;\stackrel{\delta_2}{\longrightarrow}\;\tilde{H}^k_{\Z_2}\big(X,\s{Z}\big)\;\stackrel{r}{\longrightarrow}\;\tilde{H}^k_{\Z_2}\big(X^\tau,\s{Z}\big)\;\stackrel{\delta_1}{\longrightarrow}\;H^{k+1}_{\Z_2}\big(X|X^\tau,\s{Z}\big)\;\longrightarrow\;\ldots\;
\end{equation}

With this information one can compute the equivariant cohomology of $\n{S}^{2,1}$. The fixed point set $(\n{S}^{2,1})^\theta\simeq\n{S}^1$ is a circle and the {suspension} isomorphism says
$$
\begin{aligned}
\tilde{H}^k_{\Z_2}\big(\n{S}^{2,1},\Z(1)\big)\;&\simeq\;{H}^{k-2}_{\Z_2}\big(\{\ast\},\Z\big)\;\simeq\; H^{k-2}\big(\R P^\infty,\Z\big)\\
\tilde{H}^k_{\Z_2}\big((\n{S}^{2,1})^\theta,\Z(1)\big)\;&\simeq\;{H}^{k-1}_{\Z_2}\big(\{\ast\},\Z(1)\big)\;\simeq\; H^{k-2}\big(\R P^\infty,\Z(1)\big)\;.
\end{aligned}
$$
The values of   ${H}^{k}_{\Z_2}\big(\{\ast\},\Z(j)\big)$ are listed in
\eqref{eq:gen_HPR} and \eqref{eq:gen_HPR1}. From \eqref{eq:Long_seq_red} one extracts the following exact sequence 
\begin{equation}\label{eq:exac_seq_S21}
\begin{diagram}
0&\rTo^{}&        
H^2_{\Z_2}\big(\n{S}^{2,1}|(\n{S}^{2,1})^\theta,\n{Z}(1)\big)&\rTo^{\delta_2}&   
\tilde{H}^2_{\Z_2}\big(\n{S}^{2,1},\n{Z}(1)\big)&\rTo^{r}&  
\tilde{H}^2_{\Z_2}\big((\n{S}^{2,1})^\theta,\n{Z}(1)\big)\; \vspace{-0.0mm} \\
    &&\rotatebox{-90}{$\simeq$} && \rotatebox{-90}{$\simeq$} && \rotatebox{-90}{$\simeq$}   \\
&&? && \Z &&{\Z_2}   \\                                    \end{diagram}
\end{equation}
which can be used to prove the following result:
\begin{proposition}\label{lemma:surj_map_r_2Z}
The map $r$ in \eqref{eq:exac_seq_S21} is surjective and consequently there is an isomorphism of groups
$$
H^2_{\Z_2}\big(\n{S}^{2,1}|(\n{S}^{2,1})^\theta,\n{Z}(1)\big)\;\simeq\;2\Z\;
$$
which is induced by the composition
$$
H^2_{\Z_2}\big(\n{S}^{2,1}|(\n{S}^{2,1})^\theta,\Z(1)\big)\;\stackrel{\delta_2}{\hookrightarrow}\;H^2_{\Z_2}\big(\n{S}^{2,1},\Z(1)\big)\;\stackrel{f}{\simeq}\;H^2\big(\n{S}^{2},\Z\big)\;\stackrel{c_1}{\simeq}\;\Z\;
$$
with $f$  the map that forgets the $\Z_2$-action and $c_1$ the first Chern class.
\end{proposition}
\proof
To understand the homomorphism $r$ in   \eqref{eq:exac_seq_S21} one identifies the bases of these cohomology groups by \virg{Real} line bundles according to the Kahn's isomorphism \eqref{eq:iso:eq_cohom}. The sequence of group isomorphisms
\begin{equation}\label{eq:sphe_cohom_lemma_aux01}
\tilde{H}^2_{\Z_2}\big((\n{S}^{2,1})^\theta,\n{Z}(1)\big)\;\simeq\;{H}^2_{\Z_2}\big(\n{S}^{2,0},\n{Z}(1)\big)\;\simeq\;{\rm Pic}_\R\big(\n{S}^1\big)\;\simeq\;\Z_2
\end{equation}
tells  that the non-trivial element of $\tilde{H}^2_{\Z_2}((\n{S}^{2,1})^\theta,\n{Z}(1))$ can be identified with the \emph{M\"obius bundle} on $\n{S}^1$ or, equivalently, with the  \virg{Real} line bundle $\bb{L}_M=\n{S}^1\times\C$ with \virg{Real} structure
$(k,\lambda)\mapsto(k,\phi(k)\overline{\lambda})$
where $\phi:\n{S}^1\to \n{U}(1)$ is the map given by
$\phi(k):=k_1+\ii k_2$. The first isomorphism in \eqref{eq:sphe_cohom_lemma_aux01}  can be verified directly by looking at the construction of the reduced cohomology groups (see \eg \cite[eq. 5.10]{denittis-gomi-14}). The second isomorphism is justified by the fact that 
$(\n{S}^{2,1})^\theta$ coincides with $\n{S}^1$ endowed with the trivial involution and by \cite[Proposition 4.5]{denittis-gomi-14}.
In a similar way one has the group isomorphisms 
$$
\tilde{H}^2_{\Z_2}\big(\n{S}^{2,1},\n{Z}(1)\big)\;\simeq\;{H}^2_{\Z_2}\big(\n{S}^{2,1},\n{Z}(1)\big)
\;\simeq {\rm Pic}_{\rr{R}}\big(\n{S}^{2,1}\big)\;.
$$
The isomorphism ${\rm Pic}_{\rr{R}}(\n{S}^{2,1})
\simeq\Z$ can be realized explicitely.
Consider the map $f:{H}^2_{\Z_2}(\n{S}^{2,1},\n{Z}(1))\to H^2(\n{S}^2,\Z)$ which forgets the $\Z_2$-action. Since ${H}^1_{\Z_2}(\n{S}^{2,1},\n{Z})={H}^2_{\Z_2}(\n{S}^{2,1},\n{Z})=0$ one concludes from  \cite[Proposition 2.3]{gomi-13} that $f$ is a bijection. A non trivial representative for a generator of $H^2(\n{S}^2,\Z)\simeq{\rm Pic}_\C(\n{S}^2)$ is provided by the 
 line bundle
$\bb{L}_1\to {\n{S}}^2$  described by the family of 
\emph{Hopf projections} $k\mapsto{P}_{\rm Hopf}(k)$  
in equation \eqref{eq:proj_sphe_Q_struct_line}. We know that $\bb{L}_1$ has Chern class $c_1(\bb{L}_1)=1$. 
Moreover, $\bb{L}_1$ can be endowed with a \virg{Real} structure over $\n{S}^{2,1}$ induced by 
$$
C\; {P}_{\rm Hopf}(k_0,k_1,k_2)\; C\;=\;{P}_{\rm Hopf}(k_0,-k_1,k_2)
$$
where $C$  implements the complex conjugation on $\C^2$ (\cf Section \ref{sect:non-trivial_ex}). Set the \virg{Real} line bundle $(\bb{L}_1,C)$  as a representative for a generator of ${H}^2_{\Z_2}(\n{S}^{2,1},\n{Z}(1))$.
  The inclusion  $\imath:\n{S}^{2,0}\to\n{S}^{2,1}$, realized  
by $\imath:(k_0,k_2)\mapsto(k_0,0,k_2)$,
can be used to define  the restricted (pullback) line bundle $\imath^*\bb{L}_1\in {H}^2_{\Z_2}(\n{S}^{2,0},\n{Z}(1))$. The latter is equivalently described by the family of \virg{restricted} {Hopf projections}
$$
\n{S}^1\simeq\R/2\pi\Z\;\ni\;\omega\;\longmapsto\;
\tilde{P}_{\rm Hopf}(\omega)\;:=\;\frac{1}{2}
\left(
\begin{array}{cc}
1+\cos(\omega) & \sin(\omega) \\
\sin(\omega) & 1-\cos(\omega)
\end{array}
\right)
$$
through the parameterization $k_0\equiv\sin(\omega)$ and $k_2\equiv\cos(\omega)$.
There is a continuous section of normalized eigenvectors
$$
\R/2\pi\Z\;\ni\;\omega\;\longmapsto\;s(\omega)\;:=\;\expo{-\ii\frac{\omega}{2}}\;\left(
\begin{array}{c}
\cos\left(\frac{\omega}{2}\right)  \\
\sin\left(\frac{\omega}{2}\right) 
\end{array}
\right)\;.
$$
We notice that the pre-factor $\expo{-\ii\frac{\omega}{2}}$ is need to assure the continuity of the section. One can use $s$  to fix a global trivialization
$$
\n{S}^1\times\C\;\ni\;(\omega,\lambda)\;\stackrel{\psi}{\longmapsto}\; \lambda s(\omega) \;\in\; \imath^*\bb{L}_1\;.
$$
Since the map $\psi$ is compatible with the \virg{Real} structure of  $\bb{L}_M$ it follows   that $\imath^*\bb{L}_1\simeq\bb{L}_M$ as \virg{Real} line bundles.  This fact immediately implies that the restriction map $r$ in  \eqref{eq:exac_seq_S21} is   surjective. 
\qed

\medskip

The case $\n{S}^{2,2}$ can be studied along the same lines. Again the  fixed point set $(\n{S}^{2,2})^\theta\simeq\n{S}^1$ is a circle and the {suspension} isomorphism provides
$$
\begin{aligned}
\tilde{H}^k_{\Z_2}\big(\n{S}^{2,2},\Z(1)\big)\;&\simeq\;{H}^{k-3}_{\Z_2}\big(\{\ast\},\Z(1)\big)\;\simeq\; H^{k-3}\big(\R P^\infty,\Z(1)\big)\\
\tilde{H}^k_{\Z_2}\big((\n{S}^{2,2})^\theta,\Z(1)\big)\;&\simeq\;{H}^{k-1}_{\Z_2}\big(\{\ast\},\Z(1)\big)\;\simeq\; H^{k-2}\big(\R P^\infty,\Z(1)\big)\;.
\end{aligned}
$$
In this case the  \eqref{eq:Long_seq_red} immediately provides
\begin{equation}\label{eq:rel_cohom_S22}
H^2_{\Z_2}\big(\n{S}^{2,2}|(\n{S}^{2,2})^\theta,\n{Z}(1)\big)\;=\;0\;.
\end{equation}

\medskip

The fixed point set of $\n{S}^{3,1}$ is a two-dimensional sphere $(\n{S}^{3,1})^\theta\simeq\n{S}^2$. In this case the {suspension} isomorphism provides
$$
\begin{aligned}
\tilde{H}^k_{\Z_2}\big(\n{S}^{3,1},\Z(1)\big)\;&\simeq\;{H}^{k-3}_{\Z_2}\big(\{\ast\},\Z\big)\;\simeq\; H^{k-3}\big(\R P^\infty,\Z\big)\\
\tilde{H}^k_{\Z_2}\big((\n{S}^{3,1})^\theta,\Z(1)\big)\;&\simeq\;{H}^{k-2}_{\Z_2}\big(\{\ast\},\Z(1)\big)\;\simeq\; H^{k-2}\big(\R P^\infty,\Z(1)\big)\;
\end{aligned}
$$
and  the   exact sequence \eqref{eq:Long_seq_red}  implies
\begin{equation}\label{eq:rel_cohom_S31}
H^2_{\Z_2}\big(\n{S}^{3,1}|(\n{S}^{3,1})^\theta,\n{Z}(1)\big)\;=\;0\;.
\end{equation}
%


\section{About the equivariant cohomology of involutive tori}
\label{sect:cohom_tori}
In this section we study the  equivariant cohomology
of involutive tori of type $\n{T}^{a,b,c}$  introduced in Section \ref{sect:intro}. Note that as soon as $c\geqslant 1$ the involution on  $\n{T}^{a,b,c}$ is  free. An important computational tool is provided by the Gysin exact sequence \cite[Corollary 2.11]{gomi-13} which establishes the isomorphisms of groups 
\begin{equation}\label{eq:appC_Gysin_exact}
\begin{aligned}
H^k_{\Z_2}\big(X\times \n{S}^{2,0},\Z(j)\big)\;&\simeq\;H^k_{\Z_2}\big(X,\Z(j)\big)\;\oplus\; H^{k-1}_{\Z_2}\big(X,\Z(j)\big)\\
H^k_{\Z_2}\big(X\times \n{S}^{1,1},\Z(j)\big)\;&\simeq\;H^k_{\Z_2}\big(X,\Z(j)\big)\;\oplus\; H^{k-1}_{\Z_2}\big(X,\Z(j-1)\big)\;.\\
\end{aligned}
\end{equation}

\subsection{The free-involution cases}
When $c>0$ the space $\T^{a,b,c}$ has a free involution. Due to Proposition \ref{prop:cohom_Tab1-iso} we can focus our attention on the case $c=1$.
\begin{proposition}\label{prop:cohom_Tab1}
For each $a,b\in\N\cup\{0\}$  there is a  group isomorphism
$$
H_{\Z_2}^2\big(\n{T}^{a,b,1},\Z(1)\big)\;\simeq\;{\Z_2}^a\;\oplus\;\Z^{(a+1)b}\;.
$$
\end{proposition}
\proof
An iterated use of the  Gysin exact sequences \eqref{eq:appC_Gysin_exact} provides
$$
\begin{aligned}
H_{\Z_2}^2\big(\n{T}^{a,b,1},\Z(1)\big)\;\simeq\;H_{\Z_2}^2\big(\n{T}^{0,b,1},\Z(1)\big)\;\oplus\;
H_{\Z_2}^1\big(\n{T}^{0,b,1},\Z(1)\big)^{\oplus a}\;\oplus\;
H_{\Z_2}^0\big(\n{T}^{0,b,1},\Z(1)\big)^{\oplus\binom{a}{2}}
\end{aligned}
$$
where  $H^{-j}_{\Z_2}(X,\Z(1))\equiv0$ by definition. Similarly, one has
$$
\begin{aligned}
H_{\Z_2}^2\big(\n{T}^{0,b,1},\Z(1)\big)\;&\simeq\;H_{\Z_2}^2\big(\n{S}^{0,2},\Z(1)\big)\;\oplus\;
H_{\Z_2}^1\big(\n{S}^{0,2},\Z\big)^{\oplus b}\;\oplus\;
H_{\Z_2}^0\big(\n{S}^{0,2},\Z(1)\big)^{\oplus\binom{b}{2}}\\
H_{\Z_2}^1\big(\n{T}^{0,b,1},\Z(1)\big)\;&\simeq\;H_{\Z_2}^1\big(\n{S}^{0,2},\Z(1)\big)\;\oplus\;
H_{\Z_2}^0\big(\n{S}^{0,2},\Z\big)^{\oplus b}\\
H_{\Z_2}^0\big(\n{T}^{0,b,1},\Z(1)\big)\;&\simeq\;H_{\Z_2}^0\big(\n{S}^{0,2},\Z(1)\big)
\end{aligned}
$$
due to  the equality $\n{T}^{0,0,1}=\n{S}^{0,2}$. The 
cohomology of $\n{S}^{0,2}$ has been computed in  Section \ref{set:antipodal}.
\qed

\medskip

\noindent
In low dimension one has:
 \begin{center}
 \vspace{-3mm}
 \begin{table}[h]
 \begin{tabular}{|c||c|c|c|}
\hline
 \rule[-3mm]{0mm}{9mm}
 $H_{\Z_2}^2(\n{T}^{a,b,1},\Z(1))$ & $b=0$  & $b=1$ & $b=2$  \\
\hline
 \hline
 \rule[-3mm]{0mm}{9mm}
 $a=0$& $0$ & ${\Z}$ & $\Z^2$  \\
\hline
 \rule[-3mm]{0mm}{9mm}
$a=1$ & $\Z_2$ & $\Z_2\oplus\Z^2$ & $\Z_2\oplus\Z^4$  \\
\hline
  \rule[-3mm]{0mm}{9mm}
$a=2$ & ${\Z_2}^2$ & ${\Z_2}^2\oplus\Z^3$ & ${\Z_2}^2\oplus\Z^6$  \\
\hline
 \end{tabular}
 \end{table}
 \end{center}

\subsection{The cases with non-empty fixed point sets}
The condition $c=0$ implies that $\n{T}^{a,b,0}$
has  fixed points. Let us start with  the two-dimensional torus $\n{T}^{1,1,0}$ which has a fixed point set given by the disjoint union of two circles, $(\n{T}^{1,1,0})^\tau\simeq\n{S}^{2,0}\sqcup\n{S}^{2,0}$. A repeated use of the Gysin exact sequences \eqref{eq:appC_Gysin_exact} provides
$$
\begin{aligned}
&H^k_{\Z_2}\big(\n{T}^{1,1,0},\Z(j)\big)\;\simeq\;
H^k_{\Z_2}\big(\n{S}^{2,0},\Z(j)\big)\;\oplus\;H^{k-1}_{\Z_2}\big(\n{S}^{2,0},\Z(j-1)\big)\\
&\simeq\;
H^k_{\Z_2}\big(\{\ast\},\Z(j)\big)\;\oplus\;H^{k-1}_{\Z_2}\big(\{\ast\},\Z(j)\big)\;\oplus\;H^{k-1}_{\Z_2}\big(\{\ast\},\Z(j-1)\big)\;\oplus\;H^{k-2}_{\Z_2}\big(\{\ast\},\Z(j-1)\big)\;.
\end{aligned}
$$
The equivariant cohomology of the fixed point set  is computed by
$$
\begin{aligned}
H^k_{\Z_2}\big((\n{T}^{1,1,0})^\tau,\Z(j)\big)\;&\simeq\;
H^k_{\Z_2}\big(\n{S}^{2,0},\Z(j)\big)\;\oplus\;H^{k}_{\Z_2}\big(\n{S}^{2,0},\Z(j)\big)\\
&\simeq\;
H^k_{\Z_2}\big(\{\ast\},\Z(j)\big)^{\oplus 2}\;\oplus\;H^{k-1}_{\Z_2}\big(\{\ast\},\Z(j)\big)^{\oplus 2}\;.
\end{aligned}
$$
These computations allow to extract from \eqref{eq:Long_seq} the  exact sequence 
\begin{equation}\label{eq:exac_seq_T11}
\begin{diagram}
    {\Z_2}^2&\rTo^{}&        
H^2_{\Z_2}\big(\n{T}^{1,1,0}|(\n{T}^{1,1,0})^\tau,\n{Z}(1)\big)&\rTo^{\delta_2}&   
{H}^2_{\Z_2}\big(\n{T}^{1,1,0},\n{Z}(1)\big)&\rTo^{r}&  
{H}^2_{\Z_2}\big((\n{T}^{1,1,0})^\tau,\n{Z}(1)\big)\;. \vspace{-0.0mm} \\
    &&\rotatebox{-90}{$\simeq$} && \rotatebox{-90}{$\simeq$} && \rotatebox{-90}{$\simeq$}   \\
&&? && {\Z_2}\oplus\Z &&{\Z_2}^2   \\                                    \end{diagram}
\end{equation}

\begin{proposition}\label{lemma:surj_map_r_torus_2Z}
The map $r$ in \eqref{eq:exac_seq_T11} is surjective and consequently there is an isomorphism of groups
\begin{equation}\label{eq:cohom_two_toriA}
H^2_{\Z_2}\big(\n{T}^{1,1,0}|(\n{T}^{1,1,0})^\tau,\n{Z}(1)\big)\;\simeq\;2\Z\;
\end{equation}
which is induced by the composition
\begin{equation}\label{eq:cohom_two_toriB}
H^2_{\Z_2}\big(\n{T}^{1,1,0}|(\n{T}^{1,1,0})^\tau,\n{Z}(1)\big)\;\stackrel{\delta_2}{\hookrightarrow}\;H^2_{\Z_2}\big(\n{T}^{1,1,0},\n{Z}(1)\big)\;\stackrel{f}{\longrightarrow}\;H^2\big(\n{T}^{2},\Z\big)\;\stackrel{c_1}{\simeq}\;\Z\;
\end{equation}
where $f$ is the map that forgets the $\Z_2$-action and $c_1$ is the first Chern class.
\end{proposition}
\proof
The determination of $H^2_{\Z_2}(\n{T}^{1,1,0}|(\n{T}^{1,1,0})^\tau,\n{Z}(1))$ requires an accurate knowledge of the homomorphism
$r$ in the exact sequence \eqref{eq:exac_seq_T11}. Let us start with the  $r$ in degree 1. In this case we can use the geometric identification \eqref{eq:iso:eq_cohom} to write
$$
{{H}^1_{\Z_2}\big(\n{T}^{1,1,0},\n{Z}(1)\big)}\;\simeq\;\big[\n{T}^{1,1,0}, \n{S}^{1,1}\big]_{\Z_2}\;\stackrel{r}{\longrightarrow}\;\big[(\n{T}^{1,1,0})^\tau, \n{S}^{1,1}\big]_{\Z_2}\;\simeq\;{{H}^1_{\Z_2}\big((\n{T}^{1,1,0})^\tau,\n{Z}(1)\big)}\;.
$$
A basis for 
$$
\big[(\n{T}^{1,1,0})^\tau, \n{S}^{1,1}\big]_{\Z_2}\;\simeq\;\big[\n{S}^{2,0}, \n{S}^{1,1}\big]_{\Z_2}\;\oplus\; \big[\n{S}^{2,0}, \n{S}^{1,1}\big]_{\Z_2}\;\simeq\;{\Z_2}^2
$$ 
is provided by two copies of the constant  map $g:\n{S}^{2,0}\to (-1,0)\in \n{S}^{1,1}$. A basis for 
\begin{equation}\label{eq:AABBCC}
\big[\n{T}^{1,1,0}, \n{S}^{1,1}\big]_{\Z_2}\;\simeq\;\big[\n{S}^{1,1}, \n{S}^{1,1}\big]_{\Z_2}\;\simeq\;{\Z_2}\;\oplus\;\Z
\end{equation}
is given by the constant  map $\tilde{g}:\n{S}^{2,0}\times\n{S}^{1,1}\to (-1,0)\in \n{S}^{1,1}$ for the $\Z_2$-summand 
and by the projection $\pi_2:\n{S}^{2,0}\times\n{S}^{1,1}\to  \n{S}^{1,1}$ for the $\Z$-summand. 
Note that the first isomorphism in \eqref{eq:AABBCC} is induced by
$H^1_{\Z_2}(\n{T}^{1,1,0},\n{Z}(1))\simeq H^1_{\Z_2}(\n{S}^{1,1},\n{Z}(1))$ which follows from the use of the first Gysin exact sequence \eqref{eq:appC_Gysin_exact}.
This analysis shows that the map $r$ is surjective in degree 1 and so the exact sequence \eqref{eq:exac_seq_T11} can be rewritten as
\begin{equation}\label{eq:exac_seq_T11_modifix}
0\;\longrightarrow\;H^2_{\Z_2}\big(\n{T}^{1,1,0}|(\n{T}^{1,1,0})^\tau,\n{Z}(1)\big)\;\stackrel{\delta_2}{\longrightarrow}\;{\Z_2}\;\oplus\;\Z\;\stackrel{r}{\longrightarrow}\;{ {\Z_2}^2}\;.
\end{equation}
In particular $\delta_2$ turns out to be an injection.
The Kahn's isomorphism provides an easy way to describe the basis of 
$$
{H}^2_{\Z_2}\big(\n{T}^{1,1,0},\n{Z}(1)\big)\;\simeq\;{\rm Pic}_{\rr{R}}\big(\n{S}^{2,0}\times\n{S}^{1,1}\big)\;\simeq\;{\Z_2}\;\oplus\;\Z\;.
$$
The $\Z_2$-summand in ${\rm Pic}_{\rr{R}}\big(\n{S}^{2,0}\times\n{S}^{1,1}\big)$ is given by the pullback under the projection $\pi_1:\n{S}^{2,0}\times\n{S}^{1,1}\to  \n{S}^{2,0}$ of the \virg{Real} M\"{o}bius bundle $\bb{L}_M\in {\rm Pic}_{\rr{R}}(\n{S}^{2,0})$ (\cf the proof of Proposition \ref{lemma:surj_map_r_2Z}). The construction of the generator of the $\Z$-summand requires a little of work. Let $\n{U}(1)\times\R\times\C\to \n{U}(1)\times\R$ be the trivial complex line bundle over $\n{U}(1)\times\R$ endowed with a $\Z$-action $\alpha_n:(z,x,\lambda)\mapsto(z,x+n,z^n\lambda)$ for all $n\in\Z$. Since the  $\Z$-action is free on the base space one can build the \emph{quotient} line bundle 
$\bb{L}_{1,1}:=(\n{U}(1)\times\R\times\C)/\alpha$ over $\n{U}(1)\times(\R/\Z)\simeq\T^2$.
This line bundle $\bb{L}_{1,1}\to\n{T}^2$ has Chern class $c_1(\bb{L}_{1,1})=1$. Moreover, it can be endowed 
with the \virg{Real} structure $\Theta:[(z,x,\lambda)]\mapsto [(z,-x,\overline{\lambda})]$ which converts $\bb{L}_{1,1}$ into a non-trivial element  of ${\rm Pic}_{\rr{R}}(\n{S}^{2,0}\times\n{S}^{1,1})$ with Chern class 1. 
This explicit construction proves, in particular,  that the map $f$ in \eqref{eq:cohom_two_toriB} is surjective.
A basis for 
$$
{{H}^2_{\Z_2}\big((\n{T}^{1,1,0})^\tau,\n{Z}(1)\big)}\;\simeq\;{\rm Pic}_{\rr{R}}\big(\n{S}^{2,0}\big)\;\oplus\;{\rm Pic}_{\rr{R}}\big(\n{S}^{2,0}\big)\;\simeq\; { {\Z_2}^2}
$$
is given by two copies of the M\"{o}bius bundle $\bb{L}_M$. 
At this point it is enough to note  that $\bb{L}_{1,1}$ restricts to the trivial \virg{Real} bundle on $\n{S}^{2,0}\times\{0\}\subset \n{T}^{1,1,0}$ and to the M\"{o}bius bundle on $\n{S}^{2,0}\times\{\frac{1}{2}\}\subset \n{T}^{1,1,0}$. This 
fact means  the map $r$ in \eqref{eq:exac_seq_T11_modifix} is surjective, hence 
$$
H^2_{\Z_2}\big(\n{T}^{1,1,0}|(\n{T}^{1,1,0})^\tau,\n{Z}(1)\big)\;\simeq\;{\rm Ker}\Big[r:{H}^2_{\Z_2}\big(\n{T}^{1,1,0},\n{Z}(1)\big)\to{H}^2_{\Z_2}\big((\n{T}^{1,1,0})^\tau,\n{Z}(1)\big)\Big]\;\simeq\;2\Z\;.
$$ 
Finally, the injectivity of $\delta_2$ and the surjectivity of $f$
assure that  the isomorphism  \eqref{eq:cohom_two_toriA} is realized by the composition of maps in \eqref{eq:cohom_two_toriB}.\qed

\medskip

A similar result can be proved for the torus  $\n{T}^{2,1,0}$. Let us start again with the exact sequence
\begin{equation}\label{eq:exac_seq_T21}
\begin{diagram}
    {\Z_2}^2&\rTo^{}&        
H^2_{\Z_2}\big(\n{T}^{2,1,0}|(\n{T}^{2,1,0})^\tau,\n{Z}(1)\big)&\rTo^{\delta_2}&   
{H}^2_{\Z_2}\big(\n{T}^{2,1,0},\n{Z}(1)\big)&\rTo^{r}&  
{H}^2_{\Z_2}\big((\n{T}^{2,1,0})^\tau,\n{Z}(1)\big)\;. \vspace{-0.0mm} \\
    &&\rotatebox{-90}{$\simeq$} && \rotatebox{-90}{$\simeq$} && \rotatebox{-90}{$\simeq$}   \\
&&? && {\Z_2}^2\oplus\Z^2 &&{\Z_2}^4   \\                                    \end{diagram}
\end{equation}
The computation of the cohomology groups follows by a  repeated use of the Gysin exact sequences \eqref{eq:appC_Gysin_exact}
along with the fact that
 the fixed point set $(\n{T}^{2,1,0})^\tau\simeq\n{T}^{2,0,0}\sqcup \n{T}^{2,0,0}$ is the disjoint union of two  two-dimensional tori.

\begin{proposition}\label{lemma:surj_map_r_torus_2Z_210type}
The map $r$ in \eqref{eq:exac_seq_T21} is surjective and consequently there is an isomorphism of groups
\begin{equation}\label{eq:cohom_three210_toriA}
H^2_{\Z_2}\big(\n{T}^{2,1,0}|(\n{T}^{2,1,0})^\tau,\n{Z}(1)\big)\;\simeq\;(2\Z)^2\;
\end{equation}
which is induced by the composition
\begin{equation}\label{eq:cohom_three210_toriB}
H^2_{\Z_2}\big(\n{T}^{2,1,0}|(\n{T}^{2,1,0})^\tau,\n{Z}(1)\big)\;\stackrel{\delta_2}{\hookrightarrow}\;H^2_{\Z_2}\big(\n{T}^{2,1,0},\n{Z}(1)\big)\;\stackrel{f}{\longrightarrow}\;H^2\big(\n{T}^{3},\Z\big)\;\stackrel{c_1}{\simeq}\;\Z^3\;
\end{equation}
where $f$ is the map that forgets the $\Z_2$-action and $c_1$ is the first Chern class.
\end{proposition}
\proof
The proof follows the same arguments as in Proposition \ref{lemma:surj_map_r_torus_2Z}.
One needs information about  the homomorphism
$r$ in degree 1 and 2. In degree 1 the use of the geometric identification \eqref{eq:iso:eq_cohom} 
provides
$$
{{H}^1_{\Z_2}\big(\n{T}^{2,1,0},\n{Z}(1)\big)}\;\simeq\;\big[\n{T}^{2,1,0}, \n{S}^{1,1}\big]_{\Z_2}\;\stackrel{r}{\longrightarrow}\;\big[(\n{T}^{2,1,0})^\tau, \n{S}^{1,1}\big]_{\Z_2}\;\simeq\;{{H}^1_{\Z_2}\big((\n{T}^{2,1,0})^\tau,\n{Z}(1)\big)}\;.
$$
A basis for 
$$
\big[(\n{T}^{2,1,0})^\tau, \n{S}^{1,1}\big]_{\Z_2}\;\simeq\;\big[\n{T}^{2,0,0}, \n{S}^{1,1}\big]_{\Z_2}\;\oplus\; \big[\n{T}^{2,0,0}, \n{S}^{1,1}\big]_{\Z_2}\;\simeq\;{\Z_2}^2
$$ 
is given by two copies of the constant  map $g:\n{T}^{2,0,0}\to (-1,0)\in \n{S}^{1,1}$.
A basis for 
\begin{equation}\label{eq:XXZZ}
\big[\n{T}^{2,
1,0}, \n{S}^{1,1}\big]_{\Z_2}\;\simeq\;\big[\n{S}^{1,1}, \n{S}^{1,1}\big]_{\Z_2}\;\simeq\;{\Z_2}\;\oplus\;\Z
\end{equation}
is given by the constant  map $\tilde{g}:\n{T}^{2,0,0}\times\n{S}^{1,1}\to -1\in \n{S}^{1,1}$ for the $\Z_2$-summand 
and by the projection $\pi_3:\n{T}^{2,0,0}\times\n{S}^{1,1}\to  \n{S}^{1,1}$ for the $\Z$-summand. 
Notice that the first isomorphism in \eqref{eq:XXZZ} is consequence of
$H^1_{\Z_2}(\n{T}^{2,1,0},\n{Z}(1))\simeq H^1_{\Z_2}(\n{S}^{1,1},\n{Z}(1))$ which follows from the use of the first Gysin exact sequence \eqref{eq:appC_Gysin_exact}.
This analysis says that the map $r$ is surjective in degree 1 and  the exact sequence \eqref{eq:exac_seq_T21} assumes the short form
\begin{equation}\label{eq:exac_seq_T21_modifi}
0\;\longrightarrow\;H^2_{\Z_2}\big(\n{T}^{2,1,0}|(\n{T}^{2,1,0})^\tau,\n{Z}(1)\big)\;\stackrel{\delta_2}{\longrightarrow}\;{\Z_2}^2\;\oplus\;\Z^2\;\stackrel{r}{\longrightarrow}\;{ {\Z_2}^4}\;.
\end{equation}
In particular $\delta_2$ is an injection.
In degree 2 we can use the Kahn's isomorphism
$$
{H}^2_{\Z_2}\big(\n{T}^{2,1,0},\n{Z}(1)\big)\;\simeq\;{\rm Pic}_{\rr{R}}\big(\n{S}^{2,0}\times\n{S}^{2,0}\times\n{S}^{1,1}\big)\;\simeq\;{\Z_2}^2\;\oplus\;\Z^2\;.
$$
The two $\Z_2$-summands  are given by the pullback under the first and second projection $\pi_j:\n{S}^{2,0}\times\n{S}^{2,0}\times\n{S}^{1,1}\to  \n{S}^{2,0}$ of the \virg{Real} M\"{o}bius bundle $\bb{L}_M\in {\rm Pic}_{\rr{R}}(\n{S}^{2,0})$ (\cf  the proof of Proposition \ref{lemma:surj_map_r_2Z}). 
The two $\Z$-summands  are given by the pullback under the two different projections $\tilde{\pi}_j:\n{S}^{2,0}\times\n{S}^{2,0}\times\n{S}^{1,1}\to  \n{S}^{2,0}\times\n{S}^{1,1}$ of the line bundle $\bb{L}_{1,1}\to\n{S}^{2,0}\times\n{S}^{1,1}$ constructed in the proof of Proposition
\ref{lemma:surj_map_r_torus_2Z}. This explicit construction also shows that the  $f$ in \eqref{eq:cohom_three210_toriB} restricts to a bijection from the free part $\Z^2$
of ${H}^2_{\Z_2}(\n{T}^{2,1,0},\n{Z}(1))$ to a direct summand $\Z^2$ in $H^2(\T^3,\Z)\simeq\Z^3$.
A basis for 
$$
{{H}^2_{\Z_2}\big((\n{T}^{2,1,0})^\tau,\n{Z}(1)\big)}\;\simeq\;{\rm Pic}_{\rr{R}}\big(\n{T}^{2,0,0}\big)\;\oplus\;{\rm Pic}_{\rr{R}}\big(\n{T}^{2,0,0}\big)\;\simeq\; { {\Z_2}^4}
$$
is given by two copies of $\pi_j^*\bb{L}_M$ for each $j=1,2$. 
At this point the same argument in the proof of Proposition
\ref{lemma:surj_map_r_torus_2Z} can be adapted to  show that 
the map $r$ in \eqref{eq:exac_seq_T21} is surjective, hence  
$$
H^2_{\Z_2}\big(\n{T}^{2,1,0}|(\n{T}^{2,1,0})^\tau,\n{Z}(1)\big)\;\simeq\;{\rm Ker}\Big[r:{H}^2_{\Z_2}\big(\n{T}^{2,1,0},\n{Z}(1)\big)\to{H}^2_{\Z_2}\big((\n{T}^{2,1,0})^\tau,\n{Z}(1)\big)\Big]\;\simeq\;(2\Z)^2\;.
$$ 
 Finally, the injectivity of $\delta_2$ and the properties of  $f$
assure that the isomorphism  \eqref{eq:cohom_two_toriA} is realized by the sequence of maps in \eqref{eq:cohom_three210_toriB}.\qed

\medskip

Finally, let us investigate the involutive torus  $\n{T}^{1,2,0}$. In this case the exact sequence reads
\begin{equation}\label{eq:exac_seq_T12}
\begin{diagram}
    {\Z_2}^4&\rTo^{}&        
H^2_{\Z_2}\big(\n{T}^{1,2,0}|(\n{T}^{1,2,0})^\tau,\n{Z}(1)\big)&\rTo^{\delta_2}&   
{H}^2_{\Z_2}\big(\n{T}^{1,2,0},\n{Z}(1)\big)&\rTo^{r}&  
{H}^2_{\Z_2}\big((\n{T}^{1,2,0})^\tau,\n{Z}(1)\big)\;. \vspace{-0.0mm} \\
&&\rotatebox{-90}{$\simeq$} && \rotatebox{-90}
    {$\simeq$} && \rotatebox{-90}{$\simeq$}   \\
&&? && {\Z_2}\oplus\Z^2 &&{\Z_2}^4   \\                                    \end{diagram}
\end{equation}
The computation of the cohomology groups follows from a  repeated use of  the Gysin exact sequences \eqref{eq:appC_Gysin_exact}
along with the fact
 that 
 $$
 (\n{T}^{1,2,0})^\tau\;\simeq\;\n{S}^{2,0}\;\sqcup\;\n{S}^{2,0}\;\sqcup\;\n{S}^{2,0}\;\sqcup\;\n{S}^{2,0}\;.
 $$

\begin{proposition}\label{lemma:surj_map_r_torus_2Z_120type}
There is an isomorphism of groups
\begin{equation}\label{eq:cohom_three120_toriA}
H^2_{\Z_2}\big(\n{T}^{1,2,0}|(\n{T}^{1,2,0})^\tau,\n{Z}(1)\big)\;\simeq\;\Z_2\;\oplus\;(2\Z)^2\;   
\end{equation}
which is given by the direct sum of
$$
\begin{aligned}
{\rm Coker}\Big[r:{H}^1_{\Z_2}\big(\n{T}^{1,2,0},\n{Z}(1)\big)\to{H}^1_{\Z_2}\big((\n{T}^{1,2,0})^\tau,\n{Z}(1)\big)\Big]\;&\simeq\;\Z_2
\\
{\rm Ker}\Big[r:{H}^2_{\Z_2}\big(\n{T}^{1,2,0},\n{Z}(1)\big)\to{H}^2_{\Z_2}\big((\n{T}^{1,2,0})^\tau,\n{Z}(1)\big)\Big]\;&\simeq\;(2\Z)^2\;.
\end{aligned}
$$
Moreover, in the sequence of maps 
\begin{equation}\label{eq:cohom_three120_toriB}
H^2_{\Z_2}\big(\n{T}^{1,2,0}|(\n{T}^{1,2,0})^\tau,\n{Z}(1)\big)\;\stackrel{\delta_2}{\longrightarrow}\;H^2_{\Z_2}\big(\n{T}^{1,2,0},\n{Z}(1)\big)\;\stackrel{f}{\longrightarrow}\;H^2\big(\n{T}^{3},\Z\big)\;\stackrel{c_1}{\simeq}\;\Z^3\;
\end{equation}
$\delta_2$ is an injection of the free part of 
$H^2_{\Z_2}(\n{T}^{1,2,0}|(\n{T}^{1,2,0})^\tau,\n{Z}(1))$ into the free part of $H^2_{\Z_2}(\n{T}^{1,2,0},\n{Z}(1))$ and $f$ is a bijection between the free part of $H^2_{\Z_2}(\n{T}^{1,2,0},\n{Z}(1))$ and a direct $\Z^2$-summand in $H^2(\n{T}^{3},\Z)$.
\end{proposition}
\proof
Let us study  the homomorphism
$r$ in degree 1 and 2. In degree 1 we have the geometric identification 
$$
{{H}^1_{\Z_2}\big(\n{T}^{1,2,0},\n{Z}(1)\big)}\;\simeq\;\big[\n{T}^{1,2,0}, \n{S}^{1,1}\big]_{\Z_2}\;\stackrel{r}{\longrightarrow}\;\big[(\n{T}^{1,2,0})^\tau, \n{S}^{1,1}\big]_{\Z_2}\;\simeq\;{{H}^1_{\Z_2}\big((\n{T}^{1,2,0})^\tau,\n{Z}(1)\big)}\;.
$$
A basis for 
$$
\big[(\n{T}^{1,2,0})^\tau, \n{S}^{1,1}\big]_{\Z_2}\;\simeq\;\big[\n{S}^{2,0}, \n{S}^{1,1}\big]_{\Z_2}^{\oplus 4}\;\simeq\;{\Z_2}^4
$$ 
is provided by four copies of the constant  map $g:\n{S}^{2,0}\to (-1,0)\in \n{S}^{1,1}$.
A basis for 
\begin{equation}\label{eq:XXZZyy}
\big[\n{T}^{1,
2,0}, \n{S}^{1,1}\big]_{\Z_2}\;\simeq\;H^1\big(\n{S}^{1,1},\Z(1)\big)\;\oplus\;H^0\big(\n{S}^{1,1},\Z\big)\;\simeq\;{\Z_2}\;\oplus\;\Z^2
\end{equation}
is given by the constant  map $\tilde{g}:\n{T}^{1,2,0}\to (-1,0)\in \n{S}^{1,1}$ for the $\Z_2$-summand (which comes from $H^1(\n{S}^{1,1},\Z(1))$)
and by the two distinct projections $\n{S}^{2,0}\times\n{S}^{1,1}\times\n{S}^{1,1}\to  \n{S}^{1,1}$ for the two $\Z$-summands. This proves that
$$
{\rm Coker}\Big[r:{H}^1_{\Z_2}\big(\n{T}^{1,2,0},\n{Z}(1)\big)\to{H}^1_{\Z_2}\big((\n{T}^{1,2,0})^\tau,\n{Z}(1)\big)\Big]\;\simeq\;\Z_2
$$
 and  the exact sequence \eqref{eq:exac_seq_T12} assumes the short expression
\begin{equation}\label{eq:exac_seq_T12_modifi}
0\;\longrightarrow\;\Z_2\;\longrightarrow\;H^2_{\Z_2}\big(\n{T}^{1,2,0}|(\n{T}^{1,2,0})^\tau,\n{Z}(1)\big)\;\stackrel{\delta_2}{\longrightarrow}\;{\Z_2}\;\oplus\;\Z^2\;\stackrel{r}{\longrightarrow}\;{ {\Z_2}^4}\;.
\end{equation}
Before to describe the degree 2 let us note that the forgetting map $f:{H}^1_{\Z_2}(\n{T}^{1,2,0},\n{Z}(1))\to {H}^1(\n{T}^{3},\n{Z})$ induces a bijection from the free part
$\Z^2$ of ${H}^1_{\Z_2}(\n{T}^{1,2,0},\n{Z}(1))$ 
to a $\Z^2$-summand in ${H}^1(\n{T}^{3},\n{Z})\simeq\Z^3$. In fact this last $\Z^2$-summand turns out to be generated by
the two projections $\n{S}^{2,0}\times\n{S}^{1,1}\times\n{S}^{1,1}\to  \n{S}^{1,1}$ 
upon forgetting the $\Z_2$-action.
In degree 2  the Kahn's isomorphism provides
$$
{H}^2_{\Z_2}\big(\n{T}^{1,2,0},\n{Z}(1)\big)\;\simeq\;{\rm Pic}_{\rr{R}}\big(\n{S}^{2,0}\times\n{S}^{1,1}\times\n{S}^{1,1}\big)\;\simeq\;{\Z_2}\;\oplus\;\Z^2\;.
$$
The  $\Z_2$-summand  is generated by the pullback of the \virg{Real} M\"{o}bius bundle $\bb{L}_M\in {\rm Pic}_{\rr{R}}(\n{S}^{2,0})$  under the projection $\n{S}^{2,0}\times\n{S}^{1,1}\times\n{S}^{1,1}\to\n{S}^{2,0}$. In fact 
$
{H}^2_{\Z_2}(\n{S}^{2,0},\n{Z}(1))
$
is contained in ${H}^2_{\Z_2}(\n{T}^{1,2,0},\n{Z}(1))$ as a direct summand by the splitting of the Gysin sequence \eqref{eq:appC_Gysin_exact}.
The two $\Z$-summands are generated by the pullbacks of $\bb{L}_{1,1}\in {\rm Pic}_{\rr{R}}(\n{T}^{1,1,0})$ under the two distinct projections $\n{T}^{1,2,0}\to \n{T}^{1,1,0}$.
By an exact sequence argument one can show that the forgetting map $f:{H}^2_{\Z_2}(\n{T}^{1,2,0},\n{Z}(1))\to {H}^2(\n{T}^{3},\n{Z})$ 
induces an
injection from the free part $\Z^2$ of ${H}^2_{\Z_2}(\n{T}^{1,2,0},\n{Z}(1))$ into 
$H^2(\n{T}^{3},\n{Z})\simeq\Z^3$. Actually, this injection is a bijection from the free part of ${H}^2_{\Z_2}(\n{T}^{1,2,0},\n{Z}(1))$ to a direct $\Z^2$-summand in $H^2(\n{T}^{3},\n{Z})$.
In fact this $\Z^2$-summand can be generated by the pullbacks of $\bb{L}_{1,1}\in {\rm Pic}_{\rr{R}}(\n{T}^{1,1,0})$ under the two distinct projections $\n{T}^{1,2,0}\to \n{T}^{1,1,0}$
upon forgetting the $\Z_2$-action.
A basis for 
$$
{{H}^2_{\Z_2}\big((\n{T}^{1,2,0})^\tau,\n{Z}(1)\big)}\;\simeq\;{\rm Pic}_{\rr{R}}\big(\n{S}^{2,0}\big)^{\oplus 4}\;\simeq\; { {\Z_2}^4}
$$
is given by four copies of the pullback of the line bundle $\bb{L}_M$ under the four distinct projections 
$(\n{T}^{1,2,0})^\tau\to \n{S}^{2,0}$. This analysis shows that
$$
{\rm Ker}\Big[r:{H}^2_{\Z_2}\big(\n{T}^{1,2,0},\n{Z}(1)\big)\to{H}^2_{\Z_2}\big((\n{T}^{1,2,0})^\tau,\n{Z}(1)\big)\Big]\;\simeq\;(2\Z)^2\;.
$$
The conclusions above imply that the group $H^2_{\Z_2}(\n{T}^{1,2,0}|(\n{T}^{1,2,0})^\tau,\n{Z}(1))$ fits in the exact sequence
\begin{equation}\label{eq:exac_seq_T12_modifiII}
0\;\longrightarrow\;\Z_2\;\longrightarrow\;H^2_{\Z_2}\big(\n{T}^{1,2,0}|(\n{T}^{1,2,0})^\tau,\n{Z}(1)\big)\;\stackrel{}{\longrightarrow}\;(2\Z)^2\;\stackrel{}{\longrightarrow}\;0
\end{equation}
which is splitting since $(2\Z)^2$ is a free group and this concludes the proof.
\qed


\end{document}